\documentclass[12pt]{article}

\usepackage{epsfig}
\usepackage{psfrag}
\usepackage{latexsym}
\usepackage[DIV13]{typearea}
\usepackage{amsmath}
\usepackage{amssymb}
\usepackage{amsfonts}
\usepackage{bbold}
\usepackage[footnotesize]{caption2}
\usepackage{graphicx}
\usepackage[center,footnotesize,hang]{subfigure}
\usepackage{url}
\usepackage{color}
\usepackage{cite}
\usepackage{pstricks}
\usepackage{inputenc}
\usepackage{pst-coil}
\usepackage{verbatim}
\usepackage{bbm}
\newpsobject{grilla}{psgrid}{subgriddiv=1,griddots=10,gridlabels=6pt}

\newcommand{\al}{\alpha}
\newcommand{\be}{\beta}

\newcommand{\De}{\Delta}

\newcommand{\la}{\lambda}
\newcommand{\La}{\Lambda}

\newcommand{\sig}{\sigma}
\newcommand{\vphi}{\varphi}

\def\cC{{\cal C}}

 \def\cM{{\cal M}}
 \def\cO{{\cal O}}

\def\ov{\overline}

\def\5{\overline 5}

\newcommand{\diag}{\text{diag}}

\newcommand{\unity}{\mathbb{1}}
\newcommand{\nn}{\nonumber}
\newcommand{\mean}[1]{\langle#1\rangle}
\newcommand{\derp}{\partial}

\def\eq#1{{eq.~(\ref{#1})}}

\def\vev#1{\left\langle #1\right\rangle}

\def\Tr{\mbox{Tr}\,}

\def\hbar{\hspace{0pt}\raisebox{1pt}{$-$} \hspace{-7pt} h}

\newcommand{\beq}{\begin{equation}}
\newcommand{\eeq}{\end{equation}}
\newcommand{\bac}{\beq\begin{array}}
\newcommand{\eac}{\end{array}\eeq}
\newcommand{\ba}{\begin{array}}
\newcommand{\ea}{\end{array}}
\newcommand{\bea}{\begin{eqnarray}}
\newcommand{\eea}{\end{eqnarray}}

\newcommand{\bmat}{\begin{pmatrix}}
\newcommand{\emat}{\end{pmatrix}}

\begin{document}
\begin{titlepage}
\vspace*{-1.5cm}
\phantom{hep-ph/***}
\hfil{NIKHEF 2010-007}
\hfill
\hfil{DFPD-09/TH/24}
\hfill
\hfil{CERN-PH-TH/2010-025}

\vskip 1.5cm

\begin{center}
{\Large\bf The Interplay Between GUT and Flavour Symmetries}
\vskip 0.1cm
{\Large \bf  in a Pati-Salam $\mathbf{\times}$ $\mathbf{S_4}$ Model}
\end{center}

\vskip 0.5  cm
\begin{center}
 {\large Reinier de Adelhart Toorop}~$^{a)}$\footnote{e-mail address: reintoorop@nikhef.nl}
 \vskip 0.2 cm
{\large Federica Bazzocchi}~$^{b)}$\footnote{e-mail address: fbazzoc@few.vu.nl},
and {\large Luca Merlo}~$^{c)}$\footnote{e-mail address: merlo@pd.infn.it}
\\
\vskip .2cm
$^{a)}$~Nikhef Theory Group, \\
Science Park 105, 1098 XG, Amsterdam, The Netherlands
\\
\vskip .2cm
$^{b)}$~Department of Physics and Astronomy, Vrije Universiteit Amsterdam,\\
1081 HV Amsterdam, The Netherlands
\\
\vskip .1cm
$^{c)}$~Dipartimento di Fisica `G.~Galilei', Universit\`a di Padova
\\
INFN, Sezione di Padova, Via Marzolo~8, I-35131 Padua, Italy
\\
and\\
CERN, Department of Physics, Theory Division\\
CH-1211 Geneva 23, Switzerland\\
\vskip .1cm
\end{center}
\vskip 0.7cm
\begin{abstract}
Both Grand Unified symmetries and discrete flavour symmetries are appealing ways to describe apparent structures in the gauge and flavour sectors of the Standard Model. Both symmetries put constraints on the high energy behaviour of the theory. This can give rise to unexpected interplay when building models that possess both symmetries. We investigate on the possibility to combine a Pati-Salam model with the discrete flavour symmetry $S_4$ that gives rise to quark-lepton complementarity. Under appropriate assumptions at the GUT scale, the model reproduces fermion masses and mixings both in the quark and in the lepton sectors. We show that in particular the Higgs sector and the running Yukawa couplings are strongly affected by the combined constraints of the Grand Unified and family symmetries. This in turn reduces the phenomenologically viable parameter space, with high energy mass scales confined to a small region and some parameters in the neutrino sector slightly unnatural. In the allowed regions, we can reproduce the quark masses and the CKM matrix. In the lepton sector, we reproduce the charged lepton masses, including bottom-tau unification and the Georgi-Jarlskog relation as well as the two known angles of the PMNS matrix. The neutrino mass spectrum can present a normal or an inverse hierarchy, and only allowing the neutrino parameters to spread into a range of values between $\la^{-2}$ and $\la^2$, with $\lambda\simeq0.2$. Finally, our model suggests that the reactor mixing angle is close to its current experimental bound.
\end{abstract}
\end{titlepage}
\setcounter{footnote}{0}
\vskip2truecm


\section{Introduction}

Gauge coupling unification suggests that the Standard Model (SM) gauge group is generated when a unified larger gauge group is broken at a very high energy scale compared to the electroweak (EW) one. In this picture SM fermions are accommodated in representations of the unified gauge group $G$ and an appropriate scalar Higgs sector is introduced both to trigger the spontaneous breaking of $G$ down to $SU(3)_C\times U(1)_{em}$ and to reproduce the fermion mass matrices. Typically Grand Unified Theories (GUTs) are quite constrained and non-trivial relations are obtained among the SM fermion mass matrices. On the other hand it is a common belief that fermion mass and mixing matrix patterns can be explained by appealing to a flavour symmetry $G_f$ and in the last years particular attention has been devoted to the study of discrete flavour symmetries, thanks to their simplicity in recovering realistic lepton mixing patterns.

\begin{table}[h!]
\begin{center}
\begin{tabular}{|l||cc|cc|}
\hline
&&&&\\[-2mm]
& \multicolumn{2}{c}{Ref.~\cite{Fogli:Indication}} & \multicolumn{2}{|c|}{Ref.~\cite{Maltoni:Indication}}\\[2mm]
parameter & best fit (\@$1\sig$) & 3$\sig$-interval & best fit (\@$1\sig)$ & 3$\sig$-interval\\[2mm]
\hline
&&&&\\
$\De m^2_{sol}\:[\times10^{-5}\mathrm{eV}^2]$
        & $7.67^{+0.16}_{-0.19}$ & $7.14-8.19$
        & $7.65^{+0.23}_{-0.20}$ & $7.05-8.34$\\[2mm]
$\De m^2_{atm}\: [\times10^{-3}\mathrm{eV}^2]$
        & $2.39^{+0.11}_{-0.08}$ & $2.06-2.81$
        & $2.40^{+0.12}_{-0.11}$ & $2.07-2.75$\\[2mm]
$\sin^2\theta^l_{12}$
        & $0.312^{+0.019}_{-0.018}$ & $0.26-0.37$
        & $0.304^{+0.022}_{-0.016}$ & $0.25-0.37$\\[2mm]
$\sin^2\theta^l_{23}$
        & $0.466^{+0.073}_{-0.058}$ & $0.331-0.644$
        & $0.50^{+0.07}_{-0.06}$ & $0.36-0.67$\\[2mm]
$\sin^2\theta^l_{13}$
        & $0.016^{+0.010}_{-0.010}$ & $\leq$ $0.046$
        & $0.010^{+0.016}_{-0.011}$ & $\leq$ $0.056$\\
&&&&\\
\hline
\end{tabular}
\end{center}
\caption{\it Neutrino oscillation parameters from two independent global fits \cite{Fogli:Indication, Maltoni:Indication}.}
\label{table:OscillationData}
\end{table}

The present neutrino oscillation data \cite{Maltoni:Indication,Fogli:Indication,NeutrinoData} are summarised in table \ref{table:OscillationData}, where we display the results of two independent global fits. The pattern of the mixings is characterised by two large angles and a small one: $\theta^l_{23}$ is compatible with a maximal value, but the accuracy admits relatively large deviations; $\theta^l_{12}$ is large, but about $5\sigma$ far from the maximal value; $\theta^l_{13}$ has only an upper bound. According to the type of the experiments which measured them, the mixing angle $\theta^l_{23}$ is called atmospheric, $\theta^l_{12}$ solar and $\theta^l_{13}$ reactor.
We underline that there is a tension among the two global fits presented in table \ref{table:OscillationData} on the central value of the reactor angle: in \cite{Fogli:Indication} we can read a suggestion for a positive value of $\sin^2\theta^l_{13}\simeq0.016\pm0.010$ [$1.6\sig$], while in \cite{Maltoni:Indication} a best fit value consistent with zero within less than $1\sig$ is found. Therefore we need a direct measurement by the future experiments like DOUBLE CHOOZ \cite{doublechooz}, Daya Bay \cite{dayabay}, MINOS \cite{minos}, RENO \cite{reno},T2K \cite{T2K} and NOvA \cite{NOvA }.

The closeness of the leptonic atmospheric angle $\theta^l_{23}$ to the maximal value gives relevant indications on the flavour symmetry: it is well known \cite{LV_Theorem,FeruglioSymBreaking} that a maximal $\theta^l_{23}$ is not achievable with an exact realistic symmetry. This forces to study models based on the breaking of the flavour symmetry and a promising choice is the kind of realizations based on non-Abelian discrete groups which   reproduce the lepton mixing matrix of the so-called tribimaximal (TB) pattern \cite{HPS} ($\sin^2\theta^{TB}_{12}=1/3$, $\sin^2\theta^{TB}_{23}=1/2$ and $\sin\theta^{TB}_{13}=0$) at leading order. This mixing scheme represents a very good approximation of the experimental measurements:  the TB values for the atmospheric and the reactor angles are inside the $1\sigma$ error level, while that one for the solar angle is very close to the upper $1\sigma$ value. The corrections from the symmetry breaking provide perturbations to the angles and in particular a deviation from zero for the reactor angle, in agreement with the recent indication of a positive value for $\theta^l_{13}$ \cite{Fogli:Indication}.
Despite this success of the TB mixing scheme, there are some complications, in particular when combining the flavour symmetry with GUTs. In the following, we will comment on this and motivate the viability of a competing scheme, the bimaximal (BM) mixing pattern \cite{BMmixing}.

A lot of effort has been put in reproducing the TB pattern by the use of non-Abelian discrete symmetries: the best known groups  implemented in the construction of flavour models are $A_4$ \cite{A4II,A4I}, $S_4$ \cite{S4}, $T'$ \cite{Tp} and $\Delta(27)$ \cite{SymmDelta27}. A common feature among many of these realizations is to get a spontaneous breaking scheme responsible for the TB mixing by the use of a convenient assignment of the quantum numbers to the SM particles and the introduction of a suitable set of scalar fields, the ``flavons'', which, getting non-zero vacuum expectation values (VEVs), are responsible for the symmetry breaking of $G_f$. A central aspect of the model building is the symmetry breaking chain: the flavour group is broken down to two distinct subgroups, which correspond to the low-energy flavour symmetries of the charged leptons and of neutrinos.

Further studies have been presented in which the SM is extended to a GUT scenario: in our opinion, these analyses emphasize the difficulty in the construction of a flavour GUT model \cite{FlavorGUT}  which naturally lead to realistic fermion phenomenology and to a fair explanation of the gauge symmetry breaking chain to get $SU(3)_C\times U(1)_{em}$. Concerning the case of the flavour GUT models \cite{A4TBGUT} that predicts the TB mixing in the lepton sector, the combinations of the constraints arising by the flavour symmetry and by the GUT group lead to wrong predictions for the fermion masses and mixings. The problem is usually avoided by recurring to non-minimal Higgs or flavon field content and by assuming peculiar symmetry breaking patterns of the GUT gauge symmetry and ascribing quite often at  type-II See-Saw as the origin of the neutrino masses. Moreover these patterns are often not supported by the study of the scalar Higgs potential, leaving open the question if such peculiar patterns may be actually realized or not.

Moreover, a common feature of flavour models which deal with the TB scheme is a value for the reactor angle very close to zero, in the absence of specific dynamical tricks (see \cite{Lin_LargeReactor} for a model in which such a trick is implemented). However, if the next future neutrino-appearance experiments will find a value for $\theta^l_{13}$ close to its present upper bound, about the sine of the Cabibbo angle $\lambda$, the TB mixing should be considered an accidental symmetry. In this case a new leading principle would be necessary, for which quark-lepton complementarity \cite{Complementarity}, $\theta^l_{12}+\lambda\sim \pi/4$, would be a good candidate.
This has the advantages that it can be naturally  related to GUTs and does not necessarily predict a small reactor mixing angle.

This idea was developed in a model \cite{AFM_BimaxS4} based on the $S_4$ discrete group. In this construction the PMNS matrix coincides with the bimaximal mixing \cite{BMmixing} ($\sin^2\theta^{BM}_{12}=1/2$, $\sin^2\theta^{BM}_{23}=1/2$ and $\sin\theta^{BM}_{13}=0$) in first approximation; since the BM value of the solar angle exceeds the $3\sigma$ error, large corrections are needed to make the model agree with the data; the perturbations are naturally constrained to get the ``weak'' complementarity relation, $\theta^l_{12}+\mathcal{O}(\lambda)\sim \pi/4$, and $\sin\theta^l_{13}\sim\lambda$ in most of the parameter space. The model only deals with the lepton sector and an extension in the quarks sector is lacking. In this paper we aim at revisiting the model in \cite{AFM_BimaxS4} in order to include a realistic description of quarks.

Our starting point lies in the complementarity relations:
\beq
\theta^l_{12}+\theta^q_{12} \simeq \pi/4 +\mathcal{O}(\lambda^2)\;, \qquad \qquad \theta^l_{23}+\theta^q_{23} \simeq - \pi/4 +\mathcal{O}(\lambda^2)\;.
\label{anglescomplement}
\eeq
For the third mixing angles we know that in the quark sector it is very small, $\theta^q_{13} = \mathcal{O}(\lambda^3)$, while in the lepton sector, as already mentioned, it has only an upper bound, $\sin \theta^l_{13}\lesssim\lambda$. We will see that in the model described below, we predict it to be $\mathcal{O}(\lambda)$.

In non-GUT contexts, no compelling model leading  to the exact complementarity has been produced so far and indeed in \cite{AFM_BimaxS4} a weaker version of eq. (\ref{anglescomplement}) has been used in which the quark contributions are substituted by similar terms originating from the charged lepton sector.
On the other hand, exact complementarity is possible in cases where the flavour symmetry group is combined with a GUT group, as in the Pati-Salam\footnote{Adopting a Pati-Salam context, we loose the exact gauge coupling unification, but at the same time we avoid several complications which are instead present in other GUT scenarios, as described in the text.} context \cite{PS}, where the charged lepton and the down-quark mass matrices are similar,
\beq
M_e\sim M_d.
\label{elecdownsimilar}
\eeq

Other popular GUTs are $SU(5)$ and $SO(10)$, which however are less appealing when trying to recover the QL complementarity relations. Indeed, in the minimal $SU(5)$ \cite{SU5} one has $M_e\sim M_d^T$ and as result, a correction of order $\lambda$ to the solar angle does not correspond to the Cabibbo angle of the CKM matrix.

On the other hand a reason to prefer Pati-Salam over $SO(10)$ is  related to the type-I and type-II See-Saw mechanisms.
In these two GUT contexts, we expect both left-handed (LH) and right-handed (RH) neutrino Majorana mass matrices to be present. As a result, the effective LH neutrino mass matrix will get the contributions through the type-I as well as the type-II See-Saw mechanisms. In general, and this happens also in our proposal, this interplay introduces two mass scales and a highly non-trivial flavour structure for the effective neutrino mass matrix, which difficultly leads to a realistic description of the PMNS matrix. For this reason, a hierarchy between the two contributions is usually assumed. As we will see we can reproduce quark-lepton complementarity in our model, if the type-II See-Saw is dominant.

This possibility has already been investigated in several flavour GUT models, for example in \cite{TypeII&GUT} in the context of the $SO(10)$ GUT. 
However, in \cite{Bertolini} it has been agued that the type-II dominance in the context of minimal $SO(10)$ models is highly disfavoured. Even so, when restricting to particular supersymmetric parameter space, the type-II dominance could be possible both in the minimal and non-minimal $SO(10)$ approaches, \cite{Bajc}. In the Pati-Salam context, there is much more freedom and the eventual dominance of one of the two contributions could be realized. In this paper we study the gauge Higgs potential and we verify that a type-II dominance can be justified, even if it puts strong constraints on the model building realization.

After a detailed analysis of the fermion phenomenology, we move to the study on the Higgs gauge sector, which is responsible for the gauge symmetry breaking steps to finally get $SU(3)_C\times U(1)_{em}$. Thus we consider the renormalization group equations (RGEs)  both for the gauge couplings and for  the fermion masses and mixings from the cutoff of the theory down to the low-energy scale.  Here we anticipate that the RGEs analysis is crucial: the model turns out to be viable only in a small region of the parameter space. On one side the gauge coupling RGEs analysis  is deeply modified by the non-minimal Higgs  field content required by the presence of the flavour symmetry $G_f$ and puts strong constraints on the scales of the model. On the other side the GUT flavour mass matrix structures interfere with each other, through the Yukawa RGEs, with non-negligible consequences for the neutrino phenomenology.

The paper is organized as follows. In section \ref{Sec:outline} we describe the flavour structure of the fermion mass matrices in order to recover realistic lepton and quark mixing matrices and fermion mass hierarchies. In section \ref{Sec:construction} we enter in the details of the model building construction, specifying the transformations of all the fields of the model under the gauge and flavour groups and discussing the flavon vacuum misalignment necessary to the flavour symmetry breaking chain. Afterwards we deal in section \ref{Sec:Higgs} with the study of the gauge Higgs potential and we analyze the constraints coming from the flavour symmetry on the scalar field content. In section \ref{sec:Yuk} we perform the analysis of the renormalization group running of fermion masses and mixings and of the gauge coupling constants from the GUT scale down to the low-energy scale. In section \ref{Sec:phenoanalysis} we report the phenomenological analysis after the running evolution. Finally in section \ref{Sec:conclusions} we conclude. In the appendices we report technical details.

\section{Outline of the model}
\label{Sec:outline}

In this section we present a general discussion on the flavour structure of fermion masses and mixings resulting form the analysis of the complementarity relations. Eq. (\ref{anglescomplement}) suggests that the angles in the CKM and PMNS matrices may have a common origin. Looking at their definitions
\beq
\label{relcomp}
V= V_u^\dag\, V_d, \qquad \qquad U= U_e^\dag U_\nu\;,
\eeq
where $V_u,\,V_d,\,U_e,\,U_\nu$ diagonalize $M_u\,M_u^\dag$, $M_d\,M_d^\dag$, $M_e\,M_e^\dag$ and $m_\nu$ respectively, we see that this common origin can be motivated in a GUT context where some relations among the mass matrices are present. In PS models, we have the following expression which links down-quarks and charged leptons
\beq
U_e \sim V_d\,;
\label{eqvevd}
\eeq
we will see in a while how this enters in the model construction.

In \cite{AFM_BimaxS4} the two large lepton angles in the PMNS matrix arise only from the neutrino sector: in the first order approximation, both the solar and the atmospheric angles are maximal and after that some corrections from the charged lepton sectors lower the value of the solar angle in order to accommodate the data. In our realization one maximal angle originates from the charged leptons and the other from the neutrinos: afterwards some corrections are necessary to make the solar angle agree with the measurements and we require that these corrections  come from the charged leptons. This is an other possible choice with respect to \cite{AFM_BimaxS4} and in the next section we will show that it is easily realized in our context.

In the approximation of small $\la$, the lepton mixing matrix can schematically be written as
\beq
U\; = \;R_{23}\left(-\dfrac{\pi}{4}\right) R_{13}(\lambda) R_{12}\left(\dfrac{\pi}{4} - \lambda\right) \; = \; \Big(\underbrace{R_{23}\left(-\dfrac{\pi}{4}\right) R_{13}(\lambda) R_{12}(-\lambda)}_{U_e^\dag}\Big) \underbrace{R_{12}\left(\dfrac{\pi}{4}\right)}_{U_\nu} \; ,
\label{PMNSrots}
\eeq
where $R_{ij}(\theta)$ stands for a rotation in the $(ij)$ plane of an angle $\theta$. \footnote{A coefficient of order one is in general present to multiply each angle, but we do not show them here in order to simply the discussion. We will consider the precise expressions in the following sections.} As a result, after a suitable commutation of matrices, $U_e$ can naively be written as
\beq
U_e\;=\;R_{23}\left(\dfrac{\pi}{4}\right) R_{13}(\lambda) R_{12}(\lambda)\;.
\label{Ue_outline}
\eeq

The CKM matrix is given in first approximation as
\beq
V\; = \;  R_{12}(\lambda).
\eeq
Because of eq. (\ref{eqvevd}), $V_d$ has the same structure as $U_e$ in eq. (\ref{Ue_outline}) and therefore we can obtain the CKM matrix as
\beq
V\; = \;
\Big(\underbrace{R_{12}\left(- \alpha \lambda \right) R_{13}(- \lambda) R_{23}\left(- \dfrac{\pi}{4}\right) }_{V_u^\dagger}\Big) \underbrace{R_{23}\left(\dfrac{\pi}{4}\right) R_{13}(\lambda) R_{12}(\lambda))}_{V_d}\;.
\label{CKMrots}
\eeq
We see that the angles of the $(23)$ and $(13)$ rotations in $V_u^\dagger$ should be the opposite of those in $V_d$, while the angles in  the  $(12)$ sector should be different. We have schematically indicated this via the $\alpha$ coefficient. Analogous to eq. (\ref{Ue_outline}), we write $V_u$ as
\beq
V_u\;=\;R_{23}\left(\dfrac{\pi}{4}\right) R_{13}(\lambda) R_{12}(\alpha \lambda)\;.
\label{Vu_outline}
\eeq

Moving to the explicit form of the mass matrices, the generic effective Majorana neutrino mass matrix $m_\nu$ which is diagonalized by $U_\nu$ as in eq. (\ref{PMNSrots}), $U_\nu=R_{12}\left(\frac{\pi}{4}\right)$, through $m_\nu^{diag}\;=\;U_\nu^T\,m_\nu \,U_\nu$, is given by
\beq
m_\nu \sim\left(
                \begin{array}{ccc}
                a & b & 0 \\
                b & a & 0 \\
                0 & 0 & c \\
                \end{array}
        \right)\;.
\label{Mnu1}
\eeq

In the charged lepton sector, we are looking for the most general form of the mass matrix, whose square $M_e\,M_e^\dag$ is diagonalized by the action of $U_e$ of eq. \eqref{Ue_outline},
\beq
\left(M_e \, M_e^\dag\right)^{diag}=U_e^\dag\, M_e \, M_e^\dag \,U_e\;.
\label{mlsqform}
\eeq
Inverting eq. (\ref{mlsqform}) we find in the limit $m_e\to 0$
\beq
M_e\,M_e^\dag \sim \dfrac{m_\tau^2}{2} \left(
                                            \begin{array}{ccc}
                                            0 & \lambda & \lambda \\
                                            \lambda & 1 & 1 \\
                                            \lambda & 1 & 1 \\
                                            \end{array}
                                        \right) +
\dfrac{m_\mu^2}{2} \left(
                    \begin{array}{ccc}
                    0 & \lambda & -\lambda \\
                    \lambda & 1 & -1 \\
                    -\lambda & -1 & 1 \\
                    \end{array}
                    \right) +\ldots\;,
\label{Mlsq}
\eeq
where the dots stand for suppressed contributions. This can be obtained if $M_e$ is given by
\beq
M_e  \sim \dfrac{m_\tau}{\sqrt{2}} \left(
                                        \begin{array}{ccc}
                                        0 & 0 & \lambda \\
                                        0 & 0 & 1 \\
                                        0 & 0 & 1 \\
                                        \end{array}
                                    \right) +
\dfrac{m_\mu}{\sqrt{2}} \left(
                        \begin{array}{ccc}
                        0 & \lambda & 0 \\
                        0 & 1 & 0 \\
                        0 & -1 & 0
                        \end{array}
                        \right) +\ldots\;.
\label{Ml1}
\eeq
It is interesting to note that, moving to the basis of diagonal charged leptons and considering only the leading order terms, the neutrino mass matrix results to be of the BM type, already presented in the introduction.

As already stated, the PS gauge structure implies the relation $M_e\sim M_d$ and therefore the down-quark matrix has a similar structure as in eq. \eqref{Ml1}. Looking at eq. (\ref{Vu_outline}) we find that also $M_u$ should have a similar structure, but here there is a slight difference due to the $\alpha$ coefficient.

In order to reproduce the result in eq. (\ref{CKMrots}), i.e. exact cancellations in the $(23)$ and $(13)$ sectors but not in the $(12)$ sector,
we need to study the origin of the different rotations in (\ref{Vu_outline}). It is easy to see that in the diagonalization of the mass matrices in eq. (\ref{Mlsq}), the $(12)$ rotation originates from the second families, while the $(13)$ rotation comes from the third families. We conclude that the up-quark mass matrix has the same form as (\ref{Ml1}), in which the third column is proportional to that of the down-quarks, while the second column is not.

As already stated in the introduction, we expect contributions on the effective neutrino mass matrix coming from both the type-I and the type-II See-Saw mechanisms. For the moment we just assume that the type-II is the only responsible of eq. (\ref{Mnu1}) and we verify in section \ref{Sec:Higgs} that it is indeed dominating with respect to the type-I terms.

In the next section we enter in the details of the model building, explaining the origin of the mass matrices displayed above.

\section{The flavour model building}
\label{Sec:construction}

In this section we define our framework. The model is based on the PS gauge symmetry, $SU(4)_C\times SU(2)_L\times SU(2)_R$, present at high energy where a supersymmetric context is assumed. The complete flavour group $G_f$ is the same as in \cite{AFM_BimaxS4} given by the product of the following different terms:
\beq
G_f=S_4\times Z_4\times U(1)_{FN}\times U(1)_R\;.
\eeq

The group $S_4$ is the permutation group of four distinct objects, isomorphic to the group $O$ which is the symmetry group of a regular octahedron. It has $24$ distinct elements filled in five conjugate classes and therefore it has five irreducible representations, two one-dimensional denoted as $1_1$ and $1_2$, one two-dimensional labelled as $2$ and two three-dimensional written as $3_1$ and $3_2$. Here we recall the multiplication rules, while the Clebsch-Gordan coefficients, the explicit structures of the generators, the list of the elements and further details on the group theory of $S_4$, are reported in appendix \ref{AppA}:
\beq
\begin{array}{l}
1_1\times R=R\times1_1=R\qquad \text{where $R$ stands for any representation}\\
1_2\times1_2=1_1\\
1_2\times2=2\\
1_2\times3_1=3_2\\
1_2\times3_2=3_1\\
\\
2\times2=1_1+1_2+2\\
2\times3_1=3_1+3_2\\
2\times3_2=3_1+3_2\\
\\
3_1\times3_1=3_2\times3_2=1_1+2+3_1+3_2\\
3_1\times3_2=1_2+2+3_1+3_2\;.
\end{array}
\eeq

The spontaneous symmetry breaking of $S_4$ is responsible for the flavour structure of the matrices in eqs. (\ref{Mnu1}, \ref{Ml1}): $S_4$ is broken down to two distinct subgroups and it is the presence of this mismatch at the LO which allows to construct eqs. (\ref{Mnu1}, \ref{Ml1}). More in detail, the different subgroups to which $S_4$ is broken down are the subgroups preserved by the VEVs of the flavons. In order to determine these structures, it is necessary to identify the elements of the group which leave the VEVs of the flavons invariant under their action. Doing so, we find that $S_4$ is broken down to $Z_2\times Z_2$ in the neutrino sector, originated by the elements $ST^2S$ and $T^3ST$ of the classes $\cC_2$ and $\cC_3$, respectively. In the charged fermion sector, we anticipate that the superpotential is invariant under an accidental $Z_2$ symmetry and therefore the identification of the residual subgroup must account of this additional term. We find that $S_4$ is broken down to a $Z_2\times Z_2$ group, distinct to that one in the neutrino sector, generated by the two elements $-T^2$ and $-TST^2S$ of the classes $\cC_2$ and $\cC_3$, respectively. One can argue that the group $Z_2\times Z_2$ implies a degeneracy among two of the charged fermion families, but this does not happen, because the two $Z_2$ relate different families: the accidental $Z_2$ exchanges $F_2^c$ and $F_3^c$, while the $Z_2$ subgroup of $S_4$ relates $F_1^c$ and $F_2^c$.

The other terms in $G_f$ carry out other roles: the Abelian $Z_4$ symmetry is required to avoid dangerous terms in the superpotential, it helps to keep  the  different sectors of the model separated, quarks from leptons and neutrinos from charged leptons, and it is also useful to guarantee the flavon vacuum alignment; the continuous Frogatt-Nielsen (FN) Abelian symmetry \cite{FN}, $U(1)_{FN}$, is introduced to justify charged fermion mass hierarchies; the continuous $R$-symmetry $U(1)_R$, that contains the usual $R$-parity as a subgroup, is a common feature of supersymmetric formulations and simplifies the constructions of the scalar potential. It is worth stressing that the supersymmetric context is of great utility in the discussion of the scalar potential and helps in the gauge coupling running, but a similar non-supersymmetric model can be constructed as well. In particular in the present paper we only deal with SM particle and therefore we use the same symbols for a supermultiplet and its even $R$-parity components.

\subsection{The matter, Higgs and flavon content of the model}

In the PS context, the five matter multiplets of each family of the SM  plus a RH neutrino and their superpartners are unified in only two supermultiplets: a LH and a RH ones as follows,
\beq
\ba{ccc}
{\bf SU(3)_C\times SU(2)_L\times U(1)_Y} & {\bf \rightarrow} & {\bf PS}\\[3mm]
( 3, \, 2, \, 1/6 )_{Q} +(1, \, 2, \, -1) _{L}  &\rightarrow& (4, \, 2, \, 1)\\[3mm]
(\ov{3}, \, 1, \,-2/3)_{u^c} + (\ov{3}, \,1, \,1/3 )_{d^c} +(1, \,1, \,1 )_{e^c} + (1, \,1, \,0)_{\nu^c}  &\rightarrow& ( \overline{4}, \, 1, \,2)
\ea
\eeq
The three copies of the LH supermultiplet are combined in the three-dimensional representation $3_1$ of $S_4$, while the three families of the RH supermultiplet are in $1_2$, $1_2$ and $1_1$ respectively. The fact that we can put different representations within one family in different representations of the family symmetry group is essential here. Note that this would not be possible in (minimal) $SO(10)$, where all Standard Model particles are in one sixteen dimensional representation. The first two families are also charged under $U(1)_{FN}$ by two units. This suppresses their masses with respect to the third family ones. Further suppression of the first family with respect to the second is due to their different  $Z_4$ charges.

All the properties of the matter fields are summarized in table \ref{table:matter}.

\begin{table}[!h]
\begin{center}
\begin{tabular}{|c||c|c|c|c|}
\hline
  &&&& \\
  Matter & $F_L$ & $F^c_1$ & $F^c_2$& $F^c_3$  \\
  &&&& \\[-0,3cm]
  \hline
  &&&& \\
  PS & $(4,2,1)$ & $(\ov{4},1,2)$ & $(\ov{4},1,2)$& $(\ov{4},1,2)$  \\
  &&&& \\[-0,3cm]
  $S_4$ & $3_1$ & $ 1_2$ & $1_2$ & $1_1$\\
  &&&&\\[-0,3cm]
  $U(1)_{FN}$ & 0 & 2 & 2 & 0  \\
  &&&& \\[-0,3cm]
  $Z_4 $ & $1$ & $1$ & $i$  & $-i$  \\
  &&&& \\
  \hline
  \end{tabular}
\end{center}
\caption{\it Transformation properties of the matter fields. Notice that the PS assignments should be read in agreement with $SU(4)_C\times SU(2)_L\times SU(2)_R$.}
\label{table:matter}
\end{table}

Our model contains five flavon fields: two $S_4$ triplets (${\varphi}$ and ${\varphi}'$) that, because of their $Z_4$ charge, deal at LO only with the Dirac Yukawa couplings of quarks and leptons, and two fields, one singlet ($\sig$) and one triplet ($\chi$), that, by $Z_4$, deal at LO only with the Majorana masses of neutrinos. The fifth flavon is the Froggatt-Nielsen messenger, which we indicate with $\theta$. Their properties are shown in table \ref{table:flavons}. Under the continuous $R$-symmetry, the matter superfields transform as $U(1)_R=1$, while all the flavons are neutral.

\begin{table}[!h]
\begin{center}
\begin{tabular}{|c||c||c|c|c|c|}
\hline
  && &&& \\[-0,3cm]
  Flavons &$\theta$ &  $\vphi$& $\vphi'$& $\chi$ &$\sigma$ \\
  && &&& \\[-0,3cm]
\hline
  && &&& \\
  $S_4$ &$1_1$ & $3_1$ & $3_2$ & $3_1$ &$1_1$ \\
  &&& && \\[-0,3cm]
  $U(1)_{FN}$ & -1 &  0 & 0& 0 & 0 \\
  &&& && \\[-0,3cm]
  $Z_4$ & 1 & $i$& $i$ &$1$ & $1$\\
  &&& && \\
\hline
\end{tabular}
\end{center}
\caption{\it The flavon field content and their transformation properties under the flavour symmetries. All flavon fields are singlet of the gauge group.}
\label{table:flavons}
\end{table}

Fermion masses and mixings arise from the spontaneous breaking of the flavour symmetry by means of the flavons which develop VEVs according to the following configuration: at LO we have
\bea
\mean{\varphi}=\left(
                     \begin{array}{c}
                       0 \\
                       1 \\
                       1 \\
                     \end{array}
                   \right)v_\varphi\;,
&&\mean{\varphi'}=\left(
                     \begin{array}{c}
                       0 \\
                       1 \\
                       -1 \\
                     \end{array}
                   \right)v_{\varphi'}\;,
\label{vev:charged:best}\\[3mm]
\mean{\chi}=\left(
                     \begin{array}{c}
                       0 \\
                       0 \\
                       1 \\
                     \end{array}
                   \right)v_\chi\;,
&&\mean{\sigma}=v_\sigma\;,
\label{vev:neutrinos}\\[3mm]
&\mean{\theta}=v_\theta\;.&
\label{vev:FN}
\eea
In this section we simply assume this VEV alignment and we will prove it to be a natural solution of the minimization of the scalar potential in section \ref{sec:flavonscalarpotential}. Furthermore we assume that the FN messenger and the other flavons have VEVs of the same order of magnitude: it results partly from the minimization procedure and partly from the constraints coming form the comparison with the measured mass hierarchies, as it will be clearer in the following. Without loss of generality it is useful to keep  the notation compact and write
\beq
\dfrac{VEV}{\La} \approx \lambda\;,
\eeq
where $VEV$ refers to the vacuum expectation value of any flavon of the model and $\lambda\simeq0.2$ is close to the Cabibbo angle. $\La$ is the flavon cut-off that we assume to be the largest scale in the model: it corresponds to the scale of the flavon dynamics.

The insertion of flavons in the mass terms, leads to non-renormalizable operators, as we will see in the following section. This means that we can write the superpotential as an expansion in powers of $\mathrm{flavon}/\La$ and we can stop the expansion after the first orders.

The Higgs fields of our model relevant to build the fermion mass matrices transform under the gauge group and under the $Z_4$ factor of the flavour symmetry:  in table \ref{table:higgs} we summarize their transformation rules.  The first three fields, $\phi$,  $\phi'$ and $\rho$, deal at LO only with the Dirac Yukawas. Due to the $Z_4$ charges,  $\phi$ and $\phi'$ are responsible of the third family and the charm quark masses, while $\rho$ is responsible for the strange and $\mu$ masses. The field $\rho\sim (15,2,2)$ being  in the adjoint of $SU(4)_C$  may develop VEV along the $SU(4)_C$ direction $\diag(-3,1,1,1)$. This implies that the leptons which get mass via this field are a factor $3$ heavier than the corresponding quarks and therefore this field is very useful to describe the second family, at least in the down sector, reproducing the well known Georgi-Jarlskog relation \cite{GJrelation}, $m_\mu \approx 3 m_s$, at the high energy scale. As we will see in the next sections,  in order to recover the up-quark mass hierarchies the $\rho$ projection along the light doublet  Higgses, $v_\rho^u$ and $v_\rho^d$, has to point only in the down direction: the requirement $v_\rho^u=0$ can be realized only if the Higgs field content contains two identical copy of the Higgs field $(1,2,2)$ and this justify the presence of $\phi$ and $\phi'$.

Finally, as we will see in detail in the following sections, the field $\Delta_R$ is necessary to conclude the PS symmetry breaking pattern and to recover  the SM gauge group  through its spontaneous symmetry breaking VEV. At the same time, when $\Delta_R$ develops a VEV, it gives a Majorana mass to the right-handed neutrinos thus contributing to the effective neutrino mass matrix through the usual type-I See-Saw mechanism. A second source for the neutrino mass matrix comes from $\Delta_L$, in terms of the type-II See-Saw mechanism.

\begin{table}[!h]
\begin{center}
\begin{tabular}{|c||c|c|c|c|}
\hline
  &&&&\\[-0,3cm]
 Higgses &  ${\phi},{\phi}'$& $\rho$& $\Delta_L$ &$\Delta_R$\\
  &&&&\\[-0,3cm]
  \hline
  &&&&\\
  PS & $(1,2,2)$& $(15,2,2)$ & $(\overline{10},3,1)$& $(10,1,3)$ \\
  &&&& \\[-0,3cm]
  $Z_4 $ & $1$ & $-1$ & $1$ &$-1$ \\
  &&&& \\
\hline
  \end{tabular}
\end{center}
\caption{\it The Higgs fields  responsible of generating fermion mass matrices and their transformation under the gauge and the Abelian  flavour symmetries. All Higgs fields are singlets under $S_4\times U(1)_{FN}\times U(1)_R$, while they can transform under the $Z_4$ factor.}
\label{table:higgs}
\end{table}

In our scheme the neutrino mass matrix is dominated by type-II See-Saw. As already stated in the introduction, the flavour structure of the effective neutrino mass matrix in eq. (\ref{Mnu1}) arises from an interplay between the two See-Saw sources. In the PS context $m_D\sim M_u$ and this suggests a hierarchical structure for the type-I contribution, which does not agree with the flavour structure in eq. (\ref{Mnu1}). We will show that, given $m_D\sim M_u$, the required flavour structure for the Majorana RH neutrino mass necessary to recover eq. (\ref{Mnu1}) is not allowed in our model. This suggests to find a construction in which the type-II contributions are dominating over the type-I and we will show that such a feature puts strong constraints on the model building.

\subsection{Fermion mass matrices at leading order}
In this section we study the matter superpotential, $\mathcal{W}_Y$, and the resulting mass matrices for all the fermions. The Yukawa superpotential can be written as a sum of two pieces,
\beq
\mathcal{W}_Y=\mathcal{W}_{Dir}+\mathcal{W}_{Maj}
\eeq
containing respectively terms with Dirac and Majorana structures. Furthermore, making a power expansion in terms of $\mathrm{flavon}/\Lambda$, we distinguish between leading and subleading couplings. We refer as ``leading order'' operators to those terms which provide $M_{e,d,u}$ and $m_\nu$ as in eqs. (\ref{Mnu1}, \ref{Ml1}) in the limit $\lambda\rightarrow0$. The subleading orders corresponds to operators of higher dimensions. In this way it is easier to identify each entry of the mass matrices with the corresponding operator in the superpotential and furthermore it underlines the relevance of the subleading contributions.

\subsubsection{Dirac mass terms}
We first study the Dirac matter superpotential at LO which reads
\beq
\begin{split}
\mathcal{W}^{LO}_{Dir} \;=&\;\,\phantom{+}  y_1 \dfrac{1}{\Lambda} F_L F_3^c (\phi+\phi')\vphi +\\
&+ y_2 \dfrac{1}{\Lambda^3} F_L F_2^c \theta^2 \rho \vphi' + \sum^4_{i=1} y_{3,(i)} \dfrac{1}{\Lambda^5} F_L F_2^c \theta^2 (\phi+\phi') X_i^{(1)} +\\
&+ \sum_{i=1}^3 y_{4,(i)} \dfrac{1}{\Lambda^4} F_L F_1^c \theta^2 \rho X_i^{(2)} + y_5 \dfrac{1}{\Lambda^5} F_L F_1^c \theta^2 (\phi+\phi') \chi^3\;.
\end{split}
\label{YukSupPot}
\eeq
Here we use a compact notation to avoid the proliferation of coefficients: the term $X_i^{(1,2)}$ represents a list of products defined as
\beq
\ba{rcl}
X_i^{(1)} &\equiv& \{ \vphi^3 ,\, \vphi^2\vphi' ,\, \vphi\vphi^{\prime\,2} ,\, \vphi^{\prime\,3} \}\;,\\[3mm]
X_i^{(2)} &\equiv& \{ \vphi^2 ,\, \vphi\vphi' ,\, \vphi^{\prime 2} \}\;,
\ea
\label{X1}
\eeq
where each term represents all the different $S_4$ contractions which can be constructed with those flavons; furthermore we indicate with $y_1(\phi+\phi')$ the combination $y_1^{(1)}\phi+y_1^{(2)}\phi'$ and similarly for $y_{3,\,(i)}$ and $y_5$.

When the flavour symmetry is broken the model describes a non-minimal PS model in which the Yukawa couplings present a well defined  structure.
Then, when  the PS gauge symmetry is broken to the SM gauge group $\phi$, $\phi'$ and the colour singlet component of $\rho$ mix in four light Higgses, two up-type and two down-type, $h_{u,d}$ and $h_{u,d}'$.  Thus at the EW  scale  $\phi$, $\phi'$ and $\rho$ have non-vanishing projections to the light Higgs components that acquire a VEV. We will indicate these components as $v_\phi^{u,d}$, $v_{\phi'}^{u,d}$ and $v_\rho^{u,d}$.  As already said,  we need to impose that the $\rho$ field has no projection along the up direction:  $v_\rho^{u}=0$. This can be realized because in terms of the light Higgs up-VEVs, $v_1^u=\vev{h_u}$ and $v_2^u=\vev{h_u'}$,  $v_\rho^{u}$ is given by
\beq
v_\rho^{u}= U_{13} v_1^u+U_{23} v_2^u\,,
\eeq
where the matrix $U$ is introduced in the appendix \ref{app-Higgs}.
The constraint  $v_2^u= -U_{13}/U_{23}  v_1^u$  can be imposed thanks to the freedom we have in the superpotential and in the soft potential. \footnote{This requirement, imposed \emph{by hand}, could be motivated by some symmetry argument, but to introduce a mechanism that could explain  $v_\rho^u=0$, it would be necessary deeply modifying the structure of our model. In the present paper we just assume this fine-tuning and we refer to the Appendix \ref{secD1} for further details. We just anticipate that the fine-tuning we introduce in the model is similar to the fine-tuning which is universally accepted whenever the MSSM has to be recovered.} Note that we could relax the condition $v_\rho^{u}=0$ allowing a mild hierarchy between $v_\rho^{u}$ and $v_\rho^{d}$, for example of order $\lambda^2 $, without affecting the final mass hierarchies, but in the following, for simplicity, we work under the assumption that  $v_\rho^{u}=0$.

The final Dirac fermion mass matrices we get are given by
\bea
\label{MeLO}
M^{LO}_e &=&  -3\left(
       \begin{array}{ccc}
         0 & 0 & 0 \\
         0 & y_2 & 0 \\
         0 & -y_2 & 0 \\
       \end{array}
     \right)v^d_\rho \lambda^3+
    \left(
       \begin{array}{ccc}
         0 & 0 & 0 \\
         0 & 0 & y_1 \\
         0 & 0 & y_1 \\
       \end{array}
     \right)v^d_\phi \lambda\;,\\[3mm]
\label{MdLO}
M^{LO}_d &=& \phantom{-3} \left(
       \begin{array}{ccc}
         0 & 0 & 0 \\
         0 & y_2 & 0 \\
         0 & -y_2 & 0 \\
       \end{array}
     \right)v^d_\rho \lambda^3+
    \left(
       \begin{array}{ccc}
         0 & 0 & 0 \\
         0 & 0 & y_1 \\
         0 & 0 & y_1 \\
       \end{array}
     \right)v^d_\phi \lambda\;,
\eea
\beq
\label{MuLO}
M^{LO}_u = m^{LO}_D=  \phantom{-3}\left(
       \begin{array}{ccc}
         0 & 0 & 0 \\
         0 & y_3 & 0 \\
         0 & -y_3 & 0 \\
       \end{array}
     \right)v^u_\phi\lambda^5+
    \left(
       \begin{array}{ccc}
         0 & 0 & 0 \\
         0 & 0 & y_1 \\
         0 & 0 & y_1 \\
       \end{array}
     \right)v^u_\phi \lambda\;,
\eeq
where  we used the compact notation $y_i v^{u/d}_\phi$  to indicate $ y^{(1)}_i v^{u/d}_\phi+  y^{(2)}_i v^{u/d}_{\phi'}$ and absorbed all the non-relevant CG coefficients. Note that $y_3$ is the sum of the different $y_{3,\,(i)}$ and that, by construction, $y_{1,2,3}$ can be considered complex coefficients with modulus of order 1. Note also that the different numerical coefficients between charged leptons and down-quarks originate from to the presence of $\rho$ instead of $\phi$ ($\phi'$) in the superpotential. The operators which should give contributions to the first families (those proportional to $y_4$ and $y_5$) are vanishing, thanks to the special flavon VEV alignment. As a final comment, we are neglecting at this level of approximation the contributions to $M^{LO}_{e,d}$ coming from the operators proportional to $y_3$: these terms, which preserve the anti-alignment of the second and third entries of the second columns, are $\lambda^2$ suppressed with respect to the LO ones proportional to $y_2$.

In order to identify the mass matrices in eqs. (\ref{MeLO})--(\ref{MdLO})--(\ref{MuLO}) with those in eqs. (\ref{Mnu1}, \ref{Ml1}), we need to define the (complex) fermion masses as follows:
\beq
\label{massLO}
\ba{rclcrcl}
m_\mu &\equiv&-3 y_2 v^d_\rho \lambda^3\;,  &\qquad& m_\tau &\equiv&  y_1v^d_\phi \lambda\;,\\[3mm]
m_s &\equiv& y_2v^d_\rho \lambda^3\;,  &\qquad& m_b &\equiv& y_1v^d_\phi \lambda\;,\\[3mm]
m_c &\equiv& y_3v^u_\phi \lambda^5\;,  &\qquad& m_t &\equiv&  y_1v^u_\phi \lambda\;.\\[3mm]
\ea
\eeq
Now we can comment on the masses as well as on the mixing matrices which can be driven at this approximation level. Notice that the top-quark Yukawa does not come from a renormalizable coupling and it presents the same suppression as the other third family fermion masses: as a result the dominance of the top-quark mass can be justified by the hierarchy between $v_\phi^u$ and $v_\phi^d$. The mass matrices in eqs. (\ref{MeLO})--(\ref{MdLO})--(\ref{MuLO}) are diagonalized by a maximal rotation in the sector $(23)$, i.e. $U_e=V_d=V_u=R_{23}(\pi/4)$, while the fermion mass hierarchies are given by
\beq
\left|\dfrac{m_\mu}{m_\tau}\right|\sim\left|\dfrac{m_s}{m_b}\right|\sim\cO(\lambda^2)\;,
\qquad\qquad \left|\dfrac{m_c}{m_t}\right|\sim \cO(\lambda^4)\;.
\eeq
Furthermore, at the cutoff, we recover some relations among the masses of different fermions: the $b-\tau$ unification and the Georgi-Jarlskog relation \cite{GJrelation}
\beq
|m_\tau|=|m_b|\;,
\qquad\qquad |m_\mu|=3|m_s|\;.
\label{MtauMb}
\eeq
Finally we should comment of the relative value of the top and of the bottom masses:
\beq
\left|\dfrac{m_t}{m_b}\right|=\left|\dfrac{ y_1^{(1)} v^u_\phi+ y_1^{(2)} v^u_{\phi'}}{y_1^{(1)} v^d_\phi+ y_1^{(2)} v^d_{\phi'}}\right|\;.
\eeq
Notice that usally this ratio is proportional to $\tan\beta$, the ratio between the up- and down-VEVs of the light Higgses, but this is not the case: indeed the light Higgses are combinations of $\phi$, $\phi'$ and $\rho$ and therefore we may define $\tan \beta$ as
\beq
\tan\beta\equiv\dfrac{\sqrt{ (v^u_\phi)^2 +(v^u_{\phi'})^2}}{\sqrt{ (v^d_\phi)^2 +(v^d_{\phi'})^2+(v^d_{\rho})^2}}\neq \left|\dfrac{m_t}{m_b}\right|\;.
\eeq

\subsubsection{Majorana mass terms}
We now discuss the part of the superpotential which contains the Majorana couplings. At LO\footnote{Regarding the terms which contribute to $M_R$, we consider at the LO only the first non vanishing term in the superpotential.} it is given by
\beq
\label{cnu}
\mathcal{W}^{LO}_{Maj}=\tilde k_0 \,F_L F_L \Delta_L + \sum_{i=1}^2 \dfrac{\tilde k_{1,(i)}}{\Lambda} \,F_L F_L \Delta_L X^{(3)}_i + \sum_{i=1}^3\dfrac{\tilde k_{2,(i)}}{\Lambda^2} F_L F_L \Delta_L X^{(4)}_i + z_1\, F_3^c F_3^c \Delta_R\;,
\eeq
where we used the compact notation
\beq
\ba{rcl}
X^{(3)} &\equiv& \{ \chi ,\, \sigma \}\;,\\[3mm]
X^{(4)} &\equiv& \{ \chi^2 ,\, \chi\sigma ,\, \sigma^2 \}\;.
\ea
\label{X2}
\eeq
This superpotential is responsible for giving the following Majorana LH and RH neutrino mass matrices:
\beq
{M}_L = \left(
               \begin{array}{ccc}
                 k_0 & k_1 \lambda & 0 \\
                 k_1 \lambda & k_0 & 0 \\
                 0 & 0 & k_0+k_2\lambda^2 \\
               \end{array}
             \right)v_{L}\;,\qquad\qquad
M_R = \left(
               \begin{array}{ccc}
                 0 & 0 & 0 \\
                 0 & 0 & 0 \\
                 0 & 0 & z_1 \\
               \end{array}
             \right)v_{R}\;,
\label{MnuLO}
\eeq
where $k_{0,1,2}$ and $z_1$ are coefficients of order $1$ and $v_L,v_R$ are the VEVs of $\Delta_{L,R}$ respectively. In particular $k_{0,1,2}$ are defined in terms of $\tilde k_0$, $\tilde k_{1,(i)}$ and $\tilde k_{2,(i)}$:
\beq
\ba{rcl}
k_0&\equiv& \tilde k_0+ \tilde k_{1,(2)}\lambda+ \tilde k_{2,(1)}\lambda^2+ \tilde k_{2,(3)}\lambda^2\;,\\[3mm]
k_1&\equiv& \tilde k_{1,(1)}+ \tilde k_{2,(2)}\lambda\;,\\[3mm]
k_2&\equiv& -\tilde k_{2,(1)}\;.
\ea
\eeq
While $M_L$ corresponds to the type-II contribution to the effective neutrino mass matrix, $M_R$ provides a type-I term. Even at this approximation level, we note the tension between the two See-Saw contributions: $m_\nu^\text{(type-II)}\equiv M_L$ presents a maximal rotation in the $(12)$ sector, while it is easy to verify that $m_\nu^\text{(type-I)}\equiv m_D M_R^{-1} m_D^T$ shows a democratic structure on the $(23)$ sector which corresponds to a maximal rotation in this sector. To recover the mass matrix in eq. (\ref{Mnu1}) it is necessary that the type-II contribution dominates over the type-I terms: in this case it is sufficient to identify $a$ with $k_0$, $b$ with $k_1\lambda$ and $c$ with $k_0+k_2\lambda^2$.

The neutrino masses can be written as
\bea
\left|m_{1,2}\right|^2  &=& \Big(|k_0|^2 \, \mp 2 |k_0|\, |k_1|\, \cos( \theta_{k_0} - \theta_{k_1}) \la + |k_1|^2 \, \la^2 \Big)v_{L}^2 \;,\\[3mm]
\left|m_3\right|^2 &=& \Big( |k_0|^2  \, + 2 |k_0|\, |k_2|\, \cos( \theta_{k_0} - \theta_{k_2}) \la^2 \Big) v_{L}^2\;,
\eea
with $\theta_{k_i}$ the argument of the complex number $k_i$. We need $\cos( \theta_{k_0} - \theta_{k_1}) > 0 $ in order to have $|m_1|$ smaller than $|m_2|$. We see that in most of parameter space the spectrum is quasi degenerate, as the term with $|k_0|^2$ that appears in all three masses dominates over the other terms that are $\lambda$ or $\lambda^2$ suppressed.

We also see that in most of the parameter space $|m_3|$ is the central eigenvalue: indeed $|m_1|$ ($|m_2|$) is shifted down (up) from the central value $|k_0|^2 v_L^2$ by a term proportional to $\lambda$, while $|m_3|$ stays closer to this central value as it is only shifted by a term proportional to $\la^2$. Having $|m_3|$ as the central eigenvalue is obviously in contradiction with experimental data. We conclude that our model is only viable when the term $2 |k_0|\, |k_1|\, \cos( \theta_{k_0} - \theta_{k_1}) \la$ is suppressed. This is clearly possible if either $|k_0|$ or $|k_1|$ or $\cos( \theta_{k_0} - \theta_{k_1}) $ (or a combination of them) is small. In particular, the latter condition means that $k_0$ and $k_1$ are almost perpendicular in the complex plane. We now investigate on these different scenarios, calculating the solar and atmospheric mass squared differences: taking as definition of $\De m^2_{atm}$ the mass squared difference between the heaviest  and the lightest  neutrinos, we have different results for normal (NO) and inverse (IO) mass ordering as given by
\beq
\ba{rcl}
\De m^2_{sol} &\equiv& m_2^2 - m_1^2\\[3mm]
&=& 4|k_0||k_1|\lambda\cos(\theta_{k_0}-\theta_{k_1}) v_{L}^2\;,\\[3mm]
\De m^2_{atm} &\equiv&\left \{
					\begin{array}{c} \De m^2_{atmNO} \equiv (m_3^2 - m_{1}^2)\\[3mm]
					\De m^2_{atmIO} \equiv  (m_2^2 - m_3^2)
					\end{array} \right . \\[5mm]
&=&  \underbrace{2|k_0||k_1|\cos\left(\theta_{k_0}-\theta_{k_1}\right) \la v_{L}^2}_{ \De m^2_{sol}/2}      \mp \big(|k_1|^2-2|k_0||k_2|\cos\left(\theta_{k_0}-\theta_{k_2}\right)\big) \la^2 v_{L}^2\;,
\label{Deltaatm}
\ea
\eeq
On the right-hand side of the last equation, the first term is suppressed by $\lambda$, while the second one by $\lambda^2$, and therefore we could conclude that it is the dominant contribution: it is however exactly the term which must be suppressed in order to avoid $|m_3|$ as the central eigenvalue. Furthermore, this term is equal to half of the solar mass squared difference that is about $30$ times as small as the atmospheric splitting (see table \ref{table:OscillationData}). As a result, to recover a value for $\Delta m^2_{atm}$ close to the measured one, we need that the second term on the right-hand side of eq. (\ref{Deltaatm}) is the dominant one. We can estimate the ratio between the two terms, calculating $r$, the ratio of the two mass squared differences:
\beq
\label{eqr}
\begin{split}
(r_{NO,IO})^{-1} \,&\,\equiv \frac{\De m^2_{atmNO,IO}}{\De m^2_{sol}} \\
&\,= \frac{2|k_0||k_1|\cos\left(\theta_{k_0}-\theta_{k_1}\right) \la \mp
\big(|k_1|^2-2|k_0||k_2|\cos\left(\theta_{k_0}-\theta_{k_2}\right)\big) \la^2}{4|k_0||k_1|\cos\left(\theta_{k_0}-\theta_{k_1}\right) \la} \\
&\,= \frac{1}{2} \mp \frac{|k_1|^2-2|k_0||k_2|\cos\left(\theta_{k_0}-\theta_{k_2}\right) }{4|k_0||k_1|\cos\left(\theta_{k_0}-\theta_{k_1}\right) }\la\;.
\end{split}
\eeq
The natural range of this quantity $r^{-1}$ would be something like $[0.3-0.7]$ (central value $0.5$ and corrections of order $\la$). However, measurements give $r=0.032^{+0.006}_{-0.005} \approx \la^2$, or in other words $1/r\sim 30\sim \lambda^{-2}$. We conclude that
\beq
\frac{4|k_0||k_1|\cos\left(\theta_{k_0}-\theta_{k_1}\right)}{\big||k_1|^2-2|k_0||k_2|\cos\left(\theta_{k_0}-\theta_{k_2}\right)\big|} \approx \la^3\;.
\label{eq:r}
\eeq
The most natural explanation may be assuming that $\cos\left(\theta_{k_0}-\theta_{k_1}\right)$ is very small. In that case the absolute values of all parameters can still be of order one, which was part of the naturalness requirement of the model. Neutrinos present a quasi degenerate  (QD) spectrum  and both normal and inverse ordering are possible. The typical scale of  neutrino masses $v_L$ is given in this case by
\beq
v_L^2 = \frac{\De m^2_{atm}}{\la^2 \big| |k_1|^2-2|k_0||k_2|\cos\left(\theta_{k_0}-\theta_{k_2}\right) \big|} \approx (0.1 \textrm{eV})^2\;.
\eeq

Possible alternative solutions of  \eq{eq:r} that give a non-QD  spectrum can be obtained only by admitting the parameters belong to  a larger, but less natural, range, $\lambda^2-\lambda^{-2}$. In this case  when  $k_0$ is of order $\lambda^{-2}$ and  $\cos\left(\theta_{k_0}-\theta_{k_1}\right)\sim \lambda$  while $k_{1,2}$ still of order one a inverse hierarchical spectrum can be obtained. Another possibility to get an inverse hierarchical (IH) spectrum is having $k_0$ and  $\cos\left(\theta_{k_0}-\theta_{k_1}\right)$ of order one, $k_1\sim \lambda^{-1}$ and $k_2\sim\lambda^2$.  Finally the normal hierarchical (NH) spectrum can be obtained  only in the case in which $k_2\sim \lambda^{-2}$, $\cos\left(\theta_{k_0}-\theta_{k_1}\right)$ and $k_1$ of order 1  and $k_0\sim\lambda^{2}$.

\subsubsection{Mixing matrices at  Leading Order}

Apart from the charged fermion masses of the second and third families, mixing angles and masses at leading order do not fit the experimental data: the charged fermion first families are massless, the CKM matrix is the unity matrix and the PMNS matrix can account for two maximal rotations in the $(12)$ and $(23)$ sectors:
\beq
V=\unity\;,\qquad \qquad U = R_{23}\left(-\dfrac{\pi}{4}\right)R_{12}\left(\frac{\pi}{4}\right)=\left(
         \begin{array}{ccc}
           1/\sqrt2 & -1/\sqrt2 & 0 \\
           1/2 & 1/2 & -1/\sqrt2 \\
           1/2 & 1/2 & +1/\sqrt2 \\
         \end{array}
       \right)\;.
\eeq
 At this level of approximation, the PMNS matrix corresponds to the BM pattern discussed in the introduction. As already stated, this mixing scheme does not fit the data, due to the large value of the bimaximal solar angle which is out of about $5\sigma$. Only considering the NLO contributions, which we will study in the next section, the model agrees with the measurements.

\subsection{Fermion mass matrices at higher orders}

The NLO contributions in the mass matrices originate from two sources: the first are the higher order terms in the superpotential, while the others come from the insertion of the NLO flavon VEVs in the operators in eqs. (\ref{YukSupPot}, \ref{cnu}). In section \ref{sec:flavonscalarpotential} we will show how the flavons develop VEVs and how they are corrected. Here we anticipate the results, reporting the flavon VEVs in a form which is useful for the discussion in this section:
\beq
\ba{rclrcl}
\mean{\varphi}&=&\left(
                     \begin{array}{c}
                       0 \\
                       1 \\
                       1 \\
                     \end{array}
                   \right)v_\varphi
                   +
                   \left(
                     \begin{array}{c}
                       1 \\
                       0 \\
                       0 \\
                     \end{array}
                   \right)\delta v_\varphi\;,&\qquad
\mean{\varphi'}&=&\left(
                     \begin{array}{c}
                       0 \\
                       1 \\
                       -1 \\
                     \end{array}
                   \right)v_{\varphi'}
                   +
                   \left(
                     \begin{array}{c}
                       1 \\
                       0 \\
                       0 \\
                     \end{array}
                   \right)\delta v_{\varphi'}\;,\\[3mm]
\mean{\chi}&=&\left(
                     \begin{array}{c}
                       0 \\
                       0 \\
                       1 \\
                     \end{array}
                   \right)v_\chi
                   +
                   \left(
                     \begin{array}{c}
                       0 \\
                       1 \\
                       0 \\
                     \end{array}
                  \right)\delta^2 v_\chi\;,&\qquad
\mean{\sigma}&=&v_\sigma\;.
\ea
\label{vev:NLO}
\eeq
Some comments are noteworthy: the subleading corrections are suppressed with respect the LO terms as $\delta v/v\sim \lambda$ and $\delta^2 v/v\sim \lambda^2$ for each flavon; NLO corrections to the second and third entries of $\mean{\vphi}$ and $\mean{\vphi'}$ are present, but they present the same structure of the LO terms and can be re-absorbed; similarly, the NLO corrections to the third entry of $\mean{\chi}$ and to $\mean{\sigma}$ are present, but they can be re-absorbed into the LO terms; the other entries of $\mean{\chi}$ do not receive any corrections at NLO. If we consider the NNLO approximation level, i.e. corrections of relative order $\lambda^2$ with respect the LO terms, we see that the second and the third entries of $\mean{\vphi}$ ($\mean{\vphi'}$) are not (anti-)aligned anymore and that the second entry of $\mean{\chi}$ is filled in. It is interesting that the first entry of $\mean{\chi}$ is still vanishing at this level. We will see the relevance of this structure in a while.

\subsubsection{Dirac mass terms}
The Dirac matter superpotential at NLO is given by
\beq
\begin{split}
\mathcal{W}^{NLO}_{Dir} = &\; \sum^3_{i=1} y_{6,(i)} \dfrac{1}{\Lambda^2} F_L F_3^c (\phi+\phi') X_i^{(5)} + \\
&\;+ \sum^3_{i=1} y_{7,(i)} \dfrac{1}{\Lambda^4} F_L F_2^c \theta^2 \rho X_i^{(6)}
+ \sum^8_{i=1} y_{8,(i)} \dfrac{1}{\Lambda^6} F_L F_2^c \theta^2 (\phi+\phi') X_i^{(7)} + \\
&\;+ \sum^6_{i=1} y_{9,(i)} \dfrac{1}{\Lambda^5} F_L F_1^c \theta^2 \rho X_i^{(8)}+
\sum^7_{i=1} y_{10,(i)} \dfrac{1}{\Lambda^6} F_L F_1^c \theta^2 (\phi+\phi') X_i^{(9)}+\\
&\;+ \sum^{13}_{i=1} y_{11,(i)} \dfrac{1}{\Lambda^7} F_L F_1^c \theta^2 (\phi+\phi') X_i^{(10)}\;,
\end{split}
\label{YukSupPotNLO}
\eeq
where, as for eq. (\ref{YukSupPot}), we adopt a compact notation:
\beq
\ba{lcl}
X_i^{(5)} &\equiv& \{ \vphi\sig ,\, \vphi\chi ,\, \vphi' \chi \}\;,\\[3mm]
X_i^{(6)} &\equiv& \{ \vphi'\sig ,\, \vphi\chi ,\, \vphi' \chi \}\;,\\[3mm]
X_i^{(7)} &\equiv& \{ \vphi^3\chi ,\, \vphi^2\vphi'\chi ,\, \vphi\vphi^{\prime\,2}\chi ,\, \vphi^{\prime\,3}\chi ,\, \vphi^3\sig ,\, \vphi^2\vphi'\sig ,\, \vphi\vphi^{\prime\,2}\sig ,\, \vphi^{\prime\,3}\sig \}\;,\\[3mm]
X_i^{(8)} &\equiv& \{ \vphi^2\chi ,\, \vphi\vphi'\chi ,\, \vphi^{\prime 2}\chi ,\, \vphi^2\sigma ,\, \vphi\vphi'\sigma ,\, \vphi^{\prime 2}\sigma \}\;,\\[3mm]
X_i^{(9)} &\equiv& \{ \vphi^4 ,\, \vphi^3\vphi' ,\, \vphi^2\vphi^{\prime\,2} ,\, \vphi\vphi^{\prime\,3} ,\, \vphi^{\prime\,4} ,\, \chi^4 ,\, \chi^3 \sigma \}\;,\\[3mm]
X_i^{(10)} &\equiv& \{ \vphi^4\chi ,\, \vphi^3\vphi'\chi ,\, \vphi^2\vphi^{\prime\,2}\chi ,\, \vphi\vphi^{\prime\,3}\chi ,\, \vphi^{\prime\,4}\chi ,\, \vphi^4\sigma ,\, \vphi^3\vphi'\sigma ,\, \vphi^2\vphi^{\prime\,2}\sigma ,\, \vphi\vphi^{\prime\,3}\sigma , \\[3mm]
&& \ \vphi^{\prime\,4}\sigma ,\, \chi^5 ,\, \chi^4 \sigma ,\, \chi^3 \sigma^2 \}\;.
\ea
\label{X3}
\eeq
Note that not all of these terms are non-vanishing when the flavons develop VEV: in particular $X_i^{(9)}$ for any $i$ do not give a contribution when the LO VEVs are considered: only when the corrections to the VEVs are introduced, they contribute to the mass matrices. For this reason also the terms with $X_i^{(10)}$ must be taken into account, even if they are suppressed by an additional $\Lambda$. The other terms which are vanishing at this order of approximation are $X_i^{(8)}$ for $i=4,5,6$ and $X_i^{(10)}$ for $i=6,\ldots,13$. When flavons and Higgs fields develop VEVs, we get the following Dirac mass matrices:
\bea
\label{MeNLO}
M^{NLO}_e &=&\!\!  -3
     \left(
       \begin{array}{ccc}
         0 & 0 & 0 \\
         \tilde{y}_4 & 0 & 0 \\
         \tilde{y}_9 & 0 & 0 \\
       \end{array}
     \right)v^d_\rho \lambda^5-3
     \left(
       \begin{array}{ccc}
         0 & \tilde{y}_7 \lambda & 0 \\
         0 & \tilde{y}_2 & 0 \\
         0 & -\tilde{y}_2 & 0 \\
       \end{array}
     \right)v^d_\rho \lambda^3+
    \left(
       \begin{array}{ccc}
         0 & 0 & \tilde{y}_6 \lambda \\
         0 & 0 & \tilde{y}_1 \\
         0 & 0 & \tilde{y}_1 \\
       \end{array}
     \right)v^d_\phi \lambda\;,\\[3mm]
\label{MdNLO}
M^{NLO}_d &=&\!\! \phantom{-3}\left(
       \begin{array}{ccc}
         0 & 0 & 0 \\
         \tilde{y}_4 & 0 & 0 \\
         \tilde{y}_9 & 0 & 0 \\
       \end{array}
     \right)v^d_\rho \lambda^5+
     \left(
       \begin{array}{ccc}
         0 & \tilde{y}_7 \lambda & 0 \\
         0 & \tilde{y}_2 & 0 \\
         0 & -\tilde{y}_2 & 0 \\
       \end{array}
     \right)v^d_\rho \lambda^3+
     \left(
       \begin{array}{ccc}
         0 & 0 & \tilde{y}_6 \lambda \\
         0 & 0 & \tilde{y}_1 \\
         0 & 0 & \tilde{y}_1 \\
       \end{array}
     \right)v^d_\phi \lambda\;,\\[3mm]
\label{MuNLO}
M^{NLO}_u &=&\!\! \phantom{-3} \left(
       \begin{array}{ccc}
         0 & 0 & 0 \\
         \tilde{y}_5 & 0 & 0 \\
         \tilde{y}_{10} & 0 & 0 \\
       \end{array}
     \right)v^u_\phi \lambda^7+
     \left(
       \begin{array}{ccc}
         0 & \tilde{y}_8\lambda & 0 \\
         0 & \tilde{y}_3 & 0 \\
         0 & -\tilde{y}_3 & 0 \\
       \end{array}
     \right)v^u_\phi\lambda^5+
    \left(
       \begin{array}{ccc}
         0 & 0 & \tilde{y}_6 \lambda \\
         0 & 0 & \tilde{y}_1 \\
         0 & 0 & \tilde{y}_1 \\
       \end{array}
     \right)v^u_\phi \lambda\;,
\eea
and $m_D^{NLO}=M^{NLO}_u$, where we used the definitions
\beq
\ba{lcl}
\tilde{y}_1&\equiv& y_1+y_{6,(1)}\lambda\;,\\[3mm]
\tilde{y}_2&\equiv& y_2+y_{7,(1)}\lambda\;,\\[3mm]
\tilde{y}_3&\equiv& y_3+\sum^8_{i=5} y_{6,(i)}\lambda\;,\\[3mm]
\tilde{y}_4&\equiv& {\cal F}_1[y_{9,(i)}]+ \dfrac{1}{\lambda} y_4 \left(\dfrac{\delta v_{\vphi'}}{v_{\vphi'}}-\dfrac{\delta v_\vphi}{v_\vphi}\right)  \;,\\[3mm]
\tilde{y}_5&\equiv& {\cal F}_2\left[y_{11,(i)}\right]+\dfrac{1}{\lambda} {\cal F}_3\left[y_{10,(i)};\, \dfrac{\delta v_\vphi}{v_\vphi},\, \dfrac{\delta v_{\vphi'}}{v_{\vphi'}}\right] +   \dfrac{1}{\lambda^2} {\cal F}_4 \left[y_{5};\, \dfrac{\delta^2 v_\chi}{v_\chi} \right]  \;,\\[3mm]
\tilde{y}_6&\equiv& (y_{6,(2)}+y_{6,(3)}) + \dfrac{1}{\lambda} y_1 \dfrac{\delta v_\vphi}{v_\vphi}\;,\\[3mm]
\tilde{y}_7&\equiv& (y_{7,(2)}+y_{7,(3)}) + \dfrac{1}{\lambda} y_2 \dfrac{\delta v_{\vphi'}}{v_{\vphi'}}\;,\\[3mm]
\tilde{y}_8&\equiv& \sum^4_{i=1} y_{8,(i)} + \dfrac{1}{\lambda} \sum_{i=2}^4y_{3,(i)} \dfrac{\delta v_{\vphi'}}{v_{\vphi'}}\;,\\[3mm]
\tilde{y}_9&\equiv&  {\cal F}_5[y_{9,(i)}]+ \dfrac{1}{\lambda} y_4 \left(\dfrac{\delta v_{\vphi'}}{v_{\vphi'}}+\dfrac{\delta v_\vphi}{v_\vphi}\right)  \;,\\[3mm]
\tilde{y}_{10}&\equiv& {\cal F}_6\left[y_{11,(i)}\right]+\dfrac{1}{\lambda} {\cal F}_7\left[y_{10,(i)};\, \dfrac{\delta v_\vphi}{v_\vphi},\, \dfrac{\delta v_{\vphi'}}{v_{\vphi'}}\right]\;.
\ea
\label{yTilde}
\eeq
In the previous definitions we can see that each $\tilde{y}$ is the sum of two pieces: the first refers to the terms in eq. (\ref{YukSupPotNLO}) when the LO flavon VEVs are considered; the second comes from the terms in eq. (\ref{YukSupPot}) where the NLO flavon VEVs are introduced. The only exception is ${\cal F}_3$ which refers to the term proportional to $X_i^{(9)}$ in eq. (\ref{YukSupPotNLO}) and that give contribution only when the NLO flavon VEVs are considered. These two parts are of the same order of magnitude, since $\delta v/v\sim \lambda$ and $\delta^2 v/v\sim \lambda^2$. Note that ${\cal F}_i$ are distinct linear combinations of the arguments in the squared brackets. The expressions in eqs. (\ref{MeNLO}, \ref{MdNLO}, \ref{MuNLO}) are valid at NLO level and note that the (anti-)alignment between the second and third entries of the (second) third families are still preserved. When considering higher order terms, this feature is lost and the $(1,1)$ entry of each mass matrix is filled in.

The values for the charged fermion masses given in \eq{massLO} are modified only by substituting the coefficients $y_i$ with their tilde-versions:
\beq
y_1\rightarrow \tilde{y}_1\;,\qquad\qquad
y_2\rightarrow \tilde{y}_2\;,\qquad\qquad
y_3\rightarrow \tilde{y}_3\;.
\label{massNLO}
\eeq
At this approximation level, the first family masses are not yet well-described, because they are too small.

\subsubsection{Majorana mass terms}
Moving to the Majorana part of the matter superpotential at NLO, we get
\beq
\label{WmajNLO}
\begin{split}
\mathcal{W}^{NLO}_{Maj} \;=&\; \sum_{i=1}^3\dfrac{k_{3,(i)}}{\Lambda^3}\,F_L F_L\Delta_L X^{(11)}_{i}+ \\
&\; + \dfrac{z_2}{\Lambda^4} F_2^c F_2^c \theta^4 \Delta_R + \\
&\; + \dfrac{z_{3,(1)}}{\Lambda^4} F_2^c F_3^c \theta^2 \Delta_R \vphi\vphi' + \sum_{i=2}^3 \dfrac{z_{3,(i)}}{\Lambda^5} F_2^c F_3^c \theta^2 \Delta_R X^{(12)}_{i-1} +
\sum_{i=4}^6 \dfrac{z_{3,(i)}}{\Lambda^6} F_2^c F_3^c \theta^2 \Delta_R X^{(13)}_{i-3} + \\
&\; + \sum_{i=1}^4\dfrac{z_{4,(i)}}{\Lambda^5} F_1^c F_3^c \theta^2 \Delta_R X^{(1)}_i + \sum_{i=5}^{12}\dfrac{z_{4,(i)}}{\Lambda^6} F_1^c F_3^c \theta^2 \Delta_R X^{(14)}_{i-4} + \\
&\; + \dfrac{z_5}{\Lambda^6} F_1^c F_2^c \theta^4 \Delta_R \vphi \chi+ \\
&\; + \sum_{i=1}^2\dfrac{z_{6,(i)}}{\Lambda^6} F_1^c F_1^c \theta^4 X^{(15)}_i\;,
\end{split}
\eeq
where as usual we used the compact notation with
\beq
\ba{lcl}
X_i^{(11)} &\equiv& \{ \chi^3 ,\, \chi^2\sigma ,\, \sigma^3 \}\;,\\[3mm]
X_i^{(12)} &\equiv& \{ \vphi\vphi'\chi ,\, \vphi\vphi'\sigma \}\;,\\[3mm]
X_i^{(13)} &\equiv& \{ \vphi^2\chi^2 ,\, \vphi\vphi'\chi^2 ,\, \vphi^{\prime\,2}\chi^2 \}\;,\\[3mm]
X_i^{(14)} &\equiv& \{ \vphi^3\chi ,\, \vphi^2\vphi'\chi ,\, \vphi\vphi^{\prime\,2}\chi ,\, \vphi^{\prime\,3}\chi ,\, \vphi^3\sigma ,\, \vphi^2\vphi'\sigma ,\, \vphi\vphi^{\prime\,2}\sigma ,\, \vphi^{\prime\,3}\sigma \}\;,\\[3mm]
X_i^{(15)} &\equiv& \{ \vphi^2 ,\, \vphi^{\prime\,2} \}\;.
\ea
\label{X4}
\eeq
A few comments are in place. Note that all the terms proportional to $k_3$ can be reabsorbed by a redefinition of $\tilde{k}_{0,1,2}$ and that the only new structure which corrects $\cM_L$ comes from the term $F_L F_L \Delta_L \chi$ when we consider the subleading corrections of $\mean{\chi}$: as a result the entries $(1,3)$ and $(3,1)$ of $\cM_L$ are filled in by terms proportional to $\lambda^3$. Regarding the contributions to the Majorana mass matrix for the RH neutrinos, it is important to say that the terms proportional to $z_{3,(i)}$ with $i=1,\ldots,3$ and to $z_{4,(i)}$ with $i=1,\ldots,4$ are vanishing, due to the particular flavon VEV alignment of the model. As a consequence all the NLO contributions to $\cM_R$ are of the order of $\lambda^6$, apart that one to the entry $(2,2)$ which is of the order of $\lambda^4$. Finally, we note that each entry of $\cM_R$ is independent from all the others, being $F_i^c$ singlets of the flavour symmetry, and therefore all the $z_i$ are free parameters with modulus of order 1. We listed only the dominant contributions, but the higher order terms would correspond to subleading corrections, which we can safely neglect in the following.

As a result of this analysis the Majorana masses for LH and RH neutrinos are given by
\beq
M_L^{NLO} = \left(
        \begin{array}{ccc}
          k'_0 & k'_1 \lambda & k'_3\lambda^3 \\
          k'_1 \lambda & k'_0 & 0 \\
          k'_3\lambda^3 & 0 & k'_0+k'_2\lambda^2 \\
        \end{array}
      \right)\,v_L\;,\qquad
M_R^{NLO} = \left(
        \begin{array}{ccc}
          z_6 \la^6 & z_5 \la^6 & z_4 \la^6 \\
          z_5 \la^6 &  z_2 \la^4 & z_3 \la^6  \\
          z_4 \la^6   &z_3 \la^6 & z_1 \\
        \end{array}
      \right)\,v_R\;,
\label{MnuNLO}
\eeq
where with the notation $k'_i$ we account for all the redefinitions done on the parameters. As already stated when discussing the LO mass matrices, the contributions to the effective light neutrino mass matrix come from the type-I and type-II See-Saw mechanisms. The resulting NLO $m_\nu^\text{(type-I)}$ is given by
\beq
\begin{split}
m_\nu^\text{(type-I)}=&\,m_D^{NLO}({M}_R^{NLO})^{-1}(m_D^{NLO})^T\\
=&\;\left(
  \begin{array}{ccc}
    \tilde{y}_6^2\lambda^2 & \tilde{y}_1\tilde{y}_6\lambda & \tilde{y}_1\tilde{y}_6\lambda \\
    \tilde{y}_1\tilde{y}_6\lambda & \tilde{y}_1^2 & \tilde{y}_1^2 \\
    \tilde{y}_1\tilde{y}_6\lambda & \tilde{y}_1^2 & \tilde{y}_1^2 \\
  \end{array}
\right)\dfrac{(v^{u}_\phi)^2\lambda^2}{z_1 v_R}\,,
\end{split}
\eeq
and it is diagonalized by a maximal rotation in the $(23)$ sector and not in the $(12)$ sector as demanded by eq. (\ref{PMNSrots}). As a result, we need that the type-I See-Saw contribution is at least $\mathcal{O}(\lambda^2)$ the type-II See-Saw one, which we can express using  \eq{massNLO} as
\beq
\label{eq:SSrelation}
\dfrac{m_t^2}{z_1 v_R} \leq \lambda^2 k'_0 v_L\;.
\eeq
We remember here that $v_L,v_R$ are the VEVs of $\Delta_L$ and $\Delta_R$ respectively.  In particular $v_L$ is the VEV developed by the SM $(1,3,1)$ triplet component of $\Delta_L$  and it   is induced once the EW symmetry is broken. As we will see in detail  in the next sections  the physical SM  triplet $T\sim   (1,3,1)$ arises by the mixing between the  SM $(1,3,1)$ components of $\Delta_L$ and of  two additional fields, $\Sigma$ and $\Sigma'$ transforming under the PS gauge symmetry as (1,3,3).
The VEV $\vev{T}$ of the $SU(2)_L$ triplet $T$ is related to its mass $M_T$  through the following expression
\beq
\vev{T}\simeq \alpha_{ij} \frac{v^u_i v^u_j}{M_T}\,,
\eeq
where $\alpha_{ij}$ are numerical coefficients arising by the details of the scalar potential and $v^u_1=\vev{h_u},v^u_2=\vev{h_u'}$ are the VEVs of  the two  up-type light Higgs doublets needed  for the realization of our model (see appendix \ref{app-Higgs} for details). Since $v_L$ is the projection of $\vev{T}$ along $\Delta_L$ neglecting fine tuned cases it holds that
\beq
v_L\sim  \vev{T}\,.
\eeq
From \eq{eq:SSrelation} we need therefore that $M_T$ and $v_R$ satisfy
\beq
\label{co1}
M_T\leq \lambda^2 \left( \frac{k'_0  z_1 \alpha_{ij} v^u_i v^u_j}{m_t^2}\right) v_R \,.
\eeq
At the same time neutrino mass data imply that
\beq
\label{co2}
k'_0 \alpha_{ij}\frac{v^u_i v^u_j}{M_T} \leq \cO(1)\,\,\mbox{eV}\,.
\eeq
Combining  the constraints of eqs. (\ref{co1})--(\ref{co2})  we see that in  the most  natural scenario, assuming $\alpha_{ij}=1$, $M_T$ and $v_R$ satisfy
\beq
\ba{c}
v_R\gtrsim 30\, M_T\;,\\[3mm]
10^{12}\, \mbox{GeV}\lesssim M_T\lesssim 10^{13}\,\mbox{GeV} \;.
 \ea
\label{first-range}
\eeq
Nevertheless if we allow the numerical factors $\alpha_{ij}$ laying  in  the range $0.1-10$, then  $M_T$, and  consequently $v_R$, can be reduced  even of two and one  orders of magnitude respectively
\beq
10^{10}\, \mbox{GeV}\lesssim M_T\lesssim 10^{12}\,\mbox{GeV}\;.
\label{second-range}
\eeq

In the following discussion of  lepton mixing and in the phenomenological analysis,  we will assume that indeed type-II See-Saw is dominating and we will neglect the type-I contributions. In the section devoted to the study of the scalar potential  we will justify and find out the region of the parameters space where  indeed type-II See-Saw dominates  over type-I.

\subsubsection{Mixing angles at the Next-to-Leading Order}

Looking at eqs. (\ref{MeNLO})--(\ref{MuNLO})--(\ref{MnuNLO}) we see that the fermion mass matrices are of the required form as in eqs. (\ref{Mnu1})--(\ref{Ml1}). The resulting mixing matrices are modified with respect to the LO approximation and interesting new features follow. On the quark sector, the CKM matrix receives deviations from the unity and at NLO the angle $\theta^q_{12}$ is not vanishing anymore:
\beq
\theta^q_{12} = \dfrac{\la}{\sqrt{2}} \dfrac{|\tilde{y}_7 \tilde{y}_3 - \tilde{y}_2 \tilde{y}_8|}{|\tilde{y}_2| \, |\tilde{y}_3|}\;.
\eeq
Looking at this result, the meaning of the parameter $\lambda$ is clear: it is defined as the ratio of the flavon VEVs over the cut-off of the theory, but it also determines the order of magnitude of the Cabibbo angle. This justify our initial assumption of $\lambda=0.2$.
We note that this result is in concordance with the discussion below equation (\ref{Ml1}). If the second columns of the up- and down-quark matrices are not proportional to each other, we can generate a non-vanishing Cabibbo angle $\theta^q_{12}$, while the two other angles in the CKM matrix are still vanishing.

In the lepton sector, the PMNS matrix corresponds to the BM scheme which receives the corrections as illustrated in eq. (\ref{PMNSrots}) and as a result the solar and reactor angles are given by
\bea
\theta^l_{13}&=&  \dfrac{\la}{2} \Bigg|\dfrac{\tilde{y}_6 }{\tilde{y}_1} -\dfrac{\tilde{y}_7}{\tilde{y}_2} \Bigg|,
\label{theta13LO}\\
\theta^l_{12} &=& \dfrac{\pi}{4} - \dfrac{\la}{2} \Bigg|\dfrac{\tilde{y}_6 }{\tilde{y}_1} +\dfrac{\tilde{y}_7}{\tilde{y}_2} \Bigg|\,.
\label{theta12LO}
\eea
As it is easy to see, the reactor angle and the deviation from the maximal value of the solar angle are of order $\lambda$ and therefore the model is now in agreement with the experimental data\footnote{Deriving eq. (\ref{theta12LO}), we already neglect the corrections which increase the value of the solar angle, instead of decreasing it}. This corresponds to a weaker version of eq. (\ref{anglescomplement}).  At this approximation level, the atmospheric angle remains maximal. As already stated, the forthcoming neutrino appearance experiments will provide a good test for the model, indeed they will be able to verify if $\theta^l_{13}$ is close to its present upper bound.

\subsubsection{Higher order effects}

At the next-to-next-to-leading order (NNLO) and even higher orders, many new terms appear in the superpotential. However, only few of them lead to new terms in the mass matrices, while the rest can be absorbed in redefinitions of the parameters as in eq. (\ref{yTilde}). For this reason we do not report the full list of NNLO contributions, but we just comment on the physical consequences. Three effects are worth mentioning:
\begin{itemize}

\item[-] as expected, the masses of the first families are strongly suppressed. We find that the down-quarks and the electron masses are suppressed by a factor of $\lambda^6$ and the up-quark mass by a factor of $\lambda^8$. This leads to the following mass hierarchies:
\beq
\left|\dfrac{m_d}{m_b}\right|\sim
\left|\dfrac{m_e}{m_\tau}\right|\sim\lambda^5\;,\qquad\qquad
\left|\dfrac{m_u}{m_t}\right|\sim\lambda^7\;.
\eeq
As it is easy to see, the electron and the up-quark masses perfectly fit the experimental values and there is only a small tension in the down-quark sector, where we expected a smaller suppression. However, the large number of parameters, in particular from $\phi$, $\phi'$ as well as $\rho$ couplings, which enter in the definition of the first families allows to correctly fit the down mass, without affecting other observables.

\item[-] the (anti-)alignment of the (23) and (33) elements of the Dirac mass matrices in eqs. (\ref{MeNLO}--\ref{MuNLO}) gets broken as the new terms appear. The new elements are $\lambda^2$ suppressed with respect to the older terms. As a result, the matrix that diagonalizes $M_i\,M_i^\dag$, with $i=e,u,d$, has no longer an exact maximal mixing in the (23) sector. In the lepton sector, this translates to a $\lambda^2$ deviation from maximality in the atmospheric angle of the PMNS matrix. In the quark sector, the angle  $\theta^q_{23}$ becomes of order $\la^2$. It is interesting to note that $\theta^q_{13}$ remains vanishing at this order. It only appears when even stronger suppressed terms are taken into account and is of order $\la^3$, in accordance with the Wolfenstein parametrization \cite{Wolfenstein}.

\item[-] the third columns of the mass matrices in eqs. (\ref{MeNLO})--(\ref{MdNLO}) are proportional to $v^d_\phi$, while the second column of $M_d^{NLO}$ is proportional to $v^d_\rho$ and that one of $M_e^{NLO}$ to $-3 v^d_\rho$. Therefore, also at NLO, eqs. (\ref{MtauMb}) are fulfilled. At the NNLO level, terms proportional to $v^d_\rho$ appear in the third columns of charged lepton and down-type quark matrices and terms proportional to $v^d_\phi$ in the second columns. The new terms are $\la^2 \approx 5\%$ suppressed with respect to the old entries. We thus expect deviations from the relations $|m_\tau|=|m_b|$ and $|m_\mu|=3|m_s|$ at the 5\% level.

\end{itemize}

\subsection{Flavon scalar potential}
\label{sec:flavonscalarpotential}

In this section we comment on the vacuum alignment mechanism which explains the flavon VEVs as in eqs. (\ref{vev:charged:best})--(\ref{vev:neutrinos})--(\ref{vev:FN}). It turns out that our desired alignment is exactly the one presented in  \cite{AFM_BimaxS4}, once we transform all the fields in our basis. Notice indeed that it is possible to identify each flavon of table \ref{table:flavons} with the flavons in \cite{AFM_BimaxS4}, by simply comparing the transformation properties under the full flavour group:
\beq
\vphi\longrightarrow\vphi_l\;,\qquad\quad
\vphi'\longrightarrow\chi_l\;,\qquad\quad
\chi\longrightarrow\vphi_\nu\;,\qquad\quad
\sigma\longrightarrow\xi_\nu\;.
\eeq
Using the following unitary matrix to move from the basis in \cite{AFM_BimaxS4} to our basis,
\beq
\left(
\begin{array}{ccc}
 1 & 0 & 0 \\
  0 & -i/\sqrt{2} & i/\sqrt{2} \\
 0 & 1/\sqrt{2} & 1/\sqrt{2}
\end{array}
\right)\;,
\label{MatrixTraduction}
\eeq
we find (up to irrelevant phases) the following flavon VEV alignment\footnote{The VEVs of the fields $\vphi$ and $\vphi'$ is recovered by applying the unitary matrix in eq. (\ref{MatrixTraduction}) to an equivalent configuration of eq. (51) in \cite{AFM_BimaxS4}, resulting from the application of the element $(TS)^2$ to eq. (51) and of the element $T$ to the eq. (18) of \cite{AFM_BimaxS4}}. We will comment in a while about the presence of equivalent solutions.
\beq
\vphi \propto \bmat 0 \\ 1 \\ 1 \emat \;, \qquad
\vphi' \propto \bmat 0 \\ 1 \\ -1 \emat\;, \qquad
\chi \propto \bmat 0 \\ 0 \\ 1 \emat\;,
\eeq
which correspond to eqs. (\ref{vev:charged:best})--(\ref{vev:neutrinos}).

In \cite{AFM_BimaxS4}, a set of driving superfields has been introduced to guarantee the correct flavon vacuum alignment: these new fields are gauge singlets and transform only under the flavour group, but differently from the flavons, they do not develop vacuum expectation values\footnote{This strictly holds only in the exact supersymmetric phase, while in the broken phase they develop a VEV proportional to the common soft breaking scale \cite{DrivingFields}, usually denoted as $m_{SUSY}$. This could have a relevant impact when discussing on flavour violating processes, as studied in a series of paper \cite{LFVA4} dealing with a context similar to our model.}. Under the continuous $R$-symmetry all the driving fields transform as $U(1)_R=2$ and therefore, constructing the superpotential, they  appear only linearly. For the same purpose, we introduce in our model a set of driving fields which recall those in \cite{AFM_BimaxS4}. In table \ref{table:Rflavons} we show the driving fields and their transformation properties under $S_4\times Z_4$.

\begin{table}[!h]
\begin{center}
\begin{tabular}{|c|c|c|c|c|}
\hline
  &&&& \\[-0,3cm]
  Driving & $D_R$&$\vphi_R$&$\chi_R$&$\sig_R$\\
  &&& & \\[-0,3cm]
\hline
  &&&&\\
  $S_4$ & 2&$3_2$&$3_1$&$1_1$\\
  &&& & \\[-0,3cm]
  $Z_4$ &  $-1$ & $-1$& $1$ & $1$ \\
  &&& & \\
\hline
  \end{tabular}
\end{center}
\caption{\it The driving field content and their transformation properties under $S_4\times Z_4$. They are all singlets under the gauge group and the FN symmetry, while they transform as $U(1)_R=2$ under the continuous $R$-symmetry.}
\label{table:Rflavons}
\end{table}

It is easy to determine the correspondence between our set of driving fields and those of \cite{AFM_BimaxS4}:
\beq
D_R\longrightarrow\vphi_l^0\;,\qquad\quad
\vphi_R\longrightarrow\chi_l^0\;,\qquad\quad
\chi_R\longrightarrow\vphi_\nu^0\;,\qquad\quad
\sig_R\longrightarrow\xi_\nu^0\;.
\eeq

We now construct the driving superpotential $w_d$, which contains only flavons and driving fields and in particular neither matter fields nor Higgses, and look for the conditions that minimize the scalar potential,
\beq
V=\sum_i\left|\dfrac{\derp w_d}{\derp \Phi_i}\right|^2+m_i^2|\Phi_i|^2+\ldots
\eeq
where $\Phi_i$ denote collectively all the scalar fields of the theory, $m_i^2$ are soft masses and dots stand for $D$-terms for the fields charged under gauge group and possible additional soft breaking terms. Since $m_i^2$ are expected to be much smaller that the mass scales involved in $w_d$, it is reasonable to minimize $V$ in the supersymmetric limit and to account for soft breaking effects subsequently.

Since our flavon and driving field content exactly corresponds to the one in \cite{AFM_BimaxS4}, we already know that the VEV alignment in eqs. (\ref{vev:charged:best})--(\ref{vev:neutrinos}) represents an isolate minimum of the scalar potential. We only need to identify the relations which link the VEVs $v_\vphi$, $v_{\vphi'}$, $v_\chi$ and $v_\sig$ among each other in our model. To this purpose we write the driving superpotential, which reads as
\beq
\begin{split}
w_d \;=&\; f_1 D_R \vphi \vphi + f_2 D_R \vphi' \vphi' + f_3 D_R \vphi \vphi' + f_4 \vphi_R \vphi \vphi'+\\
&+ M_1 \La \chi_R \chi + f_5 \chi_R \chi \sig + f_6 \chi_R \chi \chi +\\
&+M_2^2 \La^2 \sig_R + M_3 \La \sig_R \sig + f_7 \sig_R \sig \sig + f_8 \sig_R \chi \chi\;,
\end{split}
\eeq
where the first line deals with only the fields $\vphi$ and $\vphi'$ and the other two with $\chi$ and $\sig$. As also explained in \cite{AFM_BimaxS4}, this leads to the alignment in eqs. (\ref{vev:charged:best}, \ref{vev:neutrinos}) where the VEVs satisfy to
\beq
\ba{c}
f_1 v_\vphi^2 + f_2 v_{\vphi'}^2  + \sqrt{3} f_3 v_\vphi v_{\vphi'} = 0\;,\\[3mm]
v_\sig =-\dfrac{M_1}{f_5} \;,\qquad \qquad
v_\chi^2=\dfrac{f_5^2M_2^2 - f_5 M_1M_3 + f_7M_1^2}{2f_5^2f_8}\;.
\ea
\label{vev:values}
\eeq

The solution in eqs. (\ref{vev:charged:best}, \ref{vev:neutrinos}, \ref{vev:values}) is not unique, but it is possible to introduce a set of soft supersymmetric breaking parameters, which selects this solution as the lowest minimum of the scalar potential.

It is interesting to note the presence of an other source of uncertainty in our solution, which minimizes $V$. Given the symmetry of $w_d$ and the field configurations of eqs. (\ref{vev:charged:best}, \ref{vev:neutrinos}, \ref{vev:values}), by acting on them with elements of the flavour symmetry group $S_4\times Z_4$, we can generate other minima of the scalar potential. These alternative solutions however are physically equivalent to those of the original set and it is not restrictive to analyze the model by choosing as local minimum that one in eqs. (\ref{vev:charged:best}, \ref{vev:neutrinos}, \ref{vev:values}).

For the FN field $\theta$ to acquire a VEV, we assume that the symmetry $U(1)_{FN}$ is gauged such that $\theta$ gets its VEV through a $D$-term. The corresponding potential is of the form:
\begin{equation}
V_{D, FN}=\dfrac{1}{2}(M_{FI}^2- g_{FN}\vert\theta\vert^2+...)^2
\label{TpTB:Dterm}
\end{equation}
where $g_{FN}$ is the gauge coupling constant of $U(1)_{FN}$ and $M_{FI}^2$ denotes the contribution of the Fayet-Iliopoulos (FI) term.
Dots in eq. (\ref{TpTB:Dterm}) represent e.g. terms involving the fields $F_1^c$ and $F_2^c$ which are charged under $U(1)_{FN}$. These terms are however not relevant to calculate the VEV of the FN field and we omit them in the present discussion. $V_{D,FN}$ leads in the supersymmetric limit to
\begin{equation}
|v_\theta|^2=|\langle\theta\rangle|^2= \dfrac{M_{FI}^2}{g_{FN}}\;.
\label{vev:FNvalues}
\end{equation}

It is relevant to underline that the VEVs in eqs. (\ref{vev:values}, \ref{vev:FNvalues}), depend on Mass parameters: all these mass scales naturally have the same order of magnitude and as a result $VEVs/\La_f\sim\la$. The only exceptions are the VEVs of $\varphi$ and $\varphi'$, which depend on a flat direction. In the model, we simply assume that their VEVs have values of the same order of all the other flavon VEVs.

\subsubsection{Higher order contributions}

In this section we briefly comment on the corrections which enter in the flavon VEVs, once the higher order contributions are taken into account. We leave all the details to the appendix \ref{AppB:VEV}.

In the superpotential $w_d$, the flavons which contribute to the Dirac mass terms, $\vphi$ and $\vphi'$, and those which contribute to the Majorana mass terms, $\chi$ and $\sig$, at LO belong to two separated sectors, indeed any mixing term is prevented due to the $Z_4$ symmetry. This situation is not preserved at NLO, since the fields $\chi$ and $\sig$ are neutral under the $Z_4$ symmetry and therefore we can add each of them to all the terms in $w_d$. This leads to modifications to the LO VEV alignment of eq. (\ref{vev:charged:best}) and it turns out that the first entries of $\mean{\vphi}$ and $\mean{\vphi'}$ are filled in, while the second and third entries are corrected by terms which can be however absorbed into the LO ones, without spoiling the alignment. Also the VEVs in eq. (\ref{vev:neutrinos}) receive some corrections: the first and second entries of $\mean{\chi}$ still vanish and the NLO contributions to the third entry can again be absorbed into the LO term. This discussion justifies the results showed in eq. (\ref{vev:NLO}).

Corrections from the NNLO contributions are without particular alignments: in particular, the second and third entries of $\mean{\vphi}$ and $\mean{\vphi'}$ are no longer related, and also the second entry of $\mean{\chi}$ gets a non-zero value. It is interesting to note the the first entry of $\mean{\chi}$ remains zero.

\section{Higgs scalar potential}
\label{Sec:Higgs}

In this section we present the study of the Higgs potential in our model. It is an interesting example of how the introduction of flavour symmetries and the assumptions done to get the correct mass matrices have non-negligible consequences on the Higgs sector. As a result, the study of the Higgs scalar potential and of the gauge and the Yukawa coupling runnings in a general non-flavour PS context does not strictly hold. Notice that even if the following analysis refers to our particular choice of fields and symmetries, our conclusions can be taken as a general hint for a very large class of models that combine a discrete flavour symmetry with a grand unified scenario: indeed our model building strategy shares common features with other constructions. In particular the Higgs fields usually transform under the flavour symmetry $G_f$ and this has direct consequences on  the implementation of the grand unified symmetry breaking. Moreover type-II See-Saw dominance and particular patterns of vanishing projections of the heavy Higgs fields on the light Higgs doublets are frequently required to get the correct fermion mass matrices.

In table \ref{fullhiggs2} we list all the Higgs fields which are necessary to reproduce the correct mass matrices and to implement the desired  PS symmetry breaking pattern: in the first part of the table we report the Higgses already introduced in table \ref{table:higgs}, while the second part contains all the additional fields.

\begin{table}[!h]
\begin{center}
\begin{tabular}{|c||c|c|c|c|}
\hline
  &&&&\\[-0,3cm]
 Higgses &  ${\phi},{\phi}'$& $\rho$& $\Delta_L$ &$\Delta_R$\\
  &&&&\\[-0,3cm]
  \hline
  &&&&\\
  PS & $(1,2,2)$& $(15,2,2)$ & $(\overline{10},3,1)$& $(10,1,3)$ \\
  &&&& \\[-0,3cm]
  $Z_4 $ & $1$ & $-1$ & $1$ &$-1$ \\
  &&&& \\
\hline
  \end{tabular}
  \begin{tabular}{|c||c|c||c|c||c|c||c|}
\hline
 &&&&&&& \\[-0,3cm]
 Higgses & $\overline{\Delta}_L$ & $\overline{\Delta}_R$ & $A$ & $B$ & $\Sigma$ & $\Sigma'$ & $\xi$ \\
 &&&&&&&\\[-0,3cm]
 \hline
 &&&&&&& \\
 PS & $(10,3,1)$ & $(\overline{10},1,3)$ & $(15,1,1)$ & $(15,1,1)$ & $(1,3,3)$ & $(1,3,3)$ & $(1,1,1)$ \\
 &&&&&&& \\[-0,3cm]
 $Z_4 $ & $1$ & $-1$ & $1$ & $-1$ & $1$ & $-1$ & $-1$ \\
 &&&&&&& \\
\hline
  \end{tabular}
\end{center}
\caption{All the Higgs fields of the model and their transformation properties under the gauge group and under $Z_4$. Notice that they are invariant under the other factors of the full flavour symmetry group $G_f$.}
 \label{fullhiggs2}
\end{table}

Some of these Higgs fields are already present in the minimal version of  PS models \cite{PS_MelfoSenj}: typically a (15,1,1) multiplet -- as the fields $A$ and $B$ in table \ref{fullhiggs2} -- is used to break $SU(4)_C$ down to $SU(3)_C \times U(1)_{B-L}$ and to induce the VEVs of the couple of $(10,1,3)\oplus (\ov{10},1,3)$ -- corresponding to the fields $\Delta_R\oplus \ov{\Delta}_R$ in table \ref{fullhiggs2}. The latter VEVs breaks $SU(2)_R\times U(1)_{B-L}$ into the SM hypercharge $U(1)_Y$ concluding the symmetry breaking chain from the PS gauge group to the SM $SU(3)_C\times SU(2)_L\times U(1)_Y$. The field that triggers the EW symmetry breaking is usually a bidoublet (1,2,2) -- as $\phi$ or $\phi'$ in table  \ref{fullhiggs2}. The fields $(\ov{10},3,1)\oplus (10,3,1)$ -- the fields $\Delta_L\oplus \ov{\Delta}_L$ in table \ref{fullhiggs2} -- do not develop VEVs at tree level in the usual minimal PS model, but only when next to leading order terms are taken into account; these are typically suppressed by the Planck scale, as already stated in \cite{PS_MelfoSenj}. For this reason in the minimal PS the type-II See-Saw contributions to the effective neutrino masses are almost negligible.

We identify three main reasons for which the existent studies of the symmetry breaking patterns in the PS context \cite{PS,PS_MelfoSenj} have to be modified and this automatically justifies the presence of the new fields in table \ref{fullhiggs2}:
\begin{itemize}
\item[-] the assumption $v_{\rho}^u=0$ necessary to distinguish the up-quark sector from the others and to recover the up-quark mass hierarchies can be realized only if we include two identical copies of  bidoublet  (1,2,2), $\phi$ and $\phi'$ (see details in appendix \ref{app-Higgs}), and we then impose that four $SU(2)_L$ doublets  (2 up-type and 2 down-type) remain light;
\item[-] since the fields $\rho$, $\Delta_R$ and $\ov{\Delta}_R$ transform non-trivially under the flavour symmetry  $Z_4$, it is necessary to introduce two copies of (15,1,1) multiplets, $A$ and $B$, with opposite $Z_4$ charges, 1 and -1 respectively: $A$ is responsible of inducing the breaking of $SU(2)_R$ through its coupling with $\Delta_R$ and $\ov{\Delta}_R$; $B$ allows the coupling of the bidoublets $\phi,\phi'$  with the bidoublet $\rho$. In this way all of these three fields have a non-vanishing  projection on the light Higgs  $SU(2)_L$ doublets;
\item[-]  the component of $\Delta_L$ which corresponds to the usual SM triplet (1,3,1) can develop a VEV once the EW symmetry is broken only in the presence of  a trilinear coupling with the $SU(2)_L$ Higgs doublets. This coupling cannot be originated with only the Higgs fields $\Delta_L\oplus \ov{\Delta}_L$ and the field content given in table \ref{fullhiggs2}. For this reason we need an additional field which mediate this coupling and the simplest choice would be a bitriplet $\Sigma\sim(1,3,3)$  that can couple with the fields $\phi,\phi'$ or $\rho$ and at the same time can mixes with $\Delta_L$, when $\Delta_R$ develops VEV at the $SU(2)_R$ breaking scale. However, once more, the presence of the $Z_4$ symmetry obliges the introduction of two distinct (1,3,3) Higgs fields, $\Sigma$ and $\Sigma'$, with opposite $Z_4$ charges, $1$ and $-1$ respectively. In this way $\Sigma$ can  couple to the bidoublets $\phi,\phi'$ or $\rho$, while  $\Sigma'$ can mix with $\Delta_L$. Finally, we need a new ingredient that mixes $\Sigma $ with $\Sigma'$: a  PS  singlet  $\xi$  charged $-1$ under $Z_4$ can do the job.
\end{itemize}

The  scalar part of the superpotential is then given by\footnote{Since all the Higgs fields are neutral under the continuous $U(1)_R$, the scalar superpotential explicitly breaks it, while preserving the usual R-parity. The terms in eq. (\ref{scsuper}) could be generated from a $U(1)_R$-conserving superpotential in which the breaking is mediated by additional fields, which are $U(1)_R=2$ and develop non-vanishing VEVs. For instance the mass term $M_\phi \,\phi \phi$ could originate from a trilinear term $X \,\phi \phi$, when $\mean{X}=M_\phi$. Similarly the trilinear coupling $\lambda_\xi\Sigma \Sigma' \xi$ could originate by the non-renormalizable term $X\,\Sigma \Sigma' \xi/\La'$, when $\mean{X}/\La'=\lambda_\xi$ and $\La'$ is the energy scale of the dynamics of the field $X$. In our model we simply assume the existence of the terms in eq. (\ref{scsuper}) in the superpotential and allow for an explicit breaking of the $U(1)_R$ symmetry in this sector.}
\beq
\begin{split}
\mathcal{W} \,=&\, \frac{1}{2} M_\phi \,\phi \phi+  \frac{1}{2} M_{\phi'} \,{\phi'} {\phi'}+  M_{\phi\phi'} \,{\phi} {\phi'}+ \frac{1}{2} M_\rho\, \rho \rho+   M_{\Delta_L}\,\Delta_L\overline{\Delta}_L+\\
&\, +M_{\Delta_R}\,\Delta_R\overline{\Delta}_R+\frac{1}{2} M_A \,A A +  \frac{1}{2} M_B \,B B+ \frac{1}{2}M_\Sigma \Sigma\Sigma +  \frac{1}{2}M_{\Sigma'} {\Sigma'}{\Sigma'}+\frac{1}{2}M_\xi\xi \xi+\\
&\, + \lambda_\xi \Sigma \Sigma' \xi + \lambda_{\xi A B} A B  \xi+   \lambda_{\phi \rho} \phi B\rho+ \lambda_{\phi' \rho} \phi' B\rho+ \frac{1}{3} \lambda_A A A A+\\
&\, +\frac{1}{2} \lambda_B B B A +\lambda_L \Delta_L  \ov{\Delta}_L A + \lambda_R  \Delta_R  \ov{\Delta}_R A+\\
&\, +\frac{1}{2}\lambda_{\phi\Sigma}\phi \phi\Sigma+ \frac{1}{2}\lambda_{\phi' \Sigma}\phi' \phi'\Sigma+\lambda_{\phi\phi \Sigma}\phi' \phi\Sigma+\frac{1}{2}\lambda_{\rho\Sigma}\rho\rho\Sigma+\\
&\, +\lambda_{\Delta \Sigma' } \Delta_L  \ov{\Delta}_R \Sigma' + \ov{\lambda}_{\Delta \Sigma' } \Delta_R \ov{\Delta}_L \Sigma' +\frac{1}{3}\lambda_\Sigma \Sigma\Sigma\Sigma +\frac{1}{2}\lambda_{\Sigma'} \Sigma' \Sigma' \Sigma\;.
\end{split}
\label{scsuper}
\eeq
The vacuum configuration at the GUT breaking scale is given by\footnote{We redefine $\mean{\Delta_R}=v_R$ as $\mean{\Delta_R}=M_R$ in order to adapt to the usual notation.}
\beq
\vev{\Delta_R}=\vev{\overline{\Delta}_R}= M_R\;,\qquad \vev{A}=M_{C_1}\;,\qquad  \vev{B}=M_{C_2}\;,\qquad \vev{\xi}=V_\xi\;.
\eeq
The VEVs $\vev{A}$ and $\vev{B}$ break the $SU(4)_C$ to $SU(3)_C\times U(1)_{B-L}$, while $\vev{\Delta_R}$ and $\vev{\overline{\Delta}_R} $ break   $SU(4)_C\times SU(2)_R$ into $SU(3)_C\times U(1)_Y$.  Therefore  given our field content the colour breaking scale, $M_C=\mbox{Max}(M_{C_1},M_{C_2})$, is never smaller than the $SU(2)_R$ breaking scale, $M_R$. We may expect that   $M_{C_1}\sim M_{C_2}\equiv M_C$ but in principle they could be different.  Finally $V_\xi$ is expected to be close to the flavour breaking scale, due to the $\xi$ gauge singlet nature.  The PS breaking pattern so far sketched  is therefore summarized by
\beq
\ba{c}
SU(4)_C\times SU(2)_L\times SU(2)_R\\
\downarrow\\
SU(3)_C\times U(1)_{B-L}\times SU(2)_L\times SU(2)_R\\
\downarrow\\
SU(3)_C\times SU(2)_L\times U(1)_Y\;.
\ea
\eeq

The F-derivative system obtained by the superpotential  in \eq{scsuper}  reads as
\beq
\ba{l}
M_{\Delta_R}  +\dfrac{3}{\sqrt{2}} \lambda_R  M_{C_1}=0\;,\\[3mm]
M_B M_{C_2}  \sqrt{2} \lambda_B  M_{C_1} M_{C_2}+\lambda_{AB\xi}  M_{C_1} V_\xi=0\;,\\[3mm]
M_A  M_{C_1} + \dfrac{1}{\sqrt{2}}\lambda_B M_{C_2}^2 -\dfrac{2}{\sqrt{3}} \lambda_R M_R^2+ \sqrt{2} \lambda_A  M_{C_1}^2+\lambda_{AB\xi}   M_{C_2} V_\xi =0\;,\\[3mm]
M_\xi V_\xi +\lambda_{AB\xi} M_{C_1}M_{C_2} =0\;.
\ea
\label{Fder}
\eeq
By solving the previous equations, we can express the mass parameters that enter in the superpotential  in term of the adimensional parameters $\lambda_i$ and the physical breaking scales. All the details regarding the mass spectrum are reported in the appendix \ref{app-Higgs}, but some comments are in place. As in the minimal supersymmetric PS \cite{PS_MelfoSenj} when the singlet component of $A$ develops a VEV, there is an accidental $SU(3)$ symmetry involving $\Delta_R$ and $\ov{\Delta}_R$. When the singlet components of  these fields acquire a VEV the accidental symmetry is broken to $SU(2)$ giving rise to 5  Goldstone Bosons (GBs). At the same time $SU(2)_R\times U(1)_{B-L}$  is broken down to $U(1)_Y$, eating 3 of the 5 GBs. Therefore 2 of them, namely $\delta^{++}$ and $\ov{\delta}^{++}$ are left massless, down to the SUSY soft breaking scale $\sim 1$ TeV. This is a well known prediction of SUSY PS theories, which can be tested at LHC \cite{DeltasCharged}. On the other hand, contrary to the minimal case described in \cite{PS_MelfoSenj}, due to the mixing between the Higgs fields $A$ and $B$, no colour octet is lighter than $M_R$.

In order to assure type-II dominance and to get the correct PS symmetry breaking pattern, it is necessary that $M_T\leq M_R\leq M_C\leq V_\xi$. We will see in a while the constraints on the parameters which can be derived from this special mass ordering. For the moment, we just assume this scheme and we describe the mass spectrum in terms of the SM gauge group. Starting from the heaviest energy scale
\begin{itemize}
\item[-] at $V_\xi$\\[4mm]
2 heavy $SU(2)_L$ triplets given in sec. \ref{secT2};

\item[-] between $V_\xi$ and $M_C$
 \begin{tabbing}
{all the SM singlets given in  sec. \ref{secS1} except  one, called $\xi_0$,}\\
{the colour triplets given in sec. \ref{secS2},}\\
{the  colour octets given in sec. \ref{secS3},}\\
{the heavy doublets given in sec. \ref{secD1},}\\
{the two  heavy couples of  $SU(2)_L$  triplets  given in  sec. \ref{secT1};}
 \end{tabbing}

\item[-] at $M_C$\\[4mm]
the color scalars originating by  $\Delta_R\oplus\ov{\Delta}_R$, $\rho$ and  $\Delta_L\oplus\ov{\Delta}_L$:
\[
\begin{cases}
\text{the  $SU(2)_L$ singlets given  from sec. \ref{secS6} to sec. \ref{secS10},}\\
\text{the  $SU(2)_L$ doublets  given  from sec. \ref{secD2} to sec. \ref{secD4},}\\
\text{the $SU(2)_L$ triplets  given in sec. \ref{secT3} and  sec. \ref{secT4};}
\end{cases}
\]

\item[-]  at $M_R$\\[4mm]
the singlet  $\xi_0$;

\item[-]  at $M_T$\\[4mm]
the light  couple of  $SU(2)_L$  triplets  given in  sec.~\ref{secT1};

\item[-]  at $M_{SUSY}$
\begin{tabbing}
{the  scalar singlets   $\delta^{++}$ and $\ov{\delta}^{++}$ given in sec.~\ref{secS5},}\\
{the   $SU(2)_L$ light doublets given in sec.\,\ref{secD1}.}
 \end{tabbing}
\end{itemize}
With the Higgs field content given in tables \ref{table:higgs} and \ref{fullhiggs2} and the scalar spectrum so far sketched we can plug the gauge coupling runnings and see if the conditions given in \eq{first-range} or  those in \eq{second-range} can be satisfied.
Furthermore, in the study of  the gauge coupling runnings (see appendix \ref{AppE} for details) from the $M_{GUT}$ to the EW scale we have to impose the following constraints:
\begin{itemize}
\item[-] recover the EW values for $\alpha_3,\alpha_2$ and $\alpha_1$, related to the gauge couplings of the SM gauge group\;;
\item[-] impose that, at $M_R\leq M_C$, $U(1)_Y$ is originated by the $SU(2)_R\times U(1)_{B-L}$ breaking;
\item[-] impose $\alpha_{B-L}=\alpha_C$ at $M_C$\;;
\item[-] define the GUT scale as  the scale at which the largest $\alpha_i=1$.  In this way we are sure  to be in a perturbative regime up to the GUT scale and thus we are allowed to adopt  the one-loop renormalization group equations (RGEs).
\end{itemize}
Even using all these constraints we are left with two more freedoms, the value of the  $SU(4)_C$ and $SU(2)_R$ breaking scales, i.e. $M_C$ and  $M_R$ respectively.

We adopt two distinct approaches, that we will indicate as the \emph{more constraining} and the \emph{less constraining} ones. In the first case we define $M_C$ the scale at which the  largest $\alpha_i$ is equal or smaller than  $1/4\pi$. In this way all the gauge coupling at $M_C$ are smaller than $1$. In the second case we allow the largest $\alpha_i$ to correspond to a  gauge coupling in the range $1< g_i< 3$.  Then $M_R$ should satisfies eq. (\ref{first-range})  or eq. (\ref{second-range}), but its exact value is not fixed yet.

A few general comments are needed. The non-minimal PS field content affects  the gauge coupling runnings in a non-negligible way. In particular the presence of the charged singlets $\delta^{++}$ and $\ov{\delta}^{++}$  down to $M_{SUSY}$ deeply modifies the  $U(1)_Y$  and $SU(2)_R$ gauge coupling evolution. It turns out that the largest $\alpha_i$  above the $M_R$ scale is always $\alpha_R$. Therefore the two approaches we described can be formulated as follows:
\begin{itemize}
\item[-] More constraining approach $\Longleftrightarrow$ $\alpha_R\leq 1/4\pi$, at $M_C$\,
\item[-] Less constraining approach $\Longleftrightarrow$ $1< g_R< 3$, at $M_C$\;.
\end{itemize}

\subsubsection*{More constraining approach}

\begin{figure}[ht!]
  \centering
  \includegraphics[width=17cm]{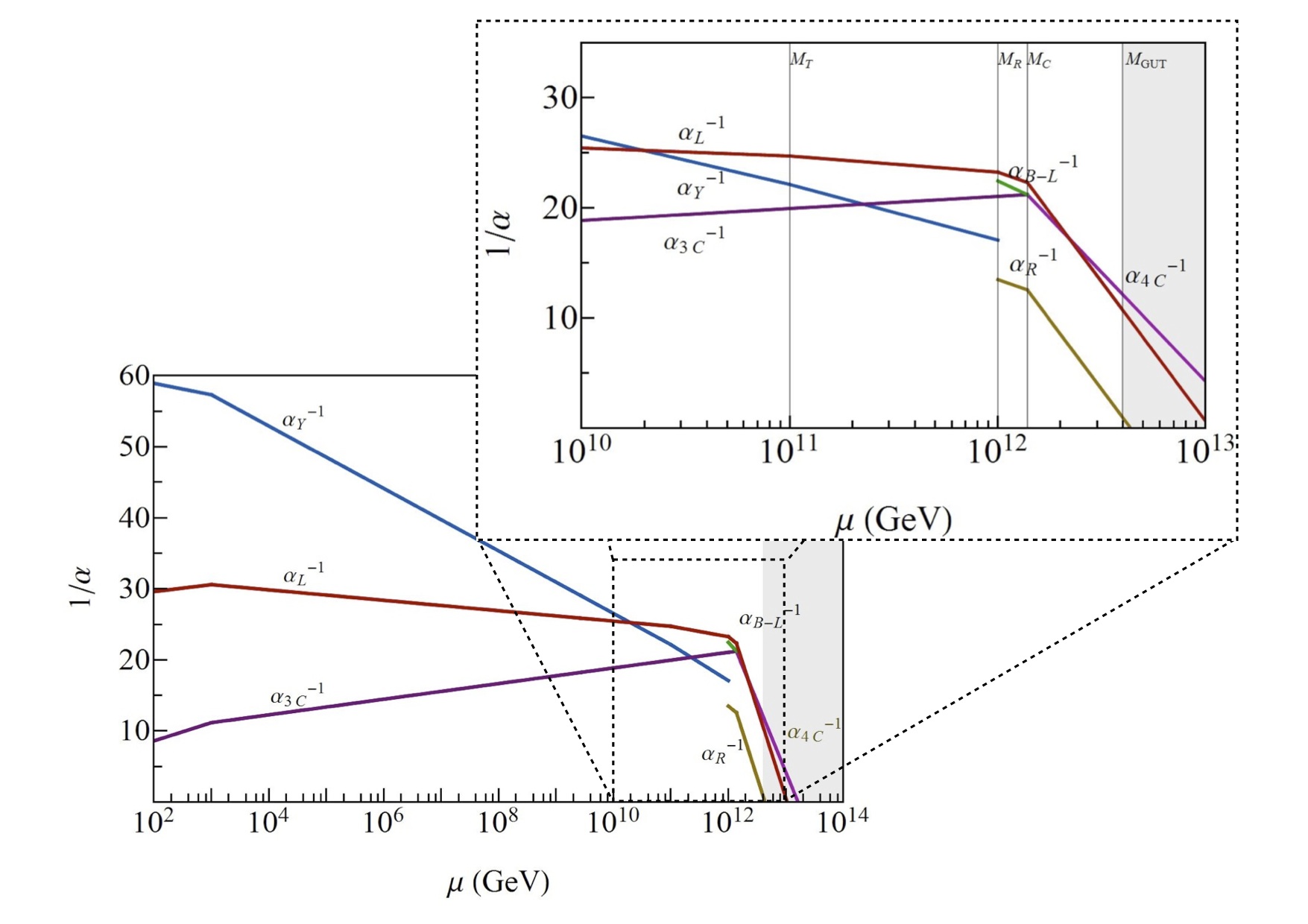}
  \caption{\it The running of the gauge coupling constants in the more constraining approach. $M_T = 10^{11}$ GeV,  $M_R = 10^{12}$ GeV, $M_C = 1.4 \times 10^{12}$ GeV (where $\alpha_R = 1/4 \pi$) and $M_{gut} = 4.0 \times 10^{12}$ GeV (where $\alpha_R = 1$). In the dotted figure, we show a detail of the full plot, restricting the energy scale inside the range $10^{10}\div10^{13}$ GeV.}
  \label{fig:gaugeRun-no}
\end{figure}

In this case there are no solutions neither for the ranges of values of $M_T$ and $M_R$ given in \eq{first-range} nor for those in  \eq{second-range}. Indeed we find that $M_R\leq 10^{12}$ GeV, as can be seen in  fig. \ref{fig:gaugeRun-no}. In other words if we adopt this constraining approach to fix the value of $M_C$, the type-I and type-II See-Saw scales require Yukawa parameters which are at least 2 order of magnitude far from their natural values to reproduce the correct neutrino mass scale. We cannot be satisfied with such a solution because in this case the type-II See-Saw dominance is obtained by increasing the Yukawa couplings in the right neutrino sector and at the same time reducing the coupling of the left-handed neutrinos with the scalar triplet.  Even if this may be considered a solution,  our challenge was to provide a justification of type-II See-Saw dominance through the analysis of the Higgs scalar  potential and not by tuning the Yukawa parameters. Moreover we introduced a FN Abelian symmetry to explain the small ($\leq 10^{-2} $) Yukawa parameters  necessary in the charged fermion sector  to reproduce the correct mass hierarchies. The presence of Yukawa parameters of this order in the purely left-handed neutrino sector makes  the introduction of the FN symmetry questionable. Notice that in the discussion of the neutrino spectrum we have enlarged the parameter range up to values not smaller than $\lambda^2$.

\subsubsection*{Less constraining approach}
\begin{figure}[th!]
  \centering
  \includegraphics[width=17cm]{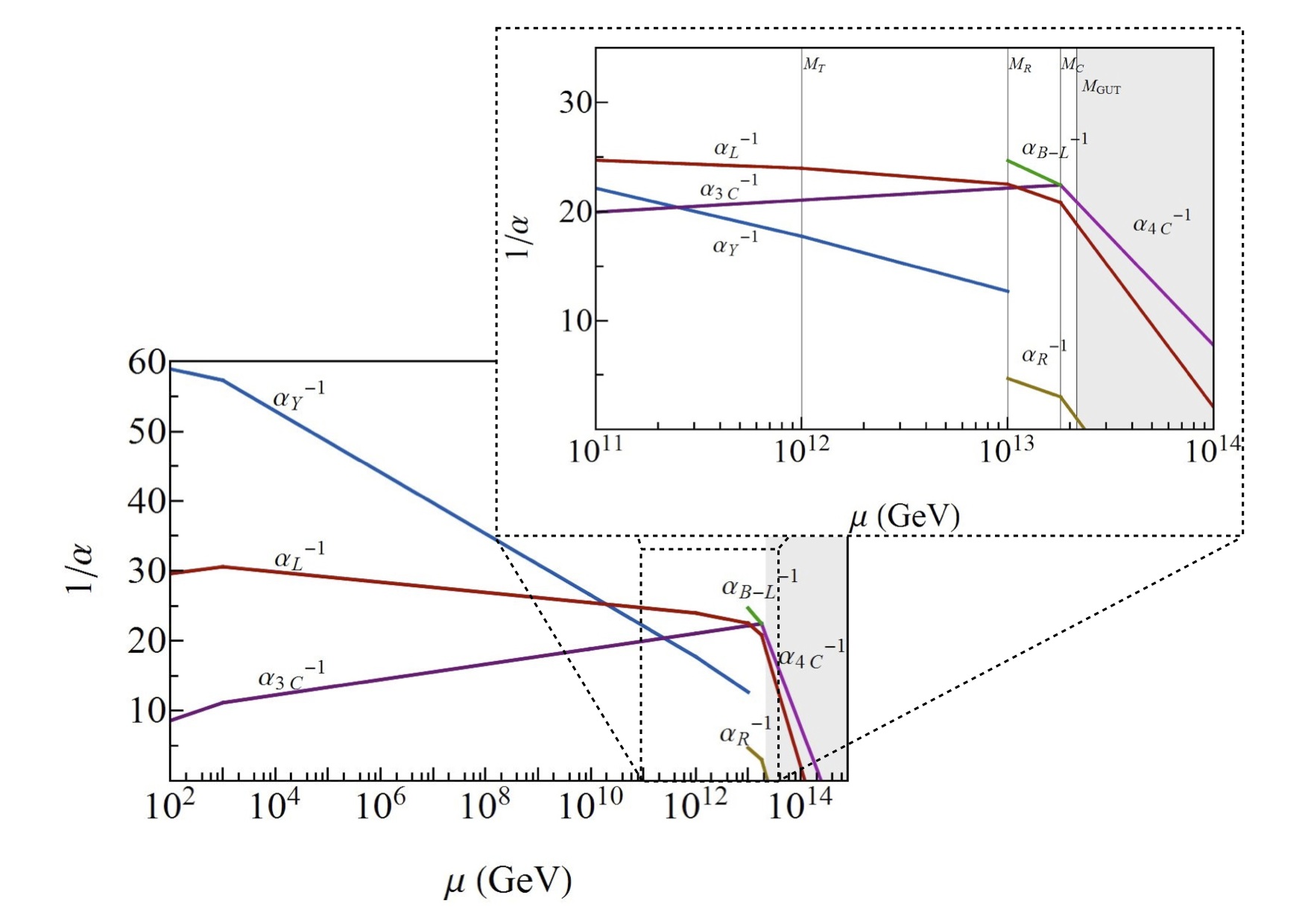}
  \caption{\it The running of the gauge coupling constants in the less constraining approach. $M_T = 10^{12}$ GeV,  $M_R = 10^{13}$ GeV, $M_C = 1.8 \times 10^{13}$ GeV (where $\alpha_R = 1/3$) and $M_{gut} = 2.2 \times 10^{13}$ GeV (where $\alpha_R = 1$). In the dotted figure, we show a detail of the full plot, restricting the energy scale inside the range $10^{11}\div10^{14}$ GeV.}
  \label{fig:gaugeRun-ok}
\end{figure}

In this case there are solutions only for the second range of values of $M_T$ and $M_R$ given in \eq{second-range}. As can be seen in  fig. \ref{fig:gaugeRun-ok}, $M_R$ can now  reach the value of $10^{13}$ GeV.  However the three scales $M_R$, $M_C$ and $M_{GUT}$ are compressed in a narrow region around $10^{13}$ GeV and therefore our model is described by an extended MSSM model almost up to the GUT scale\footnote{Above this energy scale an other gauge structure could be active. We do not take in consideration an high-energy completion of the model, but it is reasonable that larger gauge groups or particular constructions could be present at these energies: for example an $SO(10)$ inspired approach in which fermions do not belong to a unique representation.}. Nevertheless, its  PS origin is reflected  in the non-trivial relations  between the Yukawa couplings.  In conclusion, by admitting the Yukawa parameters span in a range $0.1$--$10$ and switching  $M_R$, $M_C$ and $M_{GUT}$ very close to each other,  we have found a narrow region of the parameter space where our model could still give a realistic description of fermion masses and mixings and in which  type-II See-Saw dominance is not imposed by hand. \\

To finally consider our model viable, we should study the stability of the flavour structure of the mass matrices under the RGEs from the GUT scale down to the EW one.  The study of the full set of the RGEs of the model presented  is beyond the purpose  of this paper. For this reason we will neither run the parameters of the scalar superpotential nor include and run the parameters of the soft SUSY breaking potential. Under these approximations  the EW vacuum expectation values do not change  from the GUT scale down to the EW one. However this does not affect our conclusions for what concerns the stability of the mass matrix structures since  the EW VEV shifts due to the running factorize out and  leave the Yukawa flavour structure unchanged. The study of the Yukawa  matrix running will be done in the next section.

\section{Yukawa coupling running}
\label{sec:Yuk}

In the previous section we analyzed the constraints on the scalar Higgs sector coming from the requirement of type-II See-Saw dominance and from the presence of the flavour symmetry under which the Higgs fields non-trivially transform. We found that the model is viable only in a small region of the parameter space for which $M_R$, $M_C$ and $M_{GUT}\sim 10^{13}$ GeV are very close to each other. At the same time $M_T$ lies at only one order of magnitude below $M_R$. For these reasons to study the stability of the flavour structure of the fermion mass matrices at low scale we can  consider only the running  from $M_T$ onwards, thus neglecting the running from higher energies. Furthermore we also neglect the running from $M_{SUSY}$ to the EW scale, which would introduce only minor corrections.  We work under the assumption that type-II See-Saw is dominating over type-I and moreover that the effects from the type-I terms under the RGEs are negligible. Therefore in the studying of the Yukawa coupling running we do not take into account the Weinberg operator originating by integrating out the right-handed neutrinos. The error we introduce in this way is less than $\lambda^2$ and we will see in a while that these contributions do not modify our results. Furthermore, we study the stability under the Renormalization Group (RG) running, in the approximation corresponding to the NLO, i.e. considering the mass matrices introduced in eqs. (\ref{MeNLO})--(\ref{MuNLO}).

\subsection{Yukawa matrices at $M_T$}
Since we start the renormalization group running at   $M_T$,  at this  scale we integrate out the $SU(2)_L$ scalar triplet $T$ obtaining an effective Weinberg operator responsible  of the type-II See-Saw contribution. We recall  here the origin of this effective operator. The Majorana parts of the matter superpotential given in eqs. (\ref{cnu}), (\ref{WmajNLO}) contain terms with the coupling
\beq
 F_L F_L\Delta_L\;,
\eeq
while the scalar part of the superpotential  in \eq{scsuper} contains the terms
\beq
\ba{l}
\dfrac{1}{2}\lambda_{\phi\Sigma}\phi \phi\Sigma+ \dfrac{1}{2}\lambda_{\phi' \Sigma}\phi' \phi'\Sigma+\lambda_{\phi\phi \Sigma}\phi' \phi\Sigma+ \dfrac{1}{2}\lambda_{\rho\Sigma}\rho\rho\Sigma+\\[3mm]
+\lambda_{\Delta \Sigma' } \Delta_L  \ov{\Delta}_R \Sigma' + \ov{\lambda}_{\Delta \Sigma' } \Delta_R \ov{\Delta}_L \Sigma' +\dfrac{1}{3}\lambda_\Sigma \Sigma\Sigma\Sigma +\dfrac{1}{2}\lambda_{\Sigma'} \Sigma' \Sigma' \Sigma\,,
\ea
\eeq
that ensures the mixing between the (1,3,1)  ((1,3,-1)) components of $\Delta_L$ ($\ov{\Delta}_L$), $\Sigma$ and $\Sigma'$,  whose lighter combination is identified with $T$ ($\ov{T}$), and provides the coupling of $T$ ($\ov{T}$)  with the light doublets $h_d$ and $h_d'$ ($h_u$ and $h_u'$). The effective Weinberg operator at $M_T$  is given by
\beq
 \alpha_{ij} Y_{{L}_{rs}}\, \frac{ L_r  L_s   h_{u_i} h_{u_j}}{M_T}
\eeq
where $L_i$ is the $SU(2)_L$ lepton doublets, $h_{u_{1}}=h_u$, $h_{u_{2}}=h_{u}'$ and $\alpha_{ij}$ are coefficients arising by the scalar potential and
$Y_L$ is given by
\beq
Y_L = \left(
        \begin{array}{ccc}
          k'_0 & k'_1 \lambda & 0 \\
          k'_1 \lambda & k'_0 & 0 \\
          0 & 0 & k'_0+k'_2\lambda^2 \\
        \end{array}
      \right)\;.
\label{YL}
\eeq
Notice that we are neglecting the $\lambda^3$ terms in $Y_L$, because they would be irrelevant for the following analysis.

For what concerns the charged fermion Yukawa, at $M_T$ the Dirac part of the superpotential is written as
\bea
&&Y_u Q\, U^c h_u+  Y'_u Q\, U^c h_u' + Y_d Q\, D^c h_d+  Y'_d Q\, D^c h_d'+ Y_e L\, E^c h_d+  Y'_e L\, E^c h_d'\,,
\eea
with

\bea
\label{Yuk}
Y_u &=& \frac{1}{\beta} Y_u'=   \left(
       \begin{array}{ccc}
         0 &  \tilde{y}_8\lambda^5 &  \tilde{y}_6 \lambda \\
         \tilde{y}_5 \lambda^6&  \tilde{y}_3 \lambda^4& \tilde{y}_1 \\
         \tilde{y}_{10}   \lambda^6& - \tilde{y}_3 \lambda^4 &  \tilde{y}_1 \\
       \end{array}
     \right)  \;,\nn \\[3mm]
Y_d &=&\!\! U_{13} \left(
       \begin{array}{ccc}
         0 & \tilde{y}_7 \lambda & 0 \\
         \tilde{y}_4   \lambda^2 & \tilde{y}_2 & 0 \\
         \tilde{y}_9  \lambda^2 &- \tilde{y}_2 & 0 \\
       \end{array}
     \right) \lambda^2+
     \left(
       \begin{array}{ccc}
         0 & 0 & \tilde{y}_6 \lambda \\
         0 & 0 & \tilde{y}_1 \\
         0 & 0 & \tilde{y}_1 \\
       \end{array}
     \right) \;,\nn\\[3mm]
Y_d' &=&\!\! U_{23} \left(
       \begin{array}{ccc}
         0 & \tilde{y}_7 \lambda & 0 \\
         \tilde{y}_4   \lambda^2 & \tilde{y}_2 & 0 \\
         \tilde{y}_9  \lambda^2 &- \tilde{y}_2 & 0 \\
       \end{array}
     \right) \lambda^2+\beta
     \left(
       \begin{array}{ccc}
         0 & 0 & \tilde{y}_6 \lambda \\
         0 & 0 & \tilde{y}_1 \\
         0 & 0 & \tilde{y}_1 \\
       \end{array}
     \right) \;,\\[3mm]
Y_e &=&\!\! -3 U_{13} \left(
       \begin{array}{ccc}
         0 & \tilde{y}_7 \lambda & 0 \\
         \tilde{y}_4   \lambda^2 & \tilde{y}_2 & 0 \\
         \tilde{y}_9  \lambda^2 &- \tilde{y}_2 & 0 \\
       \end{array}
     \right) \lambda^2+
     \left(
       \begin{array}{ccc}
         0 & 0 & \tilde{y}_6 \lambda \\
         0 & 0 & \tilde{y}_1 \\
         0 & 0 & \tilde{y}_1 \\
       \end{array}
     \right)\;,\nn
\eea
\bea
Y_e' &=&\!\! -3 U_{23} \left(
       \begin{array}{ccc}
         0 & \tilde{y}_7 \lambda & 0 \\
         \tilde{y}_4   \lambda^2 & \tilde{y}_2 & 0 \\
         \tilde{y}_9  \lambda^2 &- \tilde{y}_2 & 0 \\
       \end{array}
     \right) \lambda^2+\beta
     \left(
       \begin{array}{ccc}
         0 & 0 & \tilde{y}_6 \lambda \\
         0 & 0 & \tilde{y}_1 \\
         0 & 0 & \tilde{y}_1 \\
       \end{array}
     \right)\,.\nn
\eea
The $U$ matrix defines the light $SU(2)_L$ Higgses in term  of the PS Higgs field components, as explicitly written in appendix \ref{app-Higgs}, and  $\tilde{y}_i$  has  to be read as $U_{11} \tilde{y}_i^{(1)}+  U_{12} \tilde{y}_i^{(2)}$ while $\beta \tilde{y}_i$ as $  U_{21} \tilde{y}_i^{(1)}+  U_{22} \tilde{y}_i^{(2)}$.

With respect to the mass matrices given in   eqs. (\ref{MeNLO}--\ref{MuNLO}) we have reabsorbed a power of  $\lambda$.

\subsection{Analytical approximations}

In the appendix \ref{app-Yuk}, we report all the RGEs for the Yukawa matrices, while here we discuss the results. The RGEs present the general compact expressions
\bea
\frac{d Y_L}{d t'}  &=& \mathcal{F}_L\left[Y_{f'} Y^\dag_{f'}\right] Y_L+ Y_L \mathcal{F}_L^T \left[Y_{f'} Y^\dag_{f'} \right]  +\left[  \mathcal{G_L} \left[ \Tr (Y_{f'} Y_{f'}^\dag) \right]-\sum_i c^\nu_i g_i^2 \right]Y_L\,,\nn \\
\frac{d Y_f}{d t'}  &=& \mathcal{F}_f\left[Y_{f'} Y^\dag_{f'}\right] Y_f +\left[  \mathcal{G}_f\left[ \Tr (Y_{f'} Y_{f'}^\dag)\right] -\sum c^f_i g_i^2 \right]Y_f\,,
\eea
where the index $f$ runs over $\{e,u,d\}$, the parameter $t'$ is defined as $t'\equiv t/(16 \pi^2) \equiv\log \mu /(16 \pi^2) $,  $\mathcal{F}_X\left[\ldots\right]$ is a matrix written in terms of the fermion Yukawa matrices $Y_{f'}$, $\mathcal{G}_X\left[\ldots\right]$ is function of the trace in the flavour space over the Yukawa matrices $Y_{f'}$  and $c^f_i$ are the Casimir coefficients related to the group representations (see appendix \ref{app-Yuk} for the details). The generic solutions are  given by
\beq
\ba{l}
\begin{split}
Y_L(\mu)\,\sim&\, \prod_i e^{-c^\nu_i \mathcal{I}_i} \times
\exp \left[ \int_{t'(\mu_0)}^{t'(\mu)}  \mathcal{G}_L\left[ \Tr (Y_{f'} Y_{f'}^\dag)\right] dt' \right] \times
\exp \left[ \int _{t'(\mu_0)}^{t'(\mu)}  \mathcal{F}_L\left[Y_{f'} Y^\dag_{f'}\right] dt' \right] \times\\
&\,\times Y_L(\mu_0) \times
\exp \left[ \int _{t'(\mu_0)}^{t'(\mu)}  \mathcal{F}^T_L\left[Y_{f'} Y^\dag_{f'}\right] dt' \right]\;,
\end{split}\\
\\[-2mm]
\begin{split}Y_f(\mu) \,=\, \prod_i e^{-c^f_i \mathcal{I}_i} \times
\exp \left[ \int_{t'(\mu_0)}^{t'(\mu)}  \mathcal{G}_f\left[\Tr (Y_{f'} Y_{f'}^\dag)\right] dt' \right]  \times
\exp \left[ \int _{t'(\mu_0)}^{t'(\mu)}  \mathcal{F}_f\left[Y_{f'} Y^\dag_{f'}\right] dt' \right] \times
Y_f(\mu_0)\;.
\end{split}
\ea
\eeq
where $\mathcal{I}_i= \int_{t'(\mu_0)}^{t'(\mu)} g_i(t')^2 dt' $. When we fix $\mu_0\sim M_T$ and $\mu\sim M_{SUSY}$ these formulas  can be approximated by
 \beq
 \ba{l}
 \begin{split}
 Y_L(M_{SUSY})\,\simeq&\, \left(1+\mathcal{G}_L\left[\Tr (Y_{f'} Y_{f'}^\dag)\right] \Delta t' -\sum_i  c^\nu_i\mathcal{I}_i \right)Y_L(M_T)+\\
 &+\,\Bigg(\mathcal{F}_L\left[Y_{f'} Y^\dag_{f'}\right]\,Y_L(M_T) + Y_L(M_T)\mathcal{F}^T_L\left[Y_{f'} Y^\dag_{f'}\right] \Bigg)\Delta t' \;,
 \end{split}\\
 \\[-2mm]
\begin{split}
Y_f(M_{SUSY})\simeq   \left(1+ \mathcal{G}_f\left[\Tr (Y_{f'} Y_{f'}^\dag)\right] \Delta t'  -\sum_i  c^f_i\mathcal{I}_i \right) Y_f(M_T)+\mathcal{F}_f\left[Y_{f'} Y^\dag_{f'}\right]\, Y_f(M_{T})\, \Delta t'\;,
\end{split}
 \ea
 \eeq
with $\Delta t'= 1/16 \pi^2 \log (M_{SUSY}/M_{T})\approx0.13$.

For what concerns the quark sector we find the following approximated expressions for the masses of the last two families and the Cabibbo angle
\beq
\label{mass-Msusy}
\begin{array}{rcl}
m^2_t(M_{SUSY}) &\sim& m^2_t \left(1 -2 \sum_i  c^u_i\,\mathcal{I}_i +14\, \gamma\, \Delta t' \right)\;,\\[3mm]
m^2_c(M_{SUSY})&\sim& m^2_c \left(1 -2 \sum_i  c^u_i\,\mathcal{I}_i +6\,\gamma\,\Delta t' \right)\;,\\[3mm]
m^2_b(M_{SUSY})&\sim&  m^2_b \left(1 -2 \sum_i  c^d_i\,\mathcal{I}_i +16\, \gamma\, \Delta t' \right)\;,\\[3mm]
m^2_s(M_{SUSY})&\sim& m^2_s \left(1 -2 \sum_i  c^d_i\,\mathcal{I}_i +8\, \gamma\, \Delta t' \right)\;,\\[3mm]
\theta^q_{12}(M_{SUSY})&\sim& \theta^q_{12} +\dfrac{1}{6\sqrt{2}} U_{13}^2\, \lambda\, \Delta t'\;,
\end{array}
\eeq
where $\gamma= m_t^2/(v_1^u+ \beta v_2^u)^2$ and  the masses and the Cabibbo angle on the right of the previous expressions are intended at the $M_T$ scale. Note that the demand that $m^2_b$ is still positive at $M_{SUSY}$ gives an upper bound on $\gamma$ of 0.7. The charged lepton masses are very similar to the down-quark masses and indeed we have
\beq
\ba{rcl}
m^2_\tau(M_{SUSY}) &\sim&  m^2_\tau \left(1 -2 \sum_i  c^e_i\,\mathcal{I}_i +16\, \gamma\, \Delta t' \right)\;,\\[3mm]
m^2_\mu(M_{SUSY}) &\sim& m^2_\mu \left(1 -2 \sum_i  c^e_i\,\mathcal{I}_i +8\, \gamma\, \Delta t' \right)\;.
\ea
\eeq

We now consider the neutrino sector (see \cite{RunningNu} for a general approach at RGEs with or without flavour symmetries) and the modification due to the RG running. We remember that our model at the GUT scale naturally predicts the QD spectrum, with both normal and inverse ordering, while choosing a less natural range of the parameters space we may have both NH and IH spectrum. To analyze the effect of the RGEs on the neutrino mass matrix $M_\nu$,  it is worth rotate $M_\nu$ of a maximal rotation in the $(12)$ sector. Then, at the high scale, $M_\nu$ is diagonal and reads as
\beq
\label{MnuGUT}
M_\nu =\left( \begin{array}{ccc} k'_0-\lambda k'_1&0&0\\ 0&k'_0+\lambda k'_1&0\\ 0&0& k'_0+\lambda^2 k'_2\end{array}\right) v_L\,,
\eeq
where without loss of generality $k'_0$ can be taken real, by a redefinition on the phases. After the running at $M_{SUSY}$ \eq{MnuGUT} gets a correction $\Delta M_\nu$ that is given by
\beq
\label{DeltaMnu1}
\begin{split}
\dfrac{\Delta M_\nu}{v_L} \,\sim&\,\left(-k'_0 \sum_i c_i^\nu \mathcal{I}_i + \dfrac{13}{2} k'_0 \gamma\, \Delta t' \right) \unity+ \sum_i c_i^\nu \mathcal{I}_i
   \left( \begin{array}{ccc}
      \lambda k'_1 &0&0\\
      0&-\lambda k'_1&0\\
      0&0& 0\end{array}
   \right)+\\
&\,+  \Delta t'\left( \begin{array}{ccc}
                                -2 \lambda  k'_0-13 \lambda k'_1\gamma/2 & -k'_0 \gamma/2 &k'_0w_{-}  + k'_1\lambda \gamma /2 \sqrt{2} \\
                                -k'_0 \gamma/2 & 2 \lambda  k'_0+13 \lambda k'_1\gamma/2 & k'_0 w_{+} + k'_1\lambda \gamma /2 \sqrt{2} \\
                                k'_0 w_{-}+ k'_1\lambda \gamma /2 \sqrt{2} &  k'_0 w_{+}+ k'_1\lambda \gamma /2 \sqrt{2}  &7 \gamma k'_0  \end{array} \right)\;,
\end{split}
\eeq
with $w_{\pm}=(\pm \gamma/\sqrt{2}+\sqrt{2}\lambda)$, for the QD and IH case when $k'_2\sim \mathcal{O}(1)$. For the NH case  characterized by $k_2'\sim \lambda^{-2}$, $\Delta M_\nu$ assumes the following form
\beq
\label{DeltaMnu2}
\begin{split}
\dfrac{\Delta M_\nu}{v_L} \,\sim&\,\left(-k'_0 \sum_i c_i^\nu \mathcal{I}_i + \dfrac{13}{2} k'_0 \gamma\, \Delta t' \right) \unity+ \sum_i c_i^\nu \mathcal{I}_i
   \left( \begin{array}{ccc}
      \lambda k'_1 &0&0\\
      0&-\lambda k'_1&0\\
      0&0& -\lambda^2 k_2'\end{array}
   \right)+\\
&\,+  \Delta t'\left( \begin{array}{ccc}
                                -2 \lambda  k'_0-13 \lambda k'_1\gamma/2 & -k'_0 \gamma/2 &k'_0w_{-}  + k'_-\gamma /2 \sqrt{2} \\
                                -k'_0 \gamma/2 & 2 \lambda  k'_0+13 \lambda k'_1\gamma/2 & k'_0 w_{+} + k'_+ \gamma /2 \sqrt{2} \\
                                k'_0 w_{-}+ k'_-\gamma /2 \sqrt{2} &  k'_0 w_{+}+ k'_+\gamma /2 \sqrt{2}  &7 \gamma (k'_0+k_2'\lambda^2)  \end{array} \right)\;,
\end{split}
\eeq
with $k'_{\pm}=(k'_1\lambda \pm k'_2\lambda^2)$.

{We can now consider the three different cases QD, NH and IH, which the model accounts for at the GUT scale.
\begin{itemize}
\item[-] QD case $\Longrightarrow k'_0, k'_1, k'_2\sim\mathcal{O}(1)$.\\
The  correction given  by the running  induces a rotation in the $(23)$ sector characterized by
\beq
\tan2 \theta^\nu_{23} \sim -\frac{\sqrt{2}\Delta t'}{\lambda} \dfrac{k'_0}{|k'_1|}\gamma\sim -2\sqrt{2} \lambda \dfrac{k'_0}{|k'_1|}\gamma\,,
\eeq
being $\Delta t'\sim 2 \lambda^2$. This is a large contribution, which deviates the atmospheric angle from the initial maximal value, spoiling the agreement with the experimental data at $3\sigma$. A possible way-out would be if this large correction is erased by a corresponding large correction in the charged lepton mass matrix. However this is not the case, because the maximal $\theta^e_{23}$ in the charged lepton mixing matrix is stable under the RG running. As a result, the QD case is not viable.

\item[-] NH case $\Longrightarrow k'_0\sim \lambda^2$, $k'_1\sim \mathcal{O}(1)$ and $k'_2\sim\lambda^{-2}$.\\
The corrections both for  the atmospheric and the reactor angles are of order $ \Delta t' \gamma/2\sqrt{2} \sim 3 \lambda^3$ and can be safely neglected. Analogously, also the mass splittings receive deviations which can be neglected. On the other hand, the charged lepton mixing matrix is stable under the RG running. As a result the three mixing angles at $M_{SUSY}$ can be well approximated with their initial values at $M_T$.

\item[-] IH case $\Longrightarrow k'_0\sim\lambda^2$ and $k'_1, k'_2\sim\mathcal{O}(1)$.\\
All the corrections to the neutrino mixing are of order $\gamma\Delta t'/\sqrt{2}$. While the solar mass splitting receives negligible contributions, the atmospheric mass splitting is corrected as follows

\beq
\Delta m^2_{atm}(M_{SUSY})= \Delta m^2_{atm}\left(1-2\sum_i  c^\nu_i\mathcal{I}_i+13\, \gamma\, \Delta t'\right)\;.
\eeq
Combining the neutrino mixing with the charged lepton one we get the following mixing angles for the lepton sector:
\beq
\label{RGangles}
\ba{l}
\theta^l_{12}(M_{SUSY})\sim \pi/4- \theta^e_{12} +\theta^e_{13} \gamma/2 \Delta t'\;,\\[3mm]
\theta^l_{23}(M_{SUSY})\sim \,\theta^e_{23} + \gamma \Delta t'\;,\\[3mm]
\theta^l_{13}(M_{SUSY})\sim\theta^e_{13}-\theta^e_{12} \gamma/2 \Delta t'\;,
\ea
\eeq
where the corrections coming from the RG running are proportional to $ \Delta t'$. In particular the corrections to $\theta^l_{12}(M_{SUSY})$ and $\theta^l_{13}(M_{SUSY})$ come from the charged lepton sector, while that one to $\theta^l_{23}(M_{SUSY})$ arises only by the neutrino sector. This distinction strictly holds only at this order of approximation, while considering contributions of the order of $\lambda^3$ all the angles receive corrections from both the sectors.

\item[-] IH case $\Longrightarrow k'_2\sim\lambda^2$, $k'_0\sim 1$ and $k'_1\sim \lambda^{-1}$.\\
With respect to the previous case, the solar mass splitting gets a non-negligible correction, that can be written as
\beq
\begin{array}{l}
 \Delta m^2_{sol}(M_{SUSY})= \Delta m^2_{sol}\left(1-2\sum_i  c^\nu_i\,\mathcal{I}_i+14\, \gamma\, \Delta t'\right)\;.
\end{array}
\eeq
For what concerns the atmospheric mass splitting and the lepton mixing angles we recover the same shifts as in the previous case.
\end{itemize}

\section{Neutrino Phenomenological Analysis}
\label{Sec:phenoanalysis}
In the previous section, it was concluded that the quasi-degenerate spectrum is unstable under the RG running and becomes phenomenologically inviable due to too large corrections to the atmospheric mixing angle.  Only the normal and  inverse  hierarchies were shown to be stable and phenomenologically viable. This is different from \cite{AFM_BimaxS4}, where only the normal hierarchy and the quasi-degenerate spectrum with normal ordering were found.

In this section, we will discuss neutrino phenomenology in more detail, focussing on the value of the reactor mixing angle $\theta^l_{13}$ and the possibility of neutrinoless double beta decay.

The analytical expression of the reactor mixing angle is given by eq. (\ref{theta13LO}) at NLO and it is corrected by the RG running as in eq. (\ref{RGangles}) for the  two IH scenarios studied in the previous section. We see that $\theta^l_{13}$ is typically of order $\la \approx 0.2$, so $\sin^2 \theta^l_{13} \approx 0.04$, which is rather large, but still allowed at the $3 \sigma$ level as shown in table \ref{table:OscillationData}.
Here, we complete the study on $\theta^l_{13}$, performing a numerical analysis. As can be seen from eqs. (\ref{MeNLO})--(\ref{MnuNLO}), the neutrino and the charged lepton mass matrices at NLO are a function of many parameters. However the GUT nature of the model allows us to  fix most parameters that occur at LO because they enter in the low energy  expressions for quark  and charged lepton masses, as can been seen comparing    eq. (\ref{massLO})  with eq. (\ref{mass-Msusy}). Note that all dimensionless parameters, i.e. the $\tilde{y}_i$, can be fixed of order 1.

The  other free  parameters can be fixed as random numbers of order 1, except for the cases where we have argued in the previous section that they should have slightly larger or smaller values. Because of the use of random numbers, the predictions of our model are no longer single valued.

\begin{figure}[ht!]
  \centering
  \includegraphics[width=7cm]{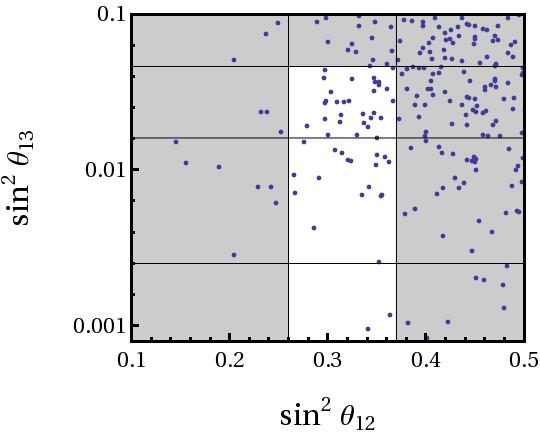}
  \includegraphics[width=7cm]{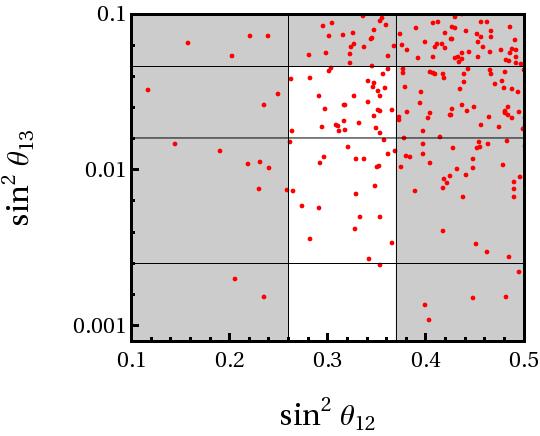}
  \includegraphics[width=7cm]{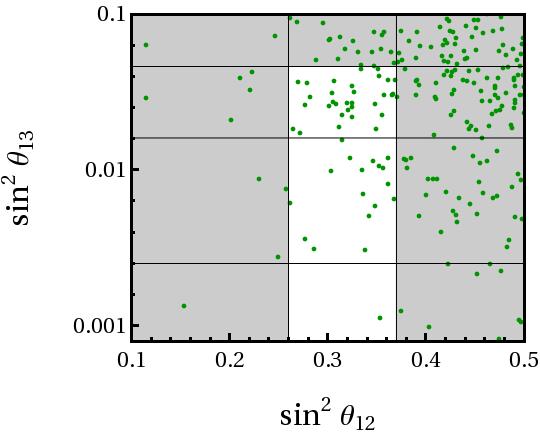}
  \caption{The solar angle versus the reactor angle. On the upper line the two IH cases of the previous section (on the left the first case and on the right the second one), while on the lower line the NH one. The two vertical lines are the 3$\sigma$ bounds for $\sin^2 \theta^l_{12}$ according to \cite{Fogli:Indication}. The upper horizontal line is the 3$\sigma$ upper bound for $\sin^2 \theta^l_{13}$ and the middle line is the best fit value. The lower line is the 95 \% exclusion confidence level after 3 years of Daya Bay data taking \cite{dayabay}.}
  \label{fig:mixingangles}
\end{figure}

We plot the reactor angle versus the solar angle in figure \ref{fig:mixingangles}. At NLO, eq.(\ref{theta12LO}), $\theta^l_{12}$ is driven away from the maximal value $\pi/4$ by a term proportional to $\la$ (note that we take only the corrections which decrease the value of the solar angle, neglecting those which increase it). We see that this deviation is not for all values of the parameters large enough to bring it in the observed region, although this happens for a significant number of them. As explained above, larger values of $\sin^2 \theta^l_{13}$ are favoured and almost all points are in the sensitive region for experiments.

To study neutrinoless double beta decay, we consider the effective $0\nu\beta\beta$ parameter $m_{ee}$, defined as
\beq
m_{ee} = [U\, \diag(m_1, \, m_2, \, m_3) \, U ]_{11}.
\eeq
In figure \ref{fig:double beta} we plot $m_{ee}$ against the lightest neutrino mass, which is $m_1$ and $m_3$ in the NH and IH case respectively. The future experiments are expected to reach good sensitivities: $90$ meV \cite{gerda} (GERDA), $20$ meV \cite{majorana} (Majorana), $50$ meV \cite{supernemo} (SuperNEMO), $15$ meV \cite{cuore} (CUORE) and $24$ meV \cite{exo} (EXO). As a result, looking at figure \ref{fig:double beta}, the whole IH band will be tested in the next future and with it the two cases of our model which allow for the IH spectrum.

\begin{figure}[ht!]
  \centering
  \includegraphics[width=7cm]{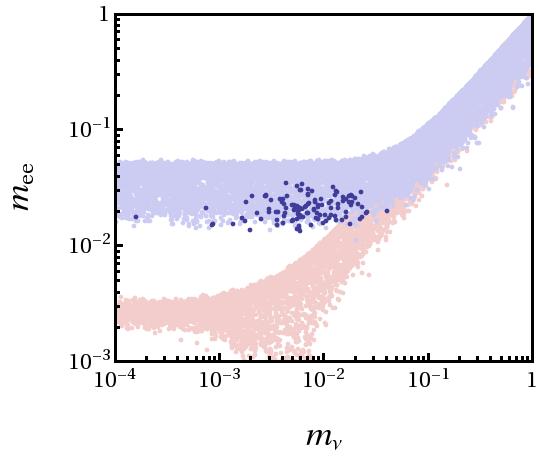}
  \includegraphics[width=7cm]{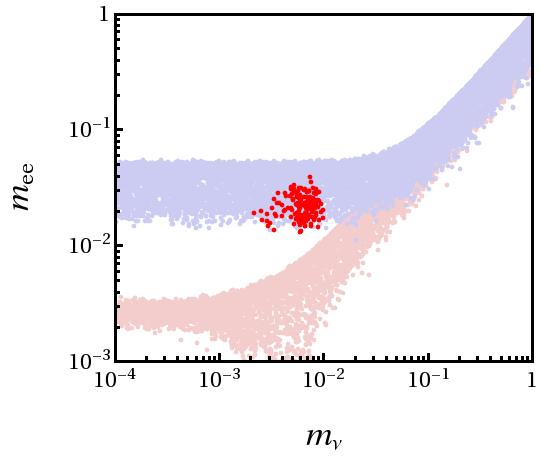}
  \includegraphics[width=7cm]{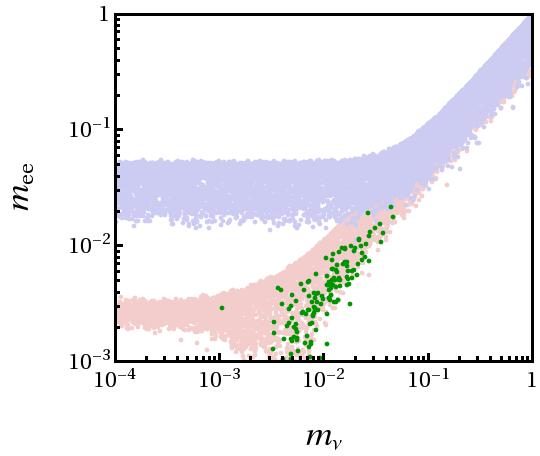}
  \caption{\it Neutrinoless double beta decay plots. On the upper line the two IH cases of the previous section (on the left the first case and on the right the second one), while on the lower line the NH one. The background red (blue) points refer to the allowed region for the NH (IH), taking into account the lepton mixing angle values with their $3\sigma$ errors.}
  \label{fig:double beta}
\end{figure}

\section{Conclusion}
\label{Sec:conclusions}

In this paper we have addressed several aspects of the interplay between a GUT based model and a discrete flavour symmetry. The paper should indeed be considered as the combination of two distinct parts: in the first one we mainly discussed the building of the model from the flavour point of view, while in the second one we faced the problem to justify the assumptions made in the first part and to achieve the correct gauge symmetry breaking chain.

More in detail, the symmetry group of our model is $PS \times G_f$, where $PS$ stands for the GUT Pati-Salam gauge group $SU(4)_C\times SU(2)_L\times SU(2)_R$ and $G_f$ for the flavour group $S_4 \times Z_4\times U(1)_{FN}\times U(1)_R$. Within this GUT context one has the relationship between the down-quark and charged leptons mass matrices, $M_d\sim M_e$, which can easily be used to revise the old idea of quark-lepton complementarity. In the model this is obtained by the use of the non-Abelian discrete flavour  symmetry $S_4$ properly broken through the VEVs of a set of flavon fields, which transform as triplets under $S_4$. The additional Abelian  symmetries, which enters in $G_f$, play different roles: $Z_4$ keeps  quarks separated from leptons and neutrinos from charged leptons and prevents dangerous couplings in the superpotential of the model; $U(1)_{FN}$ helps to justify the charged fermion mass hierarchies; $U(1)_R$ is a common ingredient of supersymmetric flavour models. It contains the discrete $R$-parity and is useful to build a suitable flavon superpotential that allows the correct $S_4$ breaking pattern.

Already at the leading order, the model shows nice features: we are able to reproduce the mass hierarchy between the third and the second charged fermion families, the bottom-tau unification, the Georgi-Jarlskog \cite{GJrelation} relation $|m_\mu|=3 | m_s|$ and, under the assumption of type-II See-Saw dominance at the GUT scale, a realistic neutrino spectrum. However at this level of approximation, both the CKM and the PMNS mixing matrices are not correct: the quark mixing matrix coincides with the identity matrix, while the lepton one is given by the BM pattern. It is worth to recall here that the BM mixing corresponds to maximal solar and atmospheric angles and to a vanishing reactor angle: only the solar angle is not in agreement with the data as it deviates from the experimental central value by a quantity close to the Cabibbo angle, $\lambda\sim0.2$.

At next-to-leading order, the wrong predictions for the fermion mixing angles are corrected: in the CKM matrix, the mixing angle $\theta^q_{12}$ receives contributions of the order of $\lambda$, fitting the value of the Cabibbo angle; analogously, in the PMNS matrix, the solar angle is corrected by the same amount and we find the nice result that $\theta^l_{12}\sim\pi/4-\cO(\lambda)$. At the same time, also the reactor angle receives significant contributions and indeed at this level of approximation it results $\theta^l_{13}\sim\cO(\lambda)$: this is an interesting feature of our model, because this value is close to its present upper bound and it will be tested in the forthcoming neutrino experiments \cite{doublechooz,dayabay,minos,reno,T2K,NOvA}.

Once we consider the higher order terms, we find the other two CKM angles of the correct order of magnitude, $\theta^q_{23}\sim\cO(\lambda^2)$ and $\theta^q_{13}\sim\cO(\lambda^3)$, and small corrections are introduced in the PMNS angles: in particular the atmospheric angle becomes $\theta^l_{23}\sim\pi/4+\cO(\lambda^2)$, justifying a small deviation from the maximality. For what concerns the masses, all the fermions are massive and the mass hierarchies fit the experimental observations.

On the other hand, the neutrino spectrum could be either quasi degenerate or normal or inverse hierarchical. Only the first case corresponds to a completely natural choice of the parameters, which, in the absence of an explanation coming form a higher energy theory, should be of order 1: in order to allow the NH and the IH, the parameters should span in a larger range of values, namely $\lambda^{-2}\div\lambda^2$.

In the second part of the paper we have studied the Higgs scalar potential and the running of both the gauge couplings and the Yukawa mass matrices under the RGEs. With this analysis we looked for the constraints which arise to justify the Higgs field VEV pattern used in the flavour section and the assumption of the type-II See-Saw dominance. The presence of the flavour group $G_f$ modifies the Higgs field content necessary to implement the classical breaking pattern of the PS gauge group and as consequence not all the results obtained by studying minimal versions of PS are recovered. In particular we need the presence of two PS multiplets $(15,1,1)$, $A$ and $B$, responsible to break the  unified colour symmetry $SU(4)_C$ to $SU(3)_C\times U(1)_{B-L}$. Two copies of the $(1,2,2)$ multiplet, $\phi$ and $\phi'$, and one $(15,2,2)$ field, $\rho$, are necessary to implement the condition $v_\rho^u=0$ . Lastly, we need the new fields  $\Sigma,\Sigma'\sim (1,3,3)$ and $\xi\sim (1,1,1)$  to have a type-II See-Saw contribution at tree level. Gauge coupling runnings are affected by the large field content and in particular we found that the requirement of having type-II See-Saw dominance constrains the model in a small region of the parameter space, in which all the heavy mass scales are sandwiched between $10^{11}$ GeV and $10^{13}$ GeV.  At the same time Yukawa mass matrix running shows that while the CKM Cabibbo angle is stable under the RGEs evolution, the PMNS mixing angles are stable only if neutrinos present a NH or an IH spectrum, ruling out the QD case. As already stated, the QD spectrum would be the most natural and probable case at the GUT scale, but the Yukawa RGEs analysis further reduces the allowed region of the  parameter space.

In section \ref{Sec:phenoanalysis}, we performed a brief phenomenological analysis on neutrino observables considering all the constraints which come from the flavour and the Higgs sectors. We have first considered the value of the reactor angle in terms of the deviations of the solar angle from the maximal value and the numerical analysis confirmed the analytical results: in our model, $\theta^l_{13}$ naturally acquires a value not far from its present upper bound.  After this, we studied the neutrinoless double beta decay effective mass, $m_{ee}$, and we have seen that the next future experiments are expected to reach sufficiently good sensitivities to test our model in the IH regime.

As a final comment, we underline that there is a strong tension in combining a GUT model with a (discrete non-Abelian)  flavour symmetry and therefore a parallel study is not only interesting but also recommended  to provide a viable model. In the first part of the paper we produced a flavour-GUT model that is quite realistic and viable. Only the study done in the second part, regarding the Higgs sector reveals that the model is restricted into a small region of the parameter space, reducing the freedom in the choice of the parameter values. Even if our results are model dependent, our construction shares many features with other models present in literature, where often a detailed discussion of the Higgs sector is missing. In our opinion, this is a serious drawback  and we would suggest to consider the interplay between GUTs and flavour symmetries in this kind of models as well.


\section*{Acknowledgments}

We thank Guido Altarelli and Ferruccio Feruglio for useful comments.
The work of RdAT and FB is part of the research program of the Foundation for Fundamental Research of Matter (FOM). The work of FB has also been partially supported by the National Organization for Scientific Research (NWO).
The work of LM has been partly supported by the European Commission under contracts MRTN-CT-2006-035505 and by the European Program ``Unification in the LHC Era'', contract PITN-GA-2009-237920 (UNILHC).

\appendix
\section{The symmetric group $S_4$}
\label{AppA}

In this appendix we report the character table and the Clebsch-Gordan coefficients of the $S_4$ discrete group in our basis.

\begin{table}[h]
\begin{center}
\begin{tabular}{|c||c|c|c|c|c|c|c||c|}
  \hline
  & $n$ & $h$ & $\chi_{1_1}$ & $\chi_{1_2}$ & $\chi_2$ & $\chi_{3_1}$ & $\chi_{3_2}$ & Elements \\
  \hline
  $\cC_1$ & 1 & 1 & 1 & 1 & 2 & 3 & 3 & $\unity$ \\
  $\cC_2$ & 3 & 2 & 1 & 1 & 2 & -1 & -1 & $T^2$, $ST^2S$, $ST^2ST^2$ \\
  $\cC_3$ & 6 & 2 & 1 & -1 & 0 & 1 & -1 & $S$, $T^3ST$, $TST^3$, $T^2ST^2$, $ST^2ST$, $TST^2S$ \\
  $\cC_4$ & 8 & 3 & 1 & 1 & -1 & 0 & 0 & $TS$, $ST$, $(TS)^2$, $(ST)^2$, $T^2ST$, $TST^2$, $T^3ST^2$, $T^2ST^3$ \\
  $\cC_5$ & 6 & 4 & 1 & -1 & 0 & -1 & 1 & $T$, $T^3$, $ST^2$, $T^2S$, $STS$, $TST$ \\
  \hline
\end{tabular}
\end{center}
\caption{\it Character table of the $S_4$ discrete group. $\cC_i$ are the conjugacy classes, $n$ the number of elements in each class, $h$ the smallest value for which $\chi^h=\unity$. In the last column we have reported the elements for each class in terms of the generators of the group.}
\end{table}

The generators, $S$ and $T$, obey to the following rules
\beq
T^4 = S^2 = (ST)^3 = (TS)^3 = \unity\;,
\eeq
and are of the following form for the five different representations:
\[
\ba{lll}
1_1:& S=1\,, & T=1\,,  \\[3mm]
1_2:& S=-1\,, & T= -1\,, \\[3mm]
2:& S=\dfrac{1}{2}\left(
         \begin{array}{cc}
            1  & \sqrt3 \\
             \sqrt3 & -1 \\
         \end{array}
	\right)\,, & 
	T=\left(
         \begin{array}{cc}
            -1  & 0 \\
             0 & 1 \\
         \end{array}
	\right)\,,  \\[5mm]
3_1:& S=\left(
        \begin{array}{ccc}
            0 & 0 & -1 \\
            0 & 1 & 0 \\
           -1 & 0 & 0 \\
        \end{array}
	\right)\,, & 
	T=\left(
        \begin{array}{ccc}
             -1 & 0 & 0 \\
             0 & 0 & -1 \\
             0 & 1  & 0 \\
        \end{array}
	\right)\,,  \\[7mm]
3_2:& S=\left(
        \begin{array}{ccc}
            0 & 0 & 1 \\
            0 & -1 & 0 \\
           1 & 0 & 0 \\
        \end{array}
	\right)\,, & 
	T=\left(
        \begin{array}{ccc}
             1 & 0 & 0 \\
             0 & 0 & 1 \\
             0 & -1 & 0 \\
        \end{array}
	\right)\,.
\ea
\]
Using these generators we calculate the Clebsch Gordan coefficient for all the Kronecker products. In the following we use $\alpha_i$ to indicate the elements of the first representation of the product and $\beta_i$ to indicate those of the second representation.\\
We start with all the multiplication rules which include the $1$-dimensional representations:
\[
\begin{array}{lcl}
1_1\otimes\eta&=&\eta\otimes1_1=\eta\quad\text{with $\eta$ any representation}\\[-10pt]
\\[8pt]
1_2\otimes1_2&=&1_1\sim\alpha\beta\\[-10pt]
\\[8pt]
1_2\otimes2&=&2\sim\left(\begin{array}{c}
                    -\alpha\beta_2 \\
                    \alpha\beta_1 \\
            \end{array}\right)\\[-10pt]
\\[8pt]
1_2\otimes3_1&=&3_2\sim\left(\begin{array}{c}
                    \alpha\beta_1 \\
                    \alpha\beta_2 \\
                    \alpha\beta_3 \\
                    \end{array}\right)\\[-10pt]
\\[8pt]
1_2\otimes3_2&=&3_1\sim\left(\begin{array}{c}
                            \alpha\beta_1 \\
                            \alpha\beta_2 \\
                            \alpha\beta_3 \\
                    \end{array}\right)
\end{array}
\]
The multiplication rules with the 2-dimensional representation are the following:
\[
\begin{array}{ll}
2\otimes2=1_1\oplus1_2\oplus2&\quad
\text{with}\quad\left\{\begin{array}{l}
                    1_1\sim\alpha_1\beta_1+\alpha_2\beta_2\\[-10pt]
                    \\[8pt]
                    1_2\sim-\alpha_1\beta_2+\alpha_2\beta_1\\[-10pt]
                    \\[8pt]
                    2\sim\left(\begin{array}{c}
                        \alpha_1\beta_2+ \alpha_2\beta_1 \\
                        \alpha_1\beta_1-\alpha_2\beta_2 \\
                    \end{array}\right)
                    \end{array}
            \right.\\[-10pt]
\\[8pt]
2\otimes3_1=3_1\oplus3_2&\quad
\text{with}\quad\left\{\begin{array}{l}
                    3_1\sim\left(\begin{array}{c}
                        \alpha_2\beta_1 \\
                        -\frac{1}{2}(\sqrt{3}\alpha_1\beta_2+\alpha_2\beta_2) \\
                       \frac{1}{2}(\sqrt{3}\alpha_1\beta_3-\alpha_2\beta_3) \\
                    \end{array}\right)\\[-10pt]
                    \\[8pt]
                    3_2\sim\left(\begin{array}{c}
                         \alpha_1\beta_1 \\
                        \frac{1}{2}(\sqrt{3}\alpha_2\beta_2-\alpha_1\beta_2) \\
                       -\frac{1}{2}(\sqrt{3}\alpha_2\beta_3+\alpha_1\beta_3) \\
                    \end{array}\right)\\
                    \end{array}
            \right.\\[-10pt]
\\[8pt]
2\otimes3_2=3_1\oplus3_2&\quad
\text{with}\quad\left\{\begin{array}{l}
                    3_1\sim\left(\begin{array}{c}
                        \alpha_1\beta_1 \\
                        \frac{1}{2}(\sqrt{3}\alpha_2\beta_2-\alpha_1\beta_2) \\
                       -\frac{1}{2}(\sqrt{3}\alpha_2\beta_3+\alpha_1\beta_3) \\
                    \end{array}\right)\\[-10pt]
                    \\[8pt]
                    3_2\sim\left(\begin{array}{c}
                         \alpha_2\beta_1 \\
                        -\frac{1}{2}(\sqrt{3}\alpha_1\beta_2+\alpha_2\beta_2) \\
                       \frac{1}{2}(\sqrt{3}\alpha_1\beta_3-\alpha_2\beta_3) \\
                    \end{array}\right)\\
                    \end{array}
            \right.\\
\end{array}
\]
The multiplication rules with the 3-dimensional representations are the following:
\[
\begin{array}{ll}
3_1\otimes3_1=3_2\otimes3_2=1_1\oplus2\oplus3_1\oplus3_2\qquad
\text{with}\quad\left\{
\begin{array}{l}
1_1\sim\alpha_1\beta_1+\alpha_2\beta_2+\alpha_3\beta_3 \\[-10pt]
                    \\[8pt]
2\sim\left(
     \begin{array}{c}
       \frac{1}{\sqrt{2}}(\al_2\be_2-\al_3\be_3) \\
       \frac{1}{\sqrt{6}}(-2\al_1\be_1+\al_2\be_2+\al_3\be_3) \\
     \end{array}
   \right)\\[-10pt]
   \\[8pt]
3_1\sim\left(\begin{array}{c}
         \al_2\be_3+\alpha_3\beta_2 \\
         \al_1\be_3+\alpha_3\beta_1  \\
         \al_1\be_2+\alpha_2\beta_1  \\
        \end{array}\right)\\[-10pt]
        \\[8pt]
3_2\sim\left(\begin{array}{c}
         \al_3\be_2-\alpha_2\beta_3 \\
         \al_1\be_3-\alpha_3\beta_1  \\
         \al_2\be_1-\alpha_1\beta_2  \\
         \end{array}\right)
\end{array}\right.
\end{array}
\]
\[
\begin{array}{ll}
3_1\otimes3_2=1_2\oplus2\oplus3_1\oplus3_2\qquad
\text{with}\quad\left\{
\begin{array}{l}
1_2\sim\alpha_1\beta_1+\alpha_2\beta_2+\alpha_3\beta_3\\[-10pt]
        \\[8pt]
2\sim\left(
     \begin{array}{c}
       \frac{1}{\sqrt{6}}(2\al_1\be_1-\al_2\be_2-\al_3\be_3) \\
       \frac{1}{\sqrt{2}}(\al_2\be_2-\al_3\be_3) \\
     \end{array}
   \right)\\[-10pt]
        \\[8pt]
3_1\sim\left(\begin{array}{c}
         \al_3\be_2-\alpha_2\beta_3 \\
         \al_1\be_3-\alpha_3\beta_1  \\
         \al_2\be_1-\alpha_1\beta_2  \\
    \end{array}\right)\\[-10pt]
        \\[8pt]
3_2\sim\left(\begin{array}{c}
         \al_2\be_3+\alpha_3\beta_2 \\
         \al_1\be_3+\alpha_3\beta_1  \\
         \al_1\be_2+\alpha_2\beta_1  \\
    \end{array}\right)\\
\end{array}\right.
\end{array}
\]

\section{Higgs scalar spectrum}
\label{app-Higgs}

We now proceed to presenting the scalar mass matrices for all the fields introduced in sec.~\ref{Sec:Higgs} according to the SM$\sim SU(3)_C\times SU(2)_L\times U(1)_Y$  representations, indicating their origin with respect to the PS and the colour-broken Pati-Salam (CbPS) phase. We recall that in the colour broken phase the symmetry group is given by $SU(3)_C\times SU(2)_L\times SU(2)_R\times U(1)_{B-L}$. For completeness for each field we also indicate the corresponding $T_{3R}$ value, with $T_{3R}$ the diagonal generator of $SU(2)_R$.  We use the same notation as in \cite{Mali}, in which we write the Dirac scalar mass matrices as they could be read directly from the superpotential  of \eq{scsuper} at the scale $M_R$. We label the mass matrices with $S$, $D$ and $T$ according to the singlet, doublet or triplet representations of the $SU(2)_L$ gauge symmetry.

\subsection {$SU(2)_L$ singlets}

\subsubsection{Singlets $(1,1,0)$}
\label{secS1}

\vspace{0.5cm}
\beq
M_{S1}= \left( \begin{array}{ccccc}
			0&0&\frac{3}{\sqrt{2}}\lambda_R M_R&0&0\\
			0&0&\frac{3}{\sqrt{2}}\lambda_R M_R&0&0\\
			\frac{3}{\sqrt{2}}\lambda_R M_R&\frac{3}{\sqrt{2}}\lambda_R M_R&\sqrt{2} M_{C_1} \lambda_A + x&\lambda_{AB\xi}V_\xi+\frac{1}				 {\sqrt{2}} \lambda_B M_{C_2} & \frac{1}{\sqrt{2}} \lambda_{AB\xi} M_{C_2}\\
			0&0& \lambda_{AB\xi}V_\xi+\frac{1}{\sqrt{2}} \lambda_B M_{C_2}& \frac{1}{\sqrt{2}} \lambda_{B\xi} M_{C_1}-y& \frac{1}{\sqrt{2}} 					 \lambda_{AB\xi} M_{C_2}\\
			0&0&  \frac{1}{\sqrt{2}} \lambda_{AB\xi} M_{C_2}& \frac{1}{\sqrt{2}} \lambda_{AB\xi} M_{C_1}& -  \frac{M_{C_1} M_{C_2}}{V_\xi 					 \sqrt{2}} \lambda_{AB\xi}
		\end{array} \right)
\eeq
with
\beq
\ba{ll}
x=\dfrac{1}{2\sqrt{2} M_{C_1}} (-4 \lambda_A M_{C_1}^2 -2\sqrt{2} \lambda_{AB\xi} M_{C_2} V_\xi -2\lambda_B M_{C_2}^2-8\lambda_R M_R^2)\\[3mm]
y=\dfrac{M_{C_1}}{\sqrt{2} M_{C_2}}(\sqrt{2}\lambda_{AB\xi} V_\xi+2 \lambda_B M_{C_2})\;.
\ea
\eeq
$M_{S1}$ has a vanishing eigenvalue that is eaten by the corresponding gauge boson. Moreover it can be checked that one of the singlets, which we call $\xi_0$, has a mass $\sim M_R$ while all the others masses appear as combination of $M_{C1},M_{C2}$ and $V_\xi$.
\beq
\begin{array}{|c||c|c|c|c|}
  \hline
   &&&&\\
  & & \mbox{PS} &\mbox{CbPS}& T_{3R}\\
  \hline
  &&&&\\
  C_1,R_1& \Delta_R& (10,1,3)&(1,1,3,-2)&1\\
  \hline
  &&&&\\
  C_2,R_2 & \ov{ \Delta}_R& (\ov{10},1,3)&(1,1,3,2)&-1\\
  \hline
  &&&&\\
    C_3,R_3& A& (15,1,1)&(1,1,1,0)&0\\
  \hline
  &&&&\\
 C_4,R_4 & B& (15,1,1)&(1,1,1,0)&0\\
   \hline
   &&&&\\
    C_5,R_5&\xi& (1,1,1)&(1,1,1,0)&0\\
  \hline
    \end{array} \nn
\eeq

\subsubsection{Singlets $(3,1,2/3)\oplus (\ov{3},1,-2/3) $}
\label{secS2}
\vspace{0.5cm}
\beq
M_{S2}= \left( \begin{array}{ccc}
			\sqrt{2} \lambda_A M_{C_1}+x &\lambda_{AB\xi} V_\xi +\frac{1}{\sqrt{2}} \lambda_B M_{C_2}& 2\lambda_R M_R\\
			\lambda_{AB\xi} V_\xi +\frac{1}{\sqrt{2}} \lambda_B M_{C_2}&\frac{1}{\sqrt{2}} \lambda_B M_{C_1}-y&0  \\
			2\lambda_R M_R &0 &-\sqrt{2} \lambda_R M_{C_1}\\
		\end{array}\right)
\eeq
$M_{S2}$ has a vanishing eigenvalue: it corresponds to the massless GBs   $(3,1,2/3)\oplus (\ov{3},1,-2/3) $ eaten by the gauge bosons.
\beq
\begin{array}{|c||c|c|c|c|}
\hline
   &&&&\\
  & & \mbox{PS} &\mbox{CbPS}& T_{3R}\\
  \hline
  &&&&\\
  C_1&A& (15,1,1)&(3,1,1,4/3)&0\\
  \hline
  &&&&\\
  C_2& B& (15,1,1)&(3,1,1,4/3)&0\\
  \hline
  &&&&\\
    C_3&  \Delta_R& (10,1,3)&(3,1,3,-2/3)&1\\
  \hline
  &&&&\\
 R_1& A& (15,1,1)&(\ov{3},1,1,-4/3)&0\\
   \hline
   &&&&\\
   R_2&B& (15,1,1)&(\ov{3},1,1,-4/3)&0\\
    \hline
   &&&&\\
   R_3&\ov{ \Delta}_R&  (\ov{10},1,3)&(\ov{3},1,3,2/3)&-1\\
  \hline
    \end{array}
\nn
\eeq

\subsubsection{Singlets $(8,1,0) $}
\label{secS3}
\vspace{0.5cm}
\beq
M_{S3}= \left( \begin{array}{cc}
			-\sqrt{2} M_{C1} \lambda_A +x& V_\xi \lambda_{AB\xi} -\frac{1}{\sqrt{2}}\lambda_B M_{C2}\\
			V_\xi \lambda_{AB\xi} -\frac{1}{\sqrt{2}}\lambda_B M_{C2} &-\frac{1}{\sqrt{2}} M_{C1}\lambda_B-y\\
		\end{array} \right)
\eeq

\beq
\begin{array}{|c||c|c|c|c|}
\hline
   &&&&\\
  & & \mbox{PS} &\mbox{CbPS}& T_{3R}\\
  \hline
  &&&&\\
  C_1,R_1&A& (15,1,1)&(8,1,1,0)&0\\
  \hline
  &&&&\\
  C_2,R_2& B& (15,1,1)&(8,1,1,0)&0\\
  \hline
     \end{array}
\nn
\eeq

\subsubsection{Singlets $(1,1,\pm 1) $}
\label{secS4}
\vspace{0.5cm}

These states correspond to the  last two  massless GBs eaten by the respective  gauge bosons.
Indeed with these last two massless states the total amount of GBs eaten  is $9= 1+ 2\times 3 + 2$,  as it should be for the  breaking $SU(4)_C\times SU(2)_L\times SU(2)_R \to  SU(3)_C\times SU(2)_L\times U(1)_Y$.
\beq
\begin{array}{|c||c|c|c|c|}
\hline
   &&&&\\
  & & \mbox{PS} &\mbox{CbPS}& T_{3R}\\
  \hline
  &&&&\\
C_1  &\Delta_R& (10,1,3)&(1,1,3,-2)&0\\
  \hline
  &&&&\\
R_1  & \ov{\Delta}_R&  (\ov{10},1,3)&(1,1,3,2)&0\\
  \hline
     \end{array}
\nn
\eeq

\subsubsection{Singlets $(1,1,\pm 2) $}
\label{secS5}
\vspace{0.5cm}
Two massless charged singlet  ($\delta^{++}$ and $\ov{\delta}^{++}$).
\beq
\begin{array}{|c||c|c|c|c|}
\hline
   &&&&\\
  & & \mbox{PS} &\mbox{CbPS}& T_{3R}\\
  \hline
  &&&&\\
 C_1 &\Delta_R& (10,1,3)&(1,1,3,-2)&-1\\
  \hline
  &&&&\\
  R_1& \ov{\Delta}_R&  (\ov{10},1,3)&(1,1,3,2)&1\\
  \hline
     \end{array}
\nn
\eeq

\subsubsection{Singlets $(3,1,-1/3)\oplus (\ov{3},1,1/3) $}
\label{secS6}
\vspace{0.5cm}
 $$-\sqrt{2} M_{C1}\lambda_R $$

\beq
\begin{array}{|c||c|c|c|c|}
\hline
   &&&&\\
  & & \mbox{PS} &\mbox{CbPS}& T_{3R}\\
  \hline
  &&&&\\
  C_1&\Delta_R& (10,1,3)&(3,1,3,-2/3)&0\\
  \hline
  &&&&\\
R_1  & \ov{\Delta}_R&  (\ov{10},1,3)&(\ov{3},1,3,2/3)&0\\
  \hline
     \end{array}
\nn
\eeq

\subsubsection{Singlets $(3,1,-4/3)\oplus (\ov{3},1,4/3) $}
\label{secS7}
\vspace{0.5cm}
 $$-\sqrt{2} M_{C1}\lambda_R $$

\beq
\begin{array}{|c||c|c|c|c|}
\hline
   &&&&\\
  & & \mbox{PS} &\mbox{CbPS}& T_{3R}\\  \hline
  &&&&\\
  C_1&\Delta_R& (10,1,3)&(3,1,3,-2/3)&-1\\
  \hline
  &&&&\\
R_1  & \ov{\Delta}_R&  (\ov{10},1,3)&(\ov{3},1,3,2/3)&1\\
  \hline
     \end{array}
\nn
\eeq

\subsubsection{Singlets $(6,1,4/3)\oplus (\ov{6},1,-4/3) $}
\label{secS8}
\vspace{0.5cm}
 $$-2\sqrt{2} M_{C1}\lambda_R $$

\beq
\begin{array}{|c||c|c|c|c|}
\hline
   &&&&\\
  & & \mbox{PS} &\mbox{CbPS}& T_{3R}\\
  \hline
  &&&&\\
C_1  &\Delta_R& (10,1,3)&(6,1,3,2/3)&1\\
  \hline
  &&&&\\
R_1  & \ov{\Delta}_R&  (\ov{10},1,3)&(\ov{6},1,3,-2/3)&-1\\
  \hline
     \end{array}
\nn
\eeq

\subsubsection{Singlets  $(6,1,1/3)\oplus (\ov{6},1-1/3) $}
\label{secS9}
\vspace{0.5cm}
 $$-2\sqrt{2} M_{C1}\lambda_R $$

\beq
\begin{array}{|c||c|c|c|c|}
\hline
   &&&&\\
  & & \mbox{PS} &\mbox{CbPS}& T_{3R}\\  \hline
  &&&&\\
C_1  &\Delta_R& (10,1,3)&(6,1,3,2/3)&0\\
  \hline
  &&&&\\
R_1  & \ov{\Delta}_R&  (\ov{10},1,3)&(\ov{6},1,3,-2/3)&0\\
  \hline
     \end{array}
\nn
\eeq

\subsubsection{Singlets $(6,1,-2/3)\oplus (\ov{6},1,2/3) $}
\label{secS10}
\vspace{0.5cm}
 $$-2\sqrt{2} M_{C1}\lambda_R $$

\beq
\begin{array}{|c||c|c|c|c|}
\hline
   &&&&\\
  & & \mbox{PS} &\mbox{CbPS}& T_{3R}\\
  \hline
  &&&&\\
C_1  &\Delta_R& (10,1,3)&(6,1,3,2/3)&-1\\
  \hline
  &&&&\\
R_1  & \ov{\Delta}_R&  (\ov{10},1,3)&(\ov{6},1,3,-2/3)&1\\
  \hline
     \end{array}
\nn
\eeq

\subsection{$SU(2)_L$ doublets}

\subsubsection{ Doublets $(1,2,\pm1/2)$}
\label{secD1}
\vspace{0.5cm}
\beq
M_{D1}= \left( \begin{array}{ccc}
			M_\phi&M_{\phi\phi'}&\frac{1}{\sqrt{2}} M_{C2}\lambda_{\phi\rho}\\
 			M_{\phi\phi'}&M_{\phi'} & \frac{1}{\sqrt{2}} M_{C2}\lambda_{\phi'\rho}   \\
			\frac{1}{\sqrt{2}} M_{C2}\lambda_{\phi\rho}&\frac{1}{\sqrt{2}} M_{C2}\lambda_{\phi'\rho}&M_\rho+ \frac{1}{\sqrt{2}}\lambda_{\rho A}				 M_{C1}
		\end{array} \right)
\eeq
When  $M_{D1} ^2$ is diagonalized   according to
\beq
 U^T\, \cdot M_{D1} ^2 \cdot U= \hat{M}_{D1} \hat{M}_{D1}
\eeq
we get three up-type (down-type)  Higgs doublets  $h_u,h_u',H_u$ ($h_d,h_d',H_d$).
\beq
\label{lightH}
\left(  \begin{array}{c}
		\phi_{u,d} \\
		\phi_{u,d}' \\
		\rho_{u,d}
	\end{array}\right)
= U^T\left(\begin{array}{c}
			h_{u,d} \\
			h_{u,d}'\\
			H_{u,d}\\
		\end{array}\right)
\eeq
Therefore  for the up (down) projections we have
\beq
\label{eqvev}
\ba{rcccl}
v_\phi^{u,d}&=&\vev{\phi_{u,d}} &=& U_{11} v^{u,d}_1+ U_{21} v^{u,d}_2 +U_{31} v^{u,d}_3\;,\\[3mm]
v_{\phi'}^{u,d}&=& \vev{\phi'_{u,d}} &=&U_{12} v^{u,d}_1+ U_{22} v^{u,d}_2 +U_{32} v^{u,d}\;,\\[3mm]
v_\rho^{u,d}&=&\vev{\rho_{u,d}} &=& U_{13} v^{u,d}_1+ U_{23} v^{u,d}_2 +U_{33} v^{u,d}_3\;,
\ea
\eeq
where  $v^{u,d}_1=\vev{h_{u,d}}, v^{u,d}_2=\vev{h'_{u,d}},v^{u,d}_3=\vev{H_{u,d}}$.  In general  a light doublet, i.e. massless at the $M_C$ scale, gets a mass term at $M_{SUSY}$ and its VEV, at the EW scale, is of order of the EW scale $\sim v_W$. On the contrary for a heavy doublet of mass $M$, its induced VEV at the EW scale is  $\sim v_W^2/M$, that  for $M\sim M_C$ is completely negligible with respect to $v_W$.  Consider now the condition $\vev{\rho_u}=0$ that we imposed to get the correct fermion mass matrices. In the standard case we would have only one up- and one down-type light Higgs doublets, being all the other doublets heavy.  Assume now that $h_{u,d}$ in \eq{lightH} are  the up- and down-type light doublets at $M_C$.  Then  we have  $v^{u,d}_1\sim v_W$ while  $v^{u,d}_2\sim v^{u,d}_3\sim0$.   From \eq{eqvev} we see that  in this case it would be  impossible to make vanishing the $\rho$ projection along the up-direction (that implies $U_{13}=0$) still maintaining   a non-vanishing $\vev{\rho_d}.$\footnote{It is worth to say that even the condition $U_{13}=0$ is not natural and  easily realized. Therefore in the most general case with only one up-type (down-type) light doublet,  the condition $\vev{\rho_u}=0$  implies that all  the other VEVs given in \eq{eqvev} vanish.} For this reason the condition  $\vev{\rho_u}=0$ implies a non-standard scenario and the presence of   two light  doublets of up-type  at the $M_C$ scale, namely $h_u$ and $h_u'$.  However the symmetric nature of $M_{D1}$ ensures that as consequence we are also left  with two down-type light  Higgs doublets, $h_d$ and $h_d'$. Nevertheless this does not imply that  $\vev{\rho_d}$ vanishes because we have
\bea
\label{eqvevrho}
\vev{\rho_u} &=& U_{13} v^u_1+ U_{23} v^u_2 =0\,,\nn\\
\vev{\rho_d} &=& U_{13} v^d_1+ U_{23} v^d_2 \neq0\,,
\eea
since $v^{u,d}_i$ depend on the soft terms.

In conclusion we have to impose two constraints on  the free parameters that enter in $M_{D1}$  corresponding to require that $M_{D1}$ has two vanishing eigenvalues. Notice that the condition of having $M_{D_1}$ of rank 1 is fine-tuned but it is not more fine-tuned than imposing $M_{D_1}$ of rank 2, which is universally accepted whenever the MSSM has to be recovered. In our case the fermion mass matrix structures impose a slightly different condition --$M_{D_1}$ of rank 1 -- but both the requirements are satisfied by fine-tuning the parameters that  enter in the mass matrix.

\beq
\begin{array}{|c||c|c|c|c|}
\hline
  &&&&\\
  & & \mbox{PS} &\mbox{CbPS}& T_{3R}\\
  \hline
  &&&&\\
  C_1&\phi& (1,2,2)&(1,2,2,0)&1/2\\
  \hline
  &&&&\\
  C_2& \phi'& (1,2,2)&(1,2,2,0)&1/2\\
  \hline
  &&&&\\
    C_3&  \rho& (15,2,2)&(1,2,2,0)&1/2\\
  \hline
  &&&&\\
 R_1&\phi& (1,2,2)&(1,2,2,0)&-1/2\\
   \hline
   &&&&\\
   R_2& \phi'& (1,2,2)&(1,2,2,0)&-1/2\\
    \hline
   &&&&\\
   R_3&  \rho& (15,2,2)&(1,2,2,0)&-1/2\\
  \hline
    \end{array}
\nn
\eeq

\subsubsection{ Doublets $(3,2,7/6)\oplus (\ov{3},2,-7/6)$}
\label{secD2}
\vspace{0.5cm}
$$M_\rho +\frac{1}{\sqrt{2}} \lambda_{\rho A} M_{C_1}$$

\beq
\begin{array}{|c||c|c|c|c|}
\hline
   &&&&\\
  & & \mbox{PS} &\mbox{CbPS}& T_{3R}\\
  \hline
  &&&&\\
C_1  &\rho& (15,2,2)&(3,2,2,4/3)&1/2\\
  \hline
  &&&&\\
 R_1 &\rho& (15,2,2)&(\ov{3},2,2,-4/3)&-1/2\\
  \hline
     \end{array}
\nn
\eeq

\subsubsection{ Doublets $(3,2,1/6)\oplus (\ov{3},2,-1/6)$}
\label{secD3}
\vspace{0.5cm}
$$M_\rho +\frac{1}{\sqrt{2}} \lambda_{\rho A} M_{C_1}$$

\beq
\begin{array}{|c||c|c|c|c|}
\hline
  &&&&\\
  & & \mbox{PS} &\mbox{CbPS}& T_{3R}\\
  \hline
  &&&&\\
C_1  &\rho& (15,2,2)&(3,2,2,4/3)&-1/2\\
  \hline
  &&&&\\
 R_1 &\rho& (15,2,2)&(\ov{3},2,2,-4/3)&1/2\\
  \hline
     \end{array}
\nn
\eeq

\subsubsection{ Doublets $(8,2,1/2)\oplus (8,2,-1/2) $}
\label{secD4}
\vspace{0.5cm}
$$M_\rho -\frac{1}{\sqrt{2}} \lambda_{\rho A} M_{C_1}$$

\beq
\begin{array}{|c||c|c|c|c|}
\hline
&&&&\\
  & & \mbox{PS} &\mbox{CbPS}& T_{3R}\\
  \hline
  &&&&\\
C_1  &\rho& (15,2,2)&(8,2,2,0)&1/2\\
  \hline
  &&&&\\
R_1  &\rho& (15,2,2)&(8,2,2,0)&-1/2\\
  \hline
     \end{array}
\nn
\eeq

\subsection{$SU(2)_L$ triplets}

\subsubsection{ Triplets $(1,3,1)\oplus (1,3,-1)$}
\label{secT1}
\vspace{0.5cm}
\beq
M_{T1}= \left( \begin{array}{ccc}
			M_\Sigma& \lambda_\xi V_\xi&0\\
			\lambda_\xi V_\xi&M_{\Sigma'} &\frac{1}{\sqrt{2}}\lambda M_R\\
			0& \frac{1}{\sqrt{2}}\ov{\lambda} M_R & M_{\Delta L}+ \frac{3}{\sqrt{2}} \lambda_L M_{C_1}
		\end{array} \right)
\eeq
with
\beq
\begin{array}{|c||c|c|c|c|}
\hline
  &&&&\\
  & & \mbox{PS} &\mbox{CbPS}& T_{3R}\\
  \hline
  &&&&\\
  C_1&\Sigma& (1,3,3)&(1,3,3,0)&1\\
  \hline
  &&&&\\
  C_2& \Sigma'& (1,3,3)&(1,3,3,0)&1\\
  \hline
  &&&&\\
  C_3& \Delta_L& (\ov{10},3,1)&(1,3,1,2)&0\\
  \hline
  &&&&\\
  R_1&\Sigma& (1,3,3)&(1,3,3,0)&-1\\
  \hline
  &&&&\\
  R_2& \Sigma'& (1,3,3)&(1,3,3,0)&-1\\
  \hline
  &&&&\\
  R_3& \ov{\Delta}_L& ({10},3,1)&(1,3,1,-2)&0\\
  \hline
     \end{array}
\nn
\eeq

\subsubsection{ Triplets $(1,3,0)$}
\label{secT2}
\vspace{0.5cm}
\beq
M_{T2}= \left( \begin{array}{cc}
			M_\Sigma& \lambda_\xi V_\xi\\
			\lambda_\xi V_\xi&M_{\Sigma'}
		\end{array} \right)
\eeq
with
\beq
\begin{array}{|c||c|c|c|c|}
\hline
   &&&&\\
  & & \mbox{PS} &\mbox{CbPS}& T_{3R}\\
  \hline
  &&&&\\
  C_1,R_1&\Sigma& (1,3,3)&(1,3,3,0)&0\\
  \hline
  &&&&\\
  C_2,R_2& \Sigma'& (1,3,3)&(1,3,3,0)&0\\
  \hline
     \end{array}
\nn
\eeq

\subsubsection{ Triplets $(3,3,-1/3)\oplus (\ov{3},3,1/3)$}
\label{secT3}
\vspace{0.5cm}
$$M_{\Delta_L}+\frac{1}{\sqrt{2}}\lambda_L M_{C1}$$
with
\beq
\begin{array}{|c||c|c|c|c|}
\hline
    &&&&\\
  & & \mbox{PS} &\mbox{CbPS}& T_{3R}\\
  \hline
  &&&&\\
C_1&{\Delta}_L& (\ov{10},3,1)&(3,3,1,2/3)&0\\
  \hline
    &&&&\\
R_1&\ov{\Delta}_L& ({10},3,1)&(\ov{3},3,1,-2/3)&0\\
  \hline
     \end{array}
\nn
\eeq

\subsubsection{ Triplets $(6,3,1/3)\oplus (\ov{6},3,-1/3)$}
\label{secT4}
\vspace{0.5cm}
$$M_{\Delta_L}-\frac{1}{\sqrt{2}}\lambda_L M_{C1}$$
with
\beq
\begin{array}{|c||c|c|c|c|}
\hline
   &&&&\\
  & & \mbox{PS} &\mbox{CbPS}& T_{3R}\\
  \hline
  &&&&\\
C_1&{\Delta}_L& (\ov{10},3,1)&(6,3,1,-2/3)&0\\
  \hline
    &&&&\\
R_1&\ov{\Delta}_L& ({10},3,1)&(\ov{6},3,1,2/3)&0\\
  \hline
     \end{array}
\nn
\eeq

\section{NLO contributions to the flavon scalar potential}
\label{AppB:VEV}

The superpotential $w_d$, linear in the driving fields $D_R$, $\varphi_R$, $\chi_R$ and $\sig_R$, is modified into:
\beq
w_d=w^0_d+\delta w_d\;.
\eeq
where $\delta w_d$ contains the NLO contributions, suppressed by one power of $1/\Lambda$ with respect to $w_d$.  The corrective term $\delta w_d$ is given by the most general quartic, $S_4\times Z_4$-invariant polynomial linear in the driving fields, and can be obtained by inserting an additional flavon field in all the LO terms. The $Z_4$-charges prevent any addition of the flavons $\vphi$ and $\vphi'$ at NLO, while a factor of $\sig$ or $\chi$ can be added to all the LO terms. The full expression of $\delta w_d$ is the following:
\beq
\delta w_d=\dfrac{1}{\La}\left(\sum_{i=1}^3x_iI_i^{\sigma_R}+\sum_{i=1}^{5}w_iI_i^{\chi_R}+\sum_{i=1}^6s_iI_i^{D_R}+
\sum_{i=1}^5v_iI_i^{\varphi_R}\right)
\eeq
where $x_i$, $w_i$, $s_i$ and $v_i$ are coefficients and $\left\{I_i^{\sigma_R},\;I_i^{\chi_R},\;I_i^{D_R},\;I_i^{\varphi_R}\right\}$ represent a basis of independent quartic invariants:
\bac{ll}
I_1^{\sigma_R}=\sigma_R\sigma\sigma\sigma\qquad\qquad&
I_3^{\sigma_R}=\sigma_R\sigma(\chi\chi)\\
I_2^{\sigma_R}=\sigma_R(\chi(\chi\chi)_{3_1})\qquad\qquad&
\eac
\bac{ll}
I_1^{\chi_R}=(\chi_R\chi)(\chi\chi)\qquad\qquad&
I_4^{\chi_R}=\left(\chi_R(\chi\chi)_{3_1}\right)\sigma\\
I_2^{\chi_R}=\left((\chi_R\chi)_2(\chi\chi)_2\right)\qquad\qquad& I_5^{\chi_R}=(\chi_R\chi)\sigma\sigma\\
I_3^{\chi_R}=\left((\chi_R\chi)_{3_1}(\chi\chi)_{3_1}\right)\qquad\qquad&
\eac
\bac{ll}
I_1^{D_R}=\left((D_R\chi)_{3_1}(\vphi\vphi')_{3_1}\right)\qquad\qquad&
I_4^{D_R}=\left(D_R(\vphi\vphi)_2\right)\sigma\\
I_2^{D_R}=\left((D_R\chi)_{3_2}(\vphi\vphi')_{3_2}\right)\qquad\qquad&
I_5^{D_R}=\left(D_R(\vphi'\vphi')_2\right)\sigma\\
I_3^{D_R}=\left((D_R\chi)_{3_1}(\vphi'\vphi')_{3_1}\right)\qquad\qquad&
I_6^{D_R}=\left(D_R(\vphi\vphi')_2\right)\sigma\\
\eac
\bac{ll}
I_1^{\varphi_R}=(\varphi_R\chi)'(\vphi\vphi')'\qquad\qquad&
I_4^{\varphi_R}=\left((\varphi_R\chi)_{3_2}(\vphi\vphi')_{3_2}\right)\\
I_2^{\varphi_R}=\left((\varphi_R\chi)_2(\vphi\vphi')_2\right)\qquad\qquad&
I_5^{\varphi_R}=\left(\varphi_R(\vphi\vphi')_{3_2}\right)\sigma\;.\\
I_3^{\varphi_R}=\left((\varphi_R\chi)_{3_1}(\vphi\vphi')_{3_1}\right)\qquad\qquad&
\eac
In the previous terms  we indicate with $(\ldots)$ the singlet $1_1$, with $(\ldots)'$ the singlet $1_2$ and with $(\ldots)_R$ ($R\,=\,2,\,3_1,\,3_2$) the representation $R$.

The NLO flavon VEVs are obtained by imposing the vanishing of the first derivative of $w_d+\delta w_d$ with respect to the driving fields $\sigma_R$, $\chi_R$, $D_R$ and $\varphi_R$. We look for a solution that perturbs eqs. (\ref{vev:charged:best}) and (\ref{vev:neutrinos}) to first order in the $1/\La$ expansion: for all components of the flavons $\Phi=(\sigma,~\chi, ~\vphi, ~\vphi')$, we denote the shifted VEV's by
\beq
\langle \Phi \rangle=\langle \Phi \rangle_{LO}+\delta \Phi
\eeq
where $\langle \Phi \rangle_{LO}$ are given by eqs. (\ref{vev:charged:best}) and (\ref{vev:neutrinos}).

It is straightforward to verify the following results. In the Majorana mass sector the shifts $\delta \sigma,~\delta \chi$ turn out to be proportional to the LO VEV's $\langle \Phi \rangle_{LO}$ and can be absorbed in a redefinition of the parameters $v_\chi$ and $v_\sig$. Instead, in the Dirac mass sector, the shifts $\delta \vphi, ~\delta  \vphi'$ have a non-trivial structure, so that the LO texture is modified:
\beq
\mean{\vphi}=\left(
                     \begin{array}{c}
                       \delta v_\vphi \\
                       v'_\vphi \\
                       v'_\vphi \\
                     \end{array}
                   \right)\qquad
\qquad\mean{\vphi'}=\left(
                     \begin{array}{c}
                       \delta v_{\vphi'} \\
                       v'_{\vphi'} \\
                       -v'_{\vphi'} \\                     
                       \end{array}
                   \right)
\label{vev:charged:nlo}
\eeq
where $v'_\vphi$ and $v'_{\vphi'}$ satisfy a relation similar to that in eq. (\ref{vev:values}) and the shifts $\delta v_\vphi$ and $\delta v_{\vphi'}$ are suppressed by a factor $\lambda$ with respect to the LO entries $v'_\vphi$ and $v'_{\vphi'}$, respectively.

\section{Gauge coupling running}
\label{AppE}

In this appendix, we provide the coefficients of the $\beta$--functions for the gauge coupling running in the different regimes. The complete matter fields run from the GUT scale down to the $M_{SUSY}$ scale, where the SUSY partners decouple. For what concerns the scalar fields in  section \ref{Sec:Higgs}, we have already outlined the scalar spectrum  according to the different scale at which the fields decouple. As a result the computation of the $\beta$--functions is straightforward. Calling $\mu$ the generic scale, we have
\begin{itemize}
\item[-] for $M_C<\mu<M_{GUT}$:  all matter is in the left and right handed multiplets $(4,\, 2,\, 1)$ and $(\ov{4} ,\, 1 \,, 2 )$ as mentioned in table \ref{table:matter}. In the Higgs sector, we have all the fields mentioned in table \ref{fullhiggs2}. This leads to the coefficients
\beq
\beta_{SU(4)_C} = 54\;, \qquad \beta_{SU(2)_L} = 69\;, \qquad \beta_{SU(2)_R} = 69\;.
\eeq
Due to the large matter content these coefficients are  very large and the $\beta$--functions are very steep.  As consequence the theory is in the Pati-Salam regime only for a very small range of energies, as can indeed be seen in figure \ref{fig:gaugeRun-ok}. Almost after passing the scale $M_C$, the $SU(2)_R$ coupling constant enters the non-perturbative regime.
\item[-] $M_R<\mu<M_{C}$:
the theory undergoes to the $SU(3)_C \times SU(2)_L \times SU(2)_R \times U(1)_{B-L}$ symmetry. Considering the matter (left and right handed doublets of quarks and leptons characterized by different $U(1)_{B-L}$ charges) and the scalar fields, we find the coefficients
\beq
\beta_{SU(3)_C} = -3\;, \qquad \beta_{SU(2)_L} = 18\;, \qquad \beta_{SU(2)_R} = 18\;, \qquad \beta_{U(1)_{B-L}} = 24\;.
\eeq

\item[-] $M_T<\mu<M_R$: in this regime we have all  the usual MSSM matter particles, four light higgs doublets (two up-type and two down-type), a couple of SM  triplets $(1,3,1)\oplus (1,3,-1)$ and two  extra charged singlets ($\delta^{++}$ and $\bar{\delta}^{++}$). The coefficients of the $\beta$--functions are
\beq
\beta_{SU(3)_C} = -3\;, \qquad \beta_{SU(2)_L} = 4\;, \qquad \beta_{U(1)_Y} = 69/5\;.
\eeq
The hypercharge that appears in the last term is related to $SU(2)_R$ and the $B-L$ charges in the previous regime by
\[
Y = T_{3R} + \frac{B-L}{2}.
\]

\item[-] $M_{SUSY}<\mu<M_T$: in this regime we have all  the usual MSSM matter particles, four light higgs doublets (two up-type and two down-type) and two  extra charged singlets ($\delta^{++}$ and $\bar{\delta}^{++}$). The $\beta$--function coefficients are
\beq
\beta_{SU(3)_C} = -3\;, \qquad \beta_{SU(2)_L} = 2\;, \qquad \beta_{U(1)_Y} = 12\;.
\eeq
This should be compared with the (-3, 1, 33/5) coefficients of the ordinary MSSM.

\item[-] $v_W<\mu<M_{SUSY}$:  we have the particle content of the standard model, with the exception that there are four  Higgs doublets. We have therefore  the following $\beta$--function coefficients
\beq
\beta_{SU(3)_C} = -7\;, \qquad \beta_{SU(2)_L} = -8/3\;, \qquad \beta_{U(1)_Y} = 22/5\;.
\eeq
This should be compared with the (-7, -19/6, 41/10) coefficients of the ordinary SM.
\end{itemize}

\section{Yukawa running}
\label{app-Yuk}

As we already reported in section \ref{sec:Yuk}, because of the closeness of the intermediate scales between $M_{GUT}$ and $M_T$  we can consider only the running between $M_T$ and $M_{SUSY}$  to provide analytical approximations for the evolution of  fermion masses and mixing under the RGEs effect. At $M_T$ the scalar $SU(2)_L$ triplet has been already integrated out giving rise to the
effective Weinberg operator  given by
\beq
 \alpha_{ij} Y_{{L}_{rs}}\, \frac{ L_r  L_s   h_{u_i} h_{u_j}}{M_T}\,,
\eeq
with $h_{u_{1}}=h_u,h_{u_{2}}=h_{u}'$ and  $\alpha_{ij}$ coefficients arising by the scalar potential and
$Y_L$ given in \eq{YL}.

For what concerns the charged fermion Yukawa, at $M_T$ the Dirac part of the superpotential is written as
\bea
&&Y_u Q\, U^c h_u+  Y'_u Q\, U^c h_u' + Y_d Q\, D^c h_d+  Y'_d Q\, D^c h_d'+ Y_e L\, E^c h_d+  Y'_e L\, E^c h_d'\,,
\eea
with the Yukawa mass matrices given in \eq{Yuk}.

The Yukawa matrices RGEs  are therefore given by

\beq
\ba{rcl}
 16 \pi^2 \,\dfrac{d Y_u}{d t}&=& \left [ 3 \,Y_u Y_u^\dag +  Y_d Y_d^\dag+  3\, Y'_u Y_u^{'\dag} + Y'_d Y_d^{'\dag}  + 3\, \mbox{Tr} ( Y_u Y_u^\dag) - \Sigma_i c_i^u g_i^2 \right ] Y_u \,, \\[3mm]
 16 \pi^2 \,\dfrac{d Y'_u}{d t}&=&  \left [ 3 \,Y_u Y_u^\dag +  Y_d Y_d^\dag+  3\, Y'_u Y_u^{'\dag} + Y'_d Y_d^{'\dag}  + 3\, \mbox{Tr} ( Y'_u Y_u^{'\dag}) - \Sigma_i c_i^u g_i^2 \right ] Y'_u \,, \\[3mm]
 16 \pi^2 \,\dfrac{d Y_d}{d t}&=&  \left [ Y_u Y_u^\dag +3\,  Y_d Y_d^\dag+  Y'_u Y_u^{'\dag} + 3\,Y'_d Y_d^{'\dag}  + 3\, \mbox{Tr} ( Y_d Y_d^\dag)+  \mbox{Tr} ( Y_e Y_e^\dag) - \Sigma_i c_i^d g_i^2 \right ] Y_d \,, \\[3mm]
 16 \pi^2 \,\dfrac{d Y'_d}{d t}&=&\left [ Y_u Y_u^\dag +3\,  Y_d Y_d^\dag+  Y'_u Y_u^{'\dag} + 3\,Y'_d Y_d^{'\dag}  + 3\, \mbox{Tr} ( Y'_d Y_d^{'\dag}) +  \mbox{Tr} ( Y'_e Y_e^{'\dag}) - \Sigma_i c_i^d g_i^2 \right ] Y'_d \,, \\[3mm]
 16 \pi^2 \,\dfrac{d Y_e}{d t}&=&  \left [ 3\,  Y_e Y_e^\dag+  3\,Y'_e Y_e^{'\dag}  + 3\, \mbox{Tr} ( Y_d Y_d^\dag)+  \mbox{Tr} ( Y_e Y_e^\dag) - \Sigma_i c_i^e g_i^2 \right ] Y_e \,, \\[3mm]
 16 \pi^2 \,\dfrac{d Y'_e}{d t}&=&  \left [ 3\,  Y_e Y_e^\dag+  3\,Y'_e Y_e^{'\dag}  + 3\, \mbox{Tr} ( Y'_d Y_d^{'\dag})+  \mbox{Tr} ( Y'_e Y_e^{'\dag}) - \Sigma_i c_i^e g_i^2 \right ] Y'_e\,, \\[3mm]
 16 \pi^2 \,\dfrac{d   Y_L}{d t}&=&   \left [ (Y_e Y_e^\dag + Y'_e Y_e^{'\dag})  Y_L  +Y_L (Y_e Y_e^\dag+  Y'_e Y_e^{'\dag})^T  - \Sigma_i c_i^\nu g_i^2)Y_L \right ](\alpha_{11}+\alpha_{12}+\alpha_{22})\\[3mm]
 &&+  \left [ 6\, \mbox{Tr} ( Y_u Y_u^\dag)\alpha_{11} + 3\, \mbox{Tr} ( Y_u Y_u^\dag)\alpha_{12}+ 3\, \mbox{Tr} ( Y'_u Y_u^{'\dag})\alpha_{12}+ 6\, \mbox{Tr} ( Y'_u Y_u^{'\dag})\alpha_{22}  \right] Y_L  \,,
\ea
\eeq
with
\beq
\ba{cccccc}
 c_1^u = \dfrac{13}{15}\;, & \quad c_2^u =3 \;, & \quad c_3^u =\dfrac{16}{3}\;, &\qquad\quad
 c_1^d =\dfrac{7}{15}\;, &  \quad c_2^d = 3\;, & \quad c_3^d =\dfrac{16}{3}\;,\\[3mm]
 c_1^e = \dfrac{9}{5}\;, & \quad c_2^e =3 \;, & \quad c_3^e =0\;, &  \qquad\qquad
 c_1^\nu = \dfrac{6}{5}\;, & \quad c_2^\nu =6 \;, & \quad c_3^\nu =0\;.
\ea
\eeq



\begin{thebibliography}{99}

\bibitem{Fogli:Indication}
G.~L.~Fogli, E.~Lisi, A.~Marrone, A.~Palazzo and A.~M.~Rotunno,
  arXiv:0809.2936 [hep-ph];
G.~L.~Fogli, E.~Lisi, A.~Marrone, A.~Palazzo and A.~M.~Rotunno,
  Phys.\ Rev.\ Lett.\  {\bf 101} (2008) 141801
  [arXiv:0806.2649 [hep-ph]].



\bibitem{Maltoni:Indication}
T.~Schwetz, M.~Tortola and J.~W.~F.~Valle,
  New J.\ Phys.\  {\bf 10} (2008) 113011
  [arXiv:0808.2016 [hep-ph]];
M.~Maltoni and T.~Schwetz,
  arXiv:0812.3161 [hep-ph].



\bibitem{NeutrinoData}
A.~Strumia and F.~Vissani,
  arXiv:hep-ph/0606054;
G.~L.~Fogli {\it et al.},
  Nucl.\ Phys.\ Proc.\ Suppl.\  {\bf 168} (2007) 341;
M.~C.~Gonzalez-Garcia and M.~Maltoni,
  Phys.\ Rept.\  {\bf 460} (2008) 1
  [arXiv:0704.1800 [hep-ph]];
T.~Schwetz,
  AIP Conf.\ Proc.\  {\bf 981} (2008) 8
  [arXiv:0710.5027 [hep-ph]];
M.~C.~Gonzalez-Garcia and M.~Maltoni,
  Phys.\ Lett.\  B {\bf 663} (2008) 405
  [arXiv:0802.3699 [hep-ph]];
A.~Bandyopadhyay, S.~Choubey, S.~Goswami, S.~T.~Petcov and D.~P.~Roy,
  arXiv:0804.4857 [hep-ph].



\bibitem{doublechooz}
F.~Ardellier {\it et al.},
  arXiv:hep-ex/0405032;
F.~Ardellier {\it et al.} [Double Chooz Collaboration],
  arXiv:hep-ex/0606025.



\bibitem{dayabay}
Y.~f.~Wang,
  arXiv:hep-ex/0610024;
X.~Guo {\it et al.}  [Daya-Bay Collaboration],
  arXiv:hep-ex/0701029.



\bibitem{minos}
A.~B.~Pereira e Sousa,
  FERMILAB-THESIS-2005-67.



\bibitem{reno}
S.~B.~Kim  [RENO Collaboration],
  AIP Conf.\ Proc.\  {\bf 981}, 205 (2008)
  [J.\ Phys.\ Conf.\ Ser.\  {\bf 120}, 052025 (2008)];



\bibitem{T2K}
Y.~Itow {\it et al.}  [The T2K Collaboration],
  arXiv:hep-ex/0106019;



\bibitem{NOvA}
D.~S.~Ayres {\it et al.}  [NOvA Collaboration],
  arXiv:hep-ex/0503053.



\bibitem{LV_Theorem}
C.~I.~Low and R.~R.~Volkas,
  Phys.\ Rev.\  D {\bf 68}, 033007 (2003)
  [arXiv:hep-ph/0305243].



\bibitem{FeruglioSymBreaking}
F.~Feruglio,
  Nucl.\ Phys.\ Proc.\ Suppl.\  {\bf 143} (2005) 184
  [Nucl.\ Phys.\ Proc.\ Suppl.\  {\bf 145} (2005) 225]
  [arXiv:hep-ph/0410131].



\bibitem{HPS}
P.~F.~Harrison, D.~H.~Perkins and W.~G.~Scott,
  Phys.\ Lett.\ B {\bf 530} (2002) 167
  [arXiv:hep-ph/0202074];
P.~F.~Harrison and W.~G.~Scott,
  Phys.\ Lett.\ B {\bf 535} (2002) 163
  [arXiv:hep-ph/0203209];
Z.~z.~Xing,
  Phys.\ Lett.\ B {\bf 533} (2002) 85
  [arXiv:hep-ph/0204049];
P.~F.~Harrison and W.~G.~Scott,
  Phys.\ Lett.\ B {\bf 547} (2002) 219
  [arXiv:hep-ph/0210197];
P.~F.~Harrison and W.~G.~Scott,
  Phys.\ Lett.\ B {\bf 557} (2003) 76
  [arXiv:hep-ph/0302025];
P.~F.~Harrison and W.~G.~Scott,
  {\it Status of tri- / bi-maximal neutrino mixing},
  arXiv:hep-ph/0402006;
P.~F.~Harrison and W.~G.~Scott,
  arXiv:hep-ph/0403278.
  
  

\bibitem{BMmixing}
V.~D.~Barger, S.~Pakvasa, T.~J.~Weiler and K.~Whisnant,
  Phys.\ Lett.\  B {\bf 437}, 107 (1998)
  [arXiv:hep-ph/9806387];
Y.~Nomura and T.~Yanagida,
  Phys.\ Rev.\  D {\bf 59} (1999) 017303
  [arXiv:hep-ph/9807325];
G.~Altarelli and F.~Feruglio,
  Phys.\ Lett.\  B {\bf 439} (1998) 112
  [arXiv:hep-ph/9807353].


\bibitem{A4I}
E.~Ma and G.~Rajasekaran,
  Phys.\ Rev.\ D {\bf 64} (2001) 113012
  [arXiv:hep-ph/0106291];
E.~Ma,
  Mod.\ Phys.\ Lett.\ A {\bf 17} (2002) 627
  [arXiv:hep-ph/0203238];
K.~S.~Babu, E.~Ma and J.~W.~F.~Valle,
  Phys.\ Lett.\ B {\bf 552} (2003) 207
  [arXiv:hep-ph/0206292];
M.~Hirsch, J.~C.~Romao, S.~Skadhauge, J.~W.~F.~Valle and A.~Villanova del Moral,
  arXiv:hep-ph/0312244;
  Phys.\ Rev.\  D {\bf 69} (2004) 093006
  [arXiv:hep-ph/0312265];
E.~Ma,
  Phys.\ Rev.\ D {\bf 70} (2004) 031901;
  Phys.\ Rev.\ D {\bf 70} (2004) 031901
  [arXiv:hep-ph/0404199];
  New J.\ Phys.\  {\bf 6} (2004) 104
  [arXiv:hep-ph/0405152];
  arXiv:hep-ph/0409075;
G.~Altarelli and F.~Feruglio,
  Nucl.\ Phys.\ B {\bf 720} (2005) 64
  [arXiv:hep-ph/0504165].
S.~L.~Chen, M.~Frigerio and E.~Ma,
  Nucl.\ Phys.\  B {\bf 724} (2005) 423
  [arXiv:hep-ph/0504181];
E.~Ma,
  Phys.\ Rev.\  D {\bf 72} (2005) 037301
  [arXiv:hep-ph/0505209];
M.~Hirsch, A.~Villanova del Moral, J.~W.~F.~Valle and E.~Ma,
   Phys.\ Rev.\  D {\bf 72} (2005) 091301
   [Erratum-ibid.\  D {\bf 72} (2005) 119904]
   [arXiv:hep-ph/0507148];
K.~S.~Babu and X.~G.~He,
  arXiv:hep-ph/0507217;
E.~Ma,
  Mod.\ Phys.\ Lett.\ A {\bf 20} (2005) 2601
  [arXiv:hep-ph/0508099];
A.~Zee,
  Phys.\ Lett.\ B {\bf 630} (2005) 58
  [arXiv:hep-ph/0508278];
E.~Ma,
  Phys.\ Rev.\  D {\bf 73} (2006) 057304
  [arXiv:hep-ph/0511133];
G.~Altarelli and F.~Feruglio,
  Nucl.\ Phys.\  B {\bf 741} (2006) 215
  [arXiv:hep-ph/0512103].
X.~G.~He, Y.~Y.~Keum and R.~R.~Volkas,
  JHEP {\bf 0604}, 039 (2006)
  [arXiv:hep-ph/0601001];
B.~Adhikary, B.~Brahmachari, A.~Ghosal, E.~Ma and M.~K.~Parida,
  Phys.\ Lett.\ B {\bf 638} (2006) 345
  [arXiv:hep-ph/0603059];
E.~Ma,
  Mod.\ Phys.\ Lett.\  A {\bf 21} (2006) 2931
  [arXiv:hep-ph/0607190];
  Mod.\ Phys.\ Lett.\  A {\bf 22} (2007) 101
  [arXiv:hep-ph/0610342];
L.~Lavoura and H.~Kuhbock,
  Mod.\ Phys.\ Lett.\  A {\bf 22} (2007) 181
  [arXiv:hep-ph/0610050];
G.~Altarelli, F.~Feruglio and Y.~Lin,
  Nucl.\ Phys.\  B {\bf 775} (2007) 31
  [arXiv:hep-ph/0610165].
X.~G.~He,
  Nucl.\ Phys.\ Proc.\ Suppl.\  {\bf 168}, 350 (2007)
  [arXiv:hep-ph/0612080];
Y.~Koide,
  Eur.\ Phys.\ J.\  C {\bf 52} (2007) 617
  [arXiv:hep-ph/0701018];
M.~Hirsch, A.~S.~Joshipura, S.~Kaneko and J.~W.~F.~Valle,
  Phys.\ Rev.\ Lett.\  {\bf 99}, 151802 (2007)
  [arXiv:hep-ph/0703046].
F.~Yin,
  Phys.\ Rev.\  D {\bf 75} (2007) 073010
  [arXiv:0704.3827 [hep-ph]];
F.~Bazzocchi, S.~Kaneko and S.~Morisi,
  JHEP {\bf 0803} (2008) 063
  [arXiv:0707.3032 [hep-ph]];
F.~Bazzocchi, S.~Morisi and M.~Picariello,
  Phys.\ Lett.\  B {\bf 659} (2008) 628
  [arXiv:0710.2928 [hep-ph]];



\bibitem{A4II}
G.~Altarelli,
  arXiv:0711.0161 [hep-ph];
A.~C.~B.~Machado and V.~Pleitez,
  Phys.\ Lett.\  B {\bf 674}, 223 (2009)
  [arXiv:0712.0781 [hep-ph]];
M.~Honda and M.~Tanimoto,
  Prog.\ Theor.\ Phys.\  {\bf 119} (2008) 583
  [arXiv:0801.0181 [hep-ph]];
B.~Brahmachari, S.~Choubey and M.~Mitra,
  Phys.\ Rev.\  D {\bf 77} (2008) 073008
  [Erratum-ibid.\  D {\bf 77} (2008) 119901]
  [arXiv:0801.3554 [hep-ph]];
B.~Adhikary and A.~Ghosal,
  Phys.\ Rev.\  D {\bf 78}, 073007 (2008)
  [arXiv:0803.3582 [hep-ph]];
M.~Hirsch, S.~Morisi and J.~W.~F.~Valle,
  Phys.\ Rev.\  D {\bf 78} (2008) 093007
  [arXiv:0804.1521 [hep-ph]];
Y.~Lin,
  Nucl.\ Phys.\  B {\bf 813}, 91 (2009)
  [arXiv:0804.2867 [hep-ph]];
C.~Csaki, C.~Delaunay, C.~Grojean and Y.~Grossman,
  JHEP {\bf 0810}, 055 (2008)
  [arXiv:0806.0356 [hep-ph]];
P.~H.~Frampton and S.~Matsuzaki,
  arXiv:0806.4592 [hep-ph];
E.~E.~Jenkins and A.~V.~Manohar,
  Phys.\ Lett.\  B {\bf 668}, 210 (2008)
  [arXiv:0807.4176 [hep-ph]];
Riazuddin,
  arXiv:0809.3648 [hep-ph];
W.~Grimus and L.~Lavoura,
  JHEP {\bf 0904}, 013 (2009)
  [arXiv:0811.4766 [hep-ph]];
S.~Morisi,
  arXiv:0901.1080 [hep-ph];
M.~C.~Chen and S.~F.~King,
  JHEP {\bf 0906}, 072 (2009)
  [arXiv:0903.0125 [hep-ph]];
G.~C.~Branco, R.~G.~Felipe, M.~N.~Rebelo and H.~Serodio,
  arXiv:0904.3076 [hep-ph];
G.~Altarelli and D.~Meloni,
  J.\ Phys.\ G {\bf 36}, 085005 (2009)
  [arXiv:0905.0620 [hep-ph]];
M.~Hirsch, S.~Morisi and J.~W.~F.~Valle,
  arXiv:0905.3056 [hep-ph];
T.~J.~Burrows and S.~F.~King,
  arXiv:0909.1433 [hep-ph];
S.~Morisi and E.~Peinado,
  arXiv:0910.4389 [hep-ph];
G.~J.~Ding and J.~F.~Liu,
  arXiv:0911.4799 [Unknown];



\bibitem{S4}
S.~Pakvasa and H.~Sugawara,
    Phys.\ Lett.\  B {\bf 82} (1979) 105;
T.~Brown, N.~Deshpande, S.~Pakvasa and H.~Sugawara,
    Phys.\ Lett.\  B {\bf 141} (1984) 95;
T.~Brown, S.~Pakvasa, H.~Sugawara and Y.~Yamanaka,
    Phys.\ Rev.\  D {\bf 30} (1984) 255;
D.~G.~Lee and R.~N.~Mohapatra,
    Phys.\ Lett.\  B {\bf 329} (1994) 463
    [arXiv:hep-ph/9403201];
R.~N.~Mohapatra, M.~K.~Parida and G.~Rajasekaran,
  Phys.\ Rev.\  D {\bf 69} (2004) 053007
  [arXiv:hep-ph/0301234];
E.~Ma,
    Phys.\ Lett.\  B {\bf 632} (2006) 352
    [arXiv:hep-ph/0508231];
C.~Hagedorn, M.~Lindner and R.~N.~Mohapatra,
    JHEP {\bf 0606} (2006) 042
    [arXiv:hep-ph/0602244];
Y.~Cai and H.~B.~Yu,
    Phys.\ Rev.\  D {\bf 74} (2006) 115005
    [arXiv:hep-ph/0608022];
F.~Caravaglios and S.~Morisi,
    Int.\ J.\ Mod.\ Phys.\  A {\bf 22} (2007) 2469
    [arXiv:hep-ph/0611078];
H.~Zhang,
    Phys.\ Lett.\  B {\bf 655} (2007) 132
    [arXiv:hep-ph/0612214];
Y.~Koide,
    JHEP {\bf 0708} (2007) 086
    [arXiv:0705.2275 [hep-ph]];
M.~K.~Parida,
  Phys.\ Rev.\  D {\bf 78} (2008) 053004
  [arXiv:0804.4571 [hep-ph]];
F.~Bazzocchi and S.~Morisi,
  Phys.\ Rev.\  D {\bf 80} (2009) 096005
  [arXiv:0811.0345 [hep-ph]];
H.~Ishimori, Y.~Shimizu and M.~Tanimoto,
    arXiv:0812.5031 [hep-ph].
F.~Bazzocchi, L.~Merlo and S.~Morisi,
  Nucl.\ Phys.\  B {\bf 816} (2009) 204
  [arXiv:0901.2086 [hep-ph]].
F.~Bazzocchi, L.~Merlo and S.~Morisi,
  Phys.\ Rev.\  D {\bf 80} (2009) 053003
  [arXiv:0902.2849 [hep-ph]];
G.~J.~Ding,
  Nucl.\ Phys.\  B {\bf 827} (2010) 82
  [arXiv:0909.2210 [hep-ph]].



\bibitem{Tp}
A.~Aranda, C.~D.~Carone and R.~F.~Lebed,
  Phys.\ Lett.\ B {\bf 474}, 170 (2000)
  [arXiv:hep-ph/9910392];
A.~Aranda, C.~D.~Carone and R.~F.~Lebed,
  Phys.\ Rev.\ D {\bf 62}, 016009 (2000)
  [arXiv:hep-ph/0002044].
F.~Feruglio, C.~Hagedorn, Y.~Lin and L.~Merlo,
  Nucl.\ Phys.\  B {\bf 775} (2007) 120
  [arXiv:hep-ph/0702194];
M.~C.~Chen and K.~T.~Mahanthappa,
  Phys.\ Lett.\  B {\bf 652} (2007) 34
  [arXiv:0705.0714 [hep-ph]];
P.~H.~Frampton and T.~W.~Kephart,
  JHEP {\bf 0709} (2007) 110
  [arXiv:0706.1186 [hep-ph]];
A.~Aranda,
  Phys.\ Rev.\  D {\bf 76} (2007) 111301
  [arXiv:0707.3661 [hep-ph]];
P.~D.~Carr and P.~H.~Frampton,
  arXiv:hep-ph/0701034;
G.~J.~Ding,
  arXiv:0803.2278 [hep-ph];
P.~H.~Frampton and S.~Matsuzaki,
  arXiv:0902.1140 [hep-ph].
M.~C.~Chen and K.~T.~Mahanthappa,
  Phys.\ Lett.\  B {\bf 681} (2009) 444
  [arXiv:0904.1721 [hep-ph]].



\bibitem{SymmDelta27}
E.~Ma,
  Mod.\ Phys.\ Lett.\  A {\bf 21} (2006) 1917
  [arXiv:hep-ph/0607056].
I.~de Medeiros Varzielas, S.~F.~King and G.~G.~Ross,
  Phys.\ Lett.\  B {\bf 648} (2007) 201
  [arXiv:hep-ph/0607045].
I.~de Medeiros Varzielas, S.~F.~King and G.~G.~Ross,
  Phys.\ Lett.\  B {\bf 644} (2007) 153
  [arXiv:hep-ph/0512313].



\bibitem{FlavorGUT}
Z.~Berezhiani,
  Nucl.\ Phys.\ Proc.\ Suppl.\  {\bf 52A}, 153 (1997)
  [arXiv:hep-ph/9607363];
C.~H.~Albright,
  in {\it Proc. of the APS/DPF/DPB Summer Study on the Future of Particle Physics (Snowmass 2001) } ed. N.~Graf,
  {\it In the Proceedings of APS / DPF / DPB Summer Study on the Future of Particle Physics (Snowmass 2001), Snowmass, Colorado, 30 Jun - 21
  Jul 2001, pp P203}
  [arXiv:hep-ph/0110259];
I.~de Medeiros Varzielas and G.~G.~Ross,
  Nucl.\ Phys.\  B {\bf 733}, 31 (2006)
  [arXiv:hep-ph/0507176];
S.~Morisi,
  arXiv:hep-ph/0605167;
F.~Caravaglios and S.~Morisi,
  {\it Prepared for IFAE 2006 (in Italian), Pavia, Italy, 19-21 Apr 2006};
I.~de Medeiros Varzielas, S.~F.~King and G.~G.~Ross,
  Phys.\ Lett.\  B {\bf 648}, 201 (2007)
  [arXiv:hep-ph/0607045];
S.~F.~King and M.~Malinsky,
  JHEP {\bf 0611}, 071 (2006)
  [arXiv:hep-ph/0608021];
Y.~Cai and H.~B.~Yu,
  Phys.\ Rev.\  D {\bf 74}, 115005 (2006)
  [arXiv:hep-ph/0608022];
F.~Caravaglios and S.~Morisi,
  Int.\ J.\ Mod.\ Phys.\  A {\bf 22}, 2469 (2007)
  [arXiv:hep-ph/0611078];
J.~R.~Ellis, M.~E.~Gomez and S.~Lola,
  JHEP {\bf 0707}, 052 (2007)
  [arXiv:hep-ph/0612292];
M.~Albrecht, W.~Altmannshofer, A.~J.~Buras, D.~Guadagnoli and D.~M.~Straub,
  JHEP {\bf 0710}, 055 (2007)
  [arXiv:0707.3954 [hep-ph]];
W.~Altmannshofer,
  arXiv:0710.1488 [hep-ph];
F.~Bazzocchi, S.~Morisi and M.~Picariello,
  Phys.\ Lett.\  B {\bf 659}, 628 (2008)
  [arXiv:0710.2928 [hep-ph]];
F.~Bazzocchi, S.~Morisi, M.~Picariello and E.~Torrente-Lujan,
  J.\ Phys.\ G {\bf 36}, 015002 (2009)
  [arXiv:0802.1693 [hep-ph]];
C.~Hagedorn, M.~A.~Schmidt and A.~Y.~Smirnov,
  Phys.\ Rev.\  D {\bf 79}, 036002 (2009)
  [arXiv:0811.2955 [hep-ph]];
H.~Ishimori, Y.~Shimizu and M.~Tanimoto,
  Prog.\ Theor.\ Phys.\  {\bf 121}, 769 (2009)
  [arXiv:0812.5031 [hep-ph]];
F.~Bazzocchi and I.~de Medeiros Varzielas,
  Phys.\ Rev.\  D {\bf 79}, 093001 (2009)
  [arXiv:0902.3250 [hep-ph]].
M.~Ishiduki, S.~G.~Kim, N.~Maekawa and K.~Sakurai,
  Phys.\ Rev.\  D {\bf 80}, 115011 (2009)
  [Erratum-ibid.\  D {\bf 81}, 039901 (2010)]
  [arXiv:0910.1336 [hep-ph]];



\bibitem{A4TBGUT}
S.~F.~King and M.~Malinsky,
  Phys.\ Lett.\  B {\bf 645} (2007) 351
  [arXiv:hep-ph/0610250];
S.~Morisi, M.~Picariello and E.~Torrente-Lujan,
  Phys.\ Rev.\  D {\bf 75} (2007) 075015
  [arXiv:hep-ph/0702034];
G.~Altarelli, F.~Feruglio and C.~Hagedorn,
  JHEP {\bf 0803} (2008) 052
  [arXiv:0802.0090 [hep-ph]];
F.~Bazzocchi, S.~Morisi, M.~Picariello and E.~Torrente-Lujan,
  J.\ Phys.\ G {\bf 36} (2009) 015002
  [arXiv:0802.1693 [hep-ph]];
F.~Bazzocchi, M.~Frigerio and S.~Morisi,
  Phys.\ Rev.\  D {\bf 78}, 116018 (2008)
  [arXiv:0809.3573 [hep-ph]];
P.~Ciafaloni, M.~Picariello, E.~Torrente-Lujan and A.~Urbano,
  Phys.\ Rev.\  D {\bf 79}, 116010 (2009)
  [arXiv:0901.2236 [hep-ph]].
C.~Hagedorn, S.~F.~King and C.~Luhn,
  arXiv:1003.4249 [hep-ph].
  
  



\bibitem{Lin_LargeReactor}
  Y.~Lin,
  arXiv:0905.3534 [hep-ph].


\bibitem{Complementarity}
P.~H.~Frampton, S.~T.~Petcov and W.~Rodejohann,
  Nucl.\ Phys.\  B {\bf 687} (2004) 31
  [arXiv:hep-ph/0401206];
G.~Altarelli, F.~Feruglio and I.~Masina,
  Nucl.\ Phys.\  B {\bf 689} (2004) 157
  [arXiv:hep-ph/0402155];
M.~Raidal,
  Phys.\ Rev.\ Lett.\  {\bf 93} (2004) 161801
  [arXiv:hep-ph/0404046];
H.~Minakata and A.~Y.~Smirnov,
  Phys.\ Rev.\  D {\bf 70} (2004) 073009
  [arXiv:hep-ph/0405088];
P.~H.~Frampton and R.~N.~Mohapatra,
  JHEP {\bf 0501}, 025 (2005), hep-ph/0407139;
J.~Ferrandis and S.~Pakvasa,
  Phys.\ Rev.\ D {\bf 71} (2005) 033004
  [arXiv:hep-ph/0412038];
S.~K.~Kang, C.~S.~Kim and J.~Lee,
  arXiv:hep-ph/0501029;
N.~Li and B.~Q.~Ma,
  arXiv:hep-ph/0501226;
K.~Cheung, S.~K.~Kang, C.~S.~Kim and J.~Lee,
  arXiv:hep-ph/0503122;
Z.~z.~Xing,
  arXiv:hep-ph/0503200;
A.~Datta, L.~Everett and P.~Ramond,
  arXiv:hep-ph/0503222;
S.~Antusch, S.~F.~King and R.~N.~Mohapatra,
  arXiv:hep-ph/0504007;
M.~Lindner, M.~A.~Schmidt and A.~Y.~Smirnov,
  arXiv:hep-ph/0505067;
H.~Minakata,
  arXiv:hep-ph/0505262;
T.~Ohlsson,
  arXiv:hep-ph/0506094;
A.~Dighe, S.~Goswami and P.~Roy,
  Phys.\ Rev.\  D {\bf 73} (2006) 071301
  [arXiv:hep-ph/0602062];
B.~C.~Chauhan, M.~Picariello, J.~Pulido and E.~Torrente-Lujan,
  Eur.\ Phys.\ J.\  C {\bf 50} (2007) 573
  [arXiv:hep-ph/0605032];
K.~A.~Hochmuth and W.~Rodejohann,
  Phys.\ Rev.\  D {\bf 75} (2007) 073001
  [arXiv:hep-ph/0607103];
M.~A.~Schmidt and A.~Y.~Smirnov,
  Phys.\ Rev.\  D {\bf 74} (2006) 113003
  [arXiv:hep-ph/0607232];
F.~Plentinger, G.~Seidl and W.~Winter,
  Nucl.\ Phys.\  B {\bf 791} (2008) 60
  [arXiv:hep-ph/0612169];
F.~Plentinger, G.~Seidl and W.~Winter,
  Phys.\ Rev.\  D {\bf 76} (2007) 113003
  [arXiv:0707.2379 [hep-ph]].




\bibitem{AFM_BimaxS4}
  G.~Altarelli, F.~Feruglio and L.~Merlo,
  JHEP {\bf 0905}, 020 (2009)
  [arXiv:0903.1940 [hep-ph]].





\bibitem{PS}
J.~C.~Pati and A.~Salam,
  Phys.\ Rev.\  D {\bf 10} (1974) 275
  [Erratum-ibid.\  D {\bf 11} (1975) 703];
V.~Elias and A.~R.~Swift,
  Phys.\ Rev.\  D {\bf 13}, 2083 (1976);
V.~Elias,
  Phys.\ Rev.\  D {\bf 14}, 1896 (1976);
V.~Elias,
  Phys.\ Rev.\  D {\bf 16}, 1586 (1977);
D.~M.~Capper,
  J.\ Phys.\ G {\bf 3}, 1317 (1977);
A.~Schorr,
  Phys.\ Rev.\  D {\bf 18}, 863 (1978);
B.~Bhuyan and B.~B.~Deo,
  Phys.\ Rev.\  D {\bf 36}, 966 (1987);
G.~K.~Leontaris and J.~Rizos,
  Phys.\ Lett.\  B {\bf 510}, 295 (2001)
  [arXiv:hep-ph/0012255];
T.~Blazek, S.~F.~King and J.~K.~Parry,
  JHEP {\bf 0305}, 016 (2003)
  [arXiv:hep-ph/0303192];
A.~Prikas and N.~D.~Tracas,
  New J.\ Phys.\  {\bf 5}, 144 (2003)
  [arXiv:hep-ph/0303258];
C.~S.~Aulakh and A.~Girdhar,
  Nucl.\ Phys.\  B {\bf 711}, 275 (2005)
  [arXiv:hep-ph/0405074];
J.~B.~Dent and T.~W.~Kephart,
  Phys.\ Rev.\  D {\bf 77}, 115008 (2008)
  [arXiv:0705.1995 [hep-ph]];
F.~Braam, J.~Reuter and D.~Wiesler,
  arXiv:0909.3081 [hep-ph].

\bibitem{SU5}
For a general review on $SU(5)$, see\\
C.~Kounnas, A.~Masiero, D.~V.~Nanopoulos and K.~A.~Olive,
 ``Grand Unification With And Without Supersymmetry And Cosmological
 Implications,''
{\it  Singapore, Singapore: World Scientific ( 1984) 425 P. ( International School For Advanced Studies Lecture Series, 2)};\\
  G.~G.~Ross,
``Grand Unified Theories,''
{\it  Reading, Usa: Benjamin/cummings ( 1984) 497 P. ( Frontiers In Physics, 60)}.


\bibitem{TypeII&GUT}
A.~S.~Joshipura, B.~P.~Kodrani and K.~M.~Patel,
  Phys.\ Rev.\  D {\bf 79} (2009) 115017
  [arXiv:0903.2161 [hep-ph]];
B.~Dutta, Y.~Mimura and R.~N.~Mohapatra,
  Phys.\ Rev.\  D {\bf 80}, 095021 (2009)
  [arXiv:0910.1043 [Unknown]];
B.~Dutta, Y.~Mimura and R.~N.~Mohapatra,
  arXiv:0911.2242 [hep-ph];
S.~F.~King and C.~Luhn,
  arXiv:0912.1344 [Unknown].



\bibitem{Bertolini}
S.~Bertolini, T.~Schwetz and M.~Malinsky,
  Phys.\ Rev.\  D {\bf 73}, 115012 (2006)
  [arXiv:hep-ph/0605006].
  
  
  
\bibitem{Bajc}
B.~Bajc, I.~Dorsner and M.~Nemevsek,
  JHEP {\bf 0811} (2008) 007
  [arXiv:0809.1069 [hep-ph]],
A.~Melfo, A.~Ramirez and G.~Senjanovic,
  arXiv:1005.0834 [Unknown].




\bibitem{FN}
C.~D.~Froggatt and H.~B.~Nielsen,
  {\it Hierarchy Of Quark Masses, Cabibbo Angles And CP Violation},
  Nucl. Phys. B {\bf 147} (1979) 277.



\bibitem{GJrelation}
  H.~Georgi and C.~Jarlskog,
  Phys.\ Lett.\  B {\bf 86}, 297 (1979).



\bibitem{Wolfenstein}
  L.~Wolfenstein,
  {\it parameterization Of The Kobayashi-Maskawa Matrix},
  Phys.\ Rev.\ Lett.\  {\bf 51} (1983) 1945.



\bibitem{DrivingFields}

G.~G.~Ross and O.~Vives,
  Phys.\ Rev.\  D {\bf 67} (2003) 095013
  [arXiv:hep-ph/0211279].
G.~G.~Ross, L.~Velasco-Sevilla and O.~Vives,
  Nucl.\ Phys.\  B {\bf 692} (2004) 50
  [arXiv:hep-ph/0401064];
F.~Feruglio, C.~Hagedorn and L.~Merlo,
  arXiv:0910.4058 [hep-ph], accepted for publication in JHEP.



\bibitem{LFVA4}
F.~Feruglio, C.~Hagedorn, Y.~Lin and L.~Merlo,
  Nucl.\ Phys.\  B {\bf 809} (2009) 218
  [arXiv:0807.3160 [hep-ph]];
H.~Ishimori, T.~Kobayashi, Y.~Omura and M.~Tanimoto,
  JHEP {\bf 0812} (2008) 082
  [arXiv:0807.4625 [hep-ph]];
F.~Feruglio, C.~Hagedorn, Y.~Lin and L.~Merlo,
  in the Proceedings of NO-VE 2008, 4th International Workshop on Neutrino Oscillations in Venice (Venice,
  Italy, 2008), University of Padua publication (Papergraf Edition, Padua, 2008), 29-43
  [arXiv:0808.0812 [hep-ph]];
A.~Hayakawa, H.~Ishimori, Y.~Shimizu and M.~Tanimoto,
  Phys.\ Lett.\  B {\bf 680} (2009) 334
  [arXiv:0904.3820 [hep-ph]];
C.~Hagedorn, E.~Molinaro and S.~T.~Petcov,
  JHEP {\bf 1002} (2010) 047
  [arXiv:0911.3605 [Unknown]];
F.~Feruglio, C.~Hagedorn, Y.~Lin and L.~Merlo,
  arXiv:0911.3874 [hep-ph], accepted for publication in Nucl.\ Phys.\ B .



\bibitem{PS_MelfoSenj}
A.~Melfo and G.~Senjanovic,
  Phys.\ Rev.\  D {\bf 68}, 035013 (2003)
  [arXiv:hep-ph/0302216].



\bibitem{DeltasCharged}
Z.~Chacko and R.~N.~Mohapatra,
  Phys.\ Rev.\  D {\bf 58} (1998) 015003
  [arXiv:hep-ph/9712359];
M.~Frank, K.~Huitu and S.~K.~Rai,
  Phys.\ Rev.\  D {\bf 77}, 015006 (2008)
  [arXiv:0710.2415 [hep-ph]].
D.~A.~Demir, M.~Frank, K.~Huitu, S.~K.~Rai and I.~Turan,
  Phys.\ Rev.\  D {\bf 78} (2008) 035013
  [arXiv:0805.4202 [hep-ph]].;
D.~A.~Demir, M.~Frank, D.~K.~Ghosh, K.~Huitu, S.~K.~Rai and I.~Turan,
  Phys.\ Rev.\  D {\bf 79} (2009) 095006
  [arXiv:0903.3955 [hep-ph]].



\bibitem{RunningNu}
S.~Antusch, J.~Kersten, M.~Lindner and M.~Ratz,
  Phys.\ Lett.\  B {\bf 544}, 1 (2002)
  [arXiv:hep-ph/0206078];
S.~Antusch, J.~Kersten, M.~Lindner and M.~Ratz,
  Nucl.\ Phys.\  B {\bf 674} (2003) 401
  [arXiv:hep-ph/0305273];
T.~Miura, T.~Shindou and E.~Takasugi,
  Phys.\ Rev.\  D {\bf 68} (2003) 093009
  [arXiv:hep-ph/0308109];
S.~Antusch, P.~Huber, J.~Kersten, T.~Schwetz and W.~Winter,
  Phys.\ Rev.\  D {\bf 70} (2004) 097302
  [arXiv:hep-ph/0404268];
S.~Antusch, J.~Kersten, M.~Lindner, M.~Ratz and M.~A.~Schmidt,
  JHEP {\bf 0503} (2005) 024
  [arXiv:hep-ph/0501272];
J.~W.~Mei,
  Phys.\ Rev.\  D {\bf 71} (2005) 073012
  [arXiv:hep-ph/0502015];
M.~Lindner, M.~A.~Schmidt and A.~Y.~Smirnov,
  JHEP {\bf 0507} (2005) 048
  [arXiv:hep-ph/0505067];
J.~R.~Ellis, A.~Hektor, M.~Kadastik, K.~Kannike and M.~Raidal,
  Phys.\ Lett.\  B {\bf 631}, 32 (2005)
  [arXiv:hep-ph/0506122];
A.~Dighe, S.~Goswami and P.~Roy,
  Phys.\ Rev.\  D {\bf 73}, 071301 (2006)
  [arXiv:hep-ph/0602062];
A.~Dighe, S.~Goswami and W.~Rodejohann,
  Phys.\ Rev.\  D {\bf 75} (2007) 073023
  [arXiv:hep-ph/0612328];
A.~Dighe, S.~Goswami and P.~Roy,
  Phys.\ Rev.\  D {\bf 76} (2007) 096005
  [arXiv:0704.3735 [hep-ph]];
S.~Boudjemaa and S.~F.~King,
  Phys.\ Rev.\  D {\bf 79} (2009) 033001
  [arXiv:0808.2782 [hep-ph]];
Y.~Lin, L.~Merlo and A.~Paris,
  arXiv:0911.3037 [Unknown].



\bibitem{Mali}
  M.~Malinsky,
  arXiv:0807.0591 [hep-ph].




\bibitem{gerda}
A.~A.~Smolnikov and f.~t.~G.~Collaboration,
  arXiv:0812.4194 [nucl-ex].



\bibitem{majorana}
Majorana Collaboration,
  arXiv:0811.2446 [nucl-ex].



\bibitem{supernemo}
H.~Ohsumi  [NEMO and SuperNEMO Collaborations],
  J.\ Phys.\ Conf.\ Ser.\  {\bf 120} (2008) 052054.



\bibitem{cuore}
A.~Giuliani  [CUORE Collaboration],
  J.\ Phys.\ Conf.\ Ser.\  {\bf 120} (2008) 052051.



\bibitem{exo}
M.~Danilov {\it et al.},
  Phys.\ Lett.\  B {\bf 480} (2000) 12
  [arXiv:hep-ex/0002003].




\end{thebibliography}
\end{document}